\documentclass{aa}

\usepackage{epsfig}
\usepackage{graphicx}
\usepackage{times}
\usepackage{color}
\usepackage{ulem}
\usepackage{hyperref}
\usepackage{url}

\usepackage{xcolor}
\hypersetup{
colorlinks,
linkcolor={red!80!black},
citecolor={blue!80!black},
urlcolor={blue!80!black}
}
\def\cm3{cm$^{-3}$}
\def\kms{km~s$^{-1}$}
\def\msunyr{M$_{\odot}$\,yr$^{-1}$}
\def\lsun{L$_{\odot}$}
\def\rsun{R$_{\odot}$}
\def\mdot{$\dot{\rm M}$}

\def\msun{M$_{\odot}$}

\def\one{\ts {\,\sc i}}
\def\two{\ts {\,\sc ii}}

\def\four{\ts {\,\sc iv}}
\def\five{\ts {\sc v}}
\def\six{\ts {\sc vi}}
\def\ndash{---}
\def\beq{\begin{equation}}
\def\eeq{\end{equation}}

\def\lesssim{\mathrel{\hbox{\rlap{\hbox{\lower4pt\hbox{$\sim$}}}\hbox{$<$}}}}
\def\gtrsim{\mathrel{\hbox{\rlap{\hbox{\lower4pt\hbox{$\sim$}}}\hbox{$>$}}}}

\def\lesssim{\mathrel{\hbox{\rlap{\hbox{\lower4pt\hbox{$\sim$}}}\hbox{$<$}}}}
\def\gtrsim{\mathrel{\hbox{\rlap{\hbox{\lower4pt\hbox{$\sim$}}}\hbox{$>$}}}}

\def\one{{\,\sc i}}
\def\two{{\,\sc ii}}

\def\four{{\,\sc iv}}
\def\five{{\sc v}}
\def\six{{\sc vi}}

\def\eight{{\sc viii}}

\def\v1d{{\sc v1d}}

\def\mesa{{\sc mesa}}
\def\cmfgen{{\sc cmfgen}}
\def\heracles{{\sc heracles}}

\def\ergs{erg\,s$^{-1}$}

\def\aj{AJ}

\def\apj{ApJ}
\def\apjs{ApJS}
\def\apjl{ApJL}
\def\aap{A\&A}
\def\araa{ARA\&A}
\def\aaps{A\&AS}
\def\mnras{MNRAS}
\def\nat{Nature}

\def\jqsrt{JQSRT}
\def\apss{Astrophysics and Space Science}

\begin{document}

\title{Explosion of red-supergiant stars: influence
  of the atmospheric structure on shock breakout and the early-time supernova radiation.}

\titlerunning{Early-time properties of Type II SNe}

\author{Luc Dessart\inst{\ref{inst1}}
  \and
  D. John Hillier\inst{\ref{inst2}}
  \and
  Edouard Audit\inst{\ref{inst3}}
  }

\institute{Unidad Mixta Internacional Franco-Chilena de Astronom\'ia (CNRS UMI 3386),
    Departamento de Astronom\'ia, Universidad de Chile,
    Camino El Observatorio 1515, Las Condes, Santiago, Chile\label{inst1}
    \and
    Department of Physics and Astronomy \& Pittsburgh Particle Physics,
    Astrophysics, and Cosmology Center (PITT PACC),  University of Pittsburgh,
    3941 O'Hara Street, Pittsburgh, PA 15260, USA.\label{inst2}
    \and
    Maison de la Simulation, CEA, CNRS, Universit\'e Paris-Sud, UVSQ,
    Universit\'e Paris-Saclay, 91191, Gif-sur-Yvette, France.\label{inst3}
  }

  \date{Accepted . Received }

\abstract{
Early-time observations of the Type II supernovae (SNe) 2013cu and 2013fs
have revealed an interaction of ejecta with material near the star surface.
Unlike the Type IIn SN\,2010jl, which interacts with a dense wind for $\sim$\,1\,yr,
the interaction ebbs after 2--3\,d, suggesting a dense and compact circumstellar envelope.
Here, we use multi-group radiation-hydrodynamics and non-local-thermodynamic-equilibrium
radiative transfer to explore the properties of red supergiant (RSG) star explosions embedded
in a variety of dense envelopes. We consider the cases of an extended static atmosphere or a
steady-state wind, adopting a range of mass loss rates.
The shock-breakout signal, the SN radiation up to 10\,d, and the ejecta dynamics are strongly
influenced by the properties of this nearby environment.  This compromises the use of early-time
observations to constrain $R_{\star}$. The presence of narrow lines for
2--3\,d in 2013fs and 2013cu require a cocoon of material of $\sim$\,0.01\,\msun\ out to 5-10$R_\star$.
Spectral lines evolve from electron-scattering to Doppler broadened, with a growing blueshift of
their emission peaks.
Recent studies propose a super-wind phase with a mass loss rate from 0.001 up
to 1\,\msunyr\ in the last months/years of the RSG life, although there is no
observational constraint that this external material is a steady-state outflow.
Alternatively, observations may be explained by the explosion of a RSG star
inside its complex atmosphere. Indeed, spatially resolved observations
reveal that RSG stars have extended atmospheres, with the presence of downflows and
upflows out to several $R_\star$, even in a standard RSG like Betelgeuse.  Mass loading
in the region intermediate between star and wind can accommodate the 0.01\,\msun\
needed to explain the observations of 2013fs. Signatures of interaction in early-time spectra of RSG
star explosions may therefore be the norm, not the exception, and a puzzling super-wind phase
prior to core-collapse may be superfluous.
}


\maketitle
\label{firstpage}


\section{Introduction}
\label{sect_intro}

   Radiation-hydrodynamic simulations of Type II supernovae (SNe) in the 1970s indicated
   that their progenitors were red-supergiant (RSG) stars \citep{grassberg_71,FA77}.
   More recently, the analysis of pre-explosion images
   has confirmed this \citep{smartt_09}.
   The RSG progenitor properties are however uncertain because of the limitation in our
   understanding of the physics of RSG envelopes (e.g., in connection to convection;
   see, e.g., \citealt{meakin_arnett_07,arnett_conv_10}),
   RSG atmospheres and environments (e.g., in connection to
   mass loss, dust/molecule formation, etc.;
   \citealt{josselin_rsg_07}), or their final core properties
   at collapse \citep{couch_3d_presn_15,mueller_3d_presn_16}.

   Single massive stars lose mass through a radiation-driven wind \citep{cak,dejager_mdot_88}.
   It is therefore unsurprising that a fraction of SNe exhibit signatures of ejecta interaction
   with circum-stellar material (CSM).
   The unambiguous signature of interaction with an optically-thick CSM moving at very slow speeds
   with respect to the SN ejecta is the observation of optical emission lines with
   narrow cores and broad electron-scattering wings.
    The frequency of such interactions is uncertain but
   is probably of the order of 5-10\% of Type II SNe (e.g., \citealt{smith_11_sn_stat}).
   Such events are classified as Type IIn SNe \citep{schlegel_90}.
   SNe IIn come, however, with a huge diversity of properties in light curves and spectra.
   Some show signatures of interaction for a few days (e.g., SN\,2013cu, \citealt{galyam_13cu_14}),
   a few weeks (SN\,1998S, \citealt{fassia_98S_00,leonard_98S_00}), or a few months (SN\,2010jl,
   \citealt{zhang_10jl_12,fransson_10jl}), together with a wide range of luminosities.
   Numerical simulations of interacting SNe connect this diversity to variations
   in CSM location and structure \citep{moriya_rsg_csm_11,D15_2n}.
   To avoid confusion, CSM refers here to circumstellar material detached from the star
   and typically located at large distances of 10--100\,$R_{\star}$.
   For events in which the signatures of interaction are seen on the discovery spectrum but disappear
   after a few hours or a few days, like in SN\,2013cu, we will
   refer to the atmosphere and/or wind  -- this material corresponds to the direct
   environment of the star, contiguous to the star surface (often referred to as the circumstellar
   envelope), rather than a standard CSM.

   SN\,2013fs is a recent example that illustrates the complexity and diversity of such interactions.
   This SN shows signatures of interaction for only $\sim$2\,d after explosion, while it resembles
   a very standard type II-P SN at $\gtrsim$10\,d (plateau light curve in visual bands;
   standard P-Cygni profiles for H\one\ lines; \citealt{yaron_13fs_17}).
   Signs of interaction have been observed at early times in other Type II SNe
   \citep{khazov_flash_16}, although most of these do not show the 100-day long
   plateau light curve of SNe II-P.
   A prototypical, well observed, event showing narrow line profiles for 10--20\,d after discovery
   and becoming a more standard Type II event with broad lines
   is SN\,1998S \citep{leonard_98S_00,fassia_98S_01}.
   Simulations suggest the early line broadening mechanism is dominated by electron scattering
   \citep{chugai_98S_01,shivvers_98S_15,D16_2n}. By 20\,d, the atmosphere/wind is optically thin and the
   spectrum forms in a fast expanding region with a steeply declining density profile (i.e.,
   at the interface between progenitor and atmosphere/wind), leading to a featureless (and blue) continuum.
   At later times during the photospheric phase, the spectrum exhibits a `standard' type II (non-interacting)
   SN spectrum with signs of line blanketing, P-Cygni profiles, and excess emission associated
   with the optically-thin cold-dense-shell.
   These three phases are well traced with radiation-hydrodynamic and non-local-thermodynamic-equilibrium
   (non-LTE) radiative-transfer
   modeling. In particular, the simultaneous presence of Doppler-broadened lines forming in the ejecta and
   narrow lines broadened by electron scattering in the atmosphere/wind is reproduced if
   the non-monotonicity of the velocity field is explicitly accounted for \citep{D16_2n}.

   The presence of an atmosphere/wind may also be inferred from the short-lived shock breakout signal.
   In a shock-driven explosion, the radiation-dominated shock will start to leak photons when the photon
   diffusion time at the shock $t_{\rm diff}= 3 \tau \Delta R / c$
   becomes comparable to the shock crossing time
   $t_{\rm shock}= \Delta R/ v_{\rm shock}$ through the outer layers of thickness $ \Delta R$
   of the progenitor star.\footnote{This diffusion time is representative for a random walk in
   a 3-D space of uniform density and constant photon-mean-free-path.}
   This shock breakout signal therefore starts at the optical
   depth $\tau \sim c / 3 v_{\rm shock}$.\footnote{Strictly
   speaking, the presence of a precursor is given by the local conditions at the shock. In the SN
   community, the term `precursor' is usually meant as the precursor to shock emergence and explosion,
   and it stands more broadly for the shock-breakout signal.}
   For the $\sim$10$^{51}$\,erg Type II SN explosions presented here, the shock speed is about
   5000\,\kms, so that the breakout signal will start from an optical depth of about 20.
   The actual duration of the shock breakout signal scales with $\Delta R$, which is the length scale between
   the photosphere and $\tau=$\,20. Depending on the RSG atmosphere/wind structure, this length scale
   may perhaps cover from 0.01 to 0.5$R_{\star}$.
   For a small scale height, the breakout signal is intrinsically short ($\lesssim$1000\,s).
   However, the timescale is stretched to at least $R_{\star}/c$, hence about 30\,min,
   for a distant observer. KSN\,2011d is the first type II SN to exhibit this shock breakout signal
    \citep{garnavich_sbo_16}, which indicates the absence of a dense wind or an extended atmosphere
    around its progenitor RSG.
   For a large scale height (as in the case of a dense wind), the duration
   can be hours, hence typically well in excess of $R_{\star}/c$.
   The observations of a 1-day long UV burst in PS1-13arp by \citet{gezari_sbo_15} suggests
   that the shock may have broken out into some dense and extended material. Observations
   as part of SHOOT by \citet{tanaka_sbo_16} reveal a very short optical rise time
   that supports such a phenomenon. By lengthening the duration of the shock-breakout signal,
   the presence of a dense atmosphere/wind may facilitate its detection.
   Although very challenging, high-cadence surveys with a high limiting magnitude can capture the fleeting
   moment of shock breakout.
   The HITS survey \citep{forster_sbo_16} or SHOOT \citep{tanaka_sbo_16}
   have not yet been successful, but the KEPLER mission has made
   one unambiguous detection \citep{garnavich_sbo_16} -- high cadence surveys that fail
   to capture shock breakout do reveal critical information on the early phase evolution
   of the SN radiation.

   Signatures of interaction with the atmosphere/wind
   may be short-lived and therefore generally missed by a delayed spectroscopic follow-up.
   Numerous events may therefore be classified as Type II-P (or II-L) while an earlier
   spectroscopic classification would have suggested a classification as Type IIn.
   One should however wonder whether a SN should be classified as a IIn because it shows
   narrow lines for a day or two, or as a II-P because it shows a 100-d long plateau with all the spectral
   characteristics of a standard SN II-P at all times past two days. Should the event be classified as
   a Type IIn if the narrow lines were seen for only one hour? This is not irrelevant. Once a SN
   has been classified as a Type IIn, it contributes to the number of events considered as interacting SNe,
   mixed in the sample with objects that show signs of interaction at all times, like SN\,2010jl.
   Hence, it has strong implications for the interpretation of SN rates.

   Based on the presence of early-time spectral signatures of interaction,
   \citet{khazov_flash_16} and \citet{yaron_13fs_17}
   propose that numerous SNe II-P may undergo a super-wind phase before core collapse.
   Recently,  \citet{morozova_2l_2p_17} argue for such super-wind phases and propose that the resulting
   ejecta/wind interaction may explain the difference between Type II-L and Type II-P light curve
   morphologies.
   Very early observations were obtained with the Type IIb SN\,2013cu, for which
   \citet{galyam_13cu_14} propose that the spectrum at 15.5\,hr forms in the dense
(\mdot$\gtrsim$0.03\,\msunyr) and fast (terminal velocity of 2500\,\kms) wind of a Wolf-Rayet progenitor star.
This interpretation rests partly on the idea that line broadening is caused by the Doppler effect.
Instead, line broadening is likely caused by electron scattering.
In this case, \citet{groh_13cu} propose a slow ($\lesssim$\,100\,\kms) and
dense (3$\times$10$^{-3}$\,\msunyr) wind.
This has been later refined by \citet{grafener_vink_13cu_16} who accounted for time delays.
They propose a significant increase on the wind mass (to be at least 0.3\,\msun), and
that the shock may be embedded within the optically-thick wind (whose outer edge is the photosphere)
for two weeks. However, the narrow line profiles
seen in SN\,2013cu are strong at 15.5\,hr after discovery, but weak at 3\,d and absent at 6\,d so the bulk of
the dense (optically-thick) material is probably swept-up already after a few days.
A total wind mass of 0.3\,\msun\ is close to what is found for SN\,1998S, and narrow lines profiles
are seen for much longer ($\gtrsim$\,10\,d). In any case, despite slight
differences in interpretation and modeling techniques, these studies argue for a super-wind phase
immediately prior to the core collapse of SN\,2013cu's progenitor.
   These observations question whether there is something fundamentally
   missing in the 1-D quasi-steady-state treatment of stellar evolution and stellar winds.
   The eruptions observed in some very massive stars like $\eta$ Car
   and the inference of ejecta interaction with a massive CSM in super-luminous SNe IIn
   give evidence in that direction.

   Here, we use 1-D radiation-hydrodynamics and 1-D non-LTE radiative transfer modeling to characterize
   the bolometric, photometric, and spectroscopic signatures of RSG explosions embedded
   in an atmosphere/wind of modest extent (within $\sim$\,10\,$R_{\star}$) and
   mass ($\lesssim$\,10$^{-1}$\,\msun).
   Previous studies that have focused on ejecta/wind interactions used either grey/multi-group
   radiation-hydrodynamics without spectral calculations \citep{moriya_rsg_csm_11,morozova_2l_2p_17},
   or spectral calculations without dynamics \citep{yaron_13fs_17}.

   The structure of the paper is as follows.
   In the next section, we discuss the properties of the environment of RSG stars.
   We then present our numerical approach in Section~\ref{sect_method}.
   In Section~\ref{sect_heracles}, we discuss the
   results from the multi-group radiation-hydrodynamics simulations with \heracles, focusing
   on the properties of the bolometric light curves and the dynamic structure of the
   interaction between the SN ejecta and the progenitor atmosphere/wind
   (velocity, density, temperature, optical depth). These
   simulations are post-processed with \cmfgen\ using a non-LTE steady-state approach
   that works for an arbitrary velocity field. The resulting multi-band light curves and spectra
   are discussed in Section~\ref{sect_cmfgen}.
   We compare our results to observations in Section~\ref{sect_obs}.
   We  present our conclusions in Section~\ref{sect_conc}.

\begin{figure}
\epsfig{file=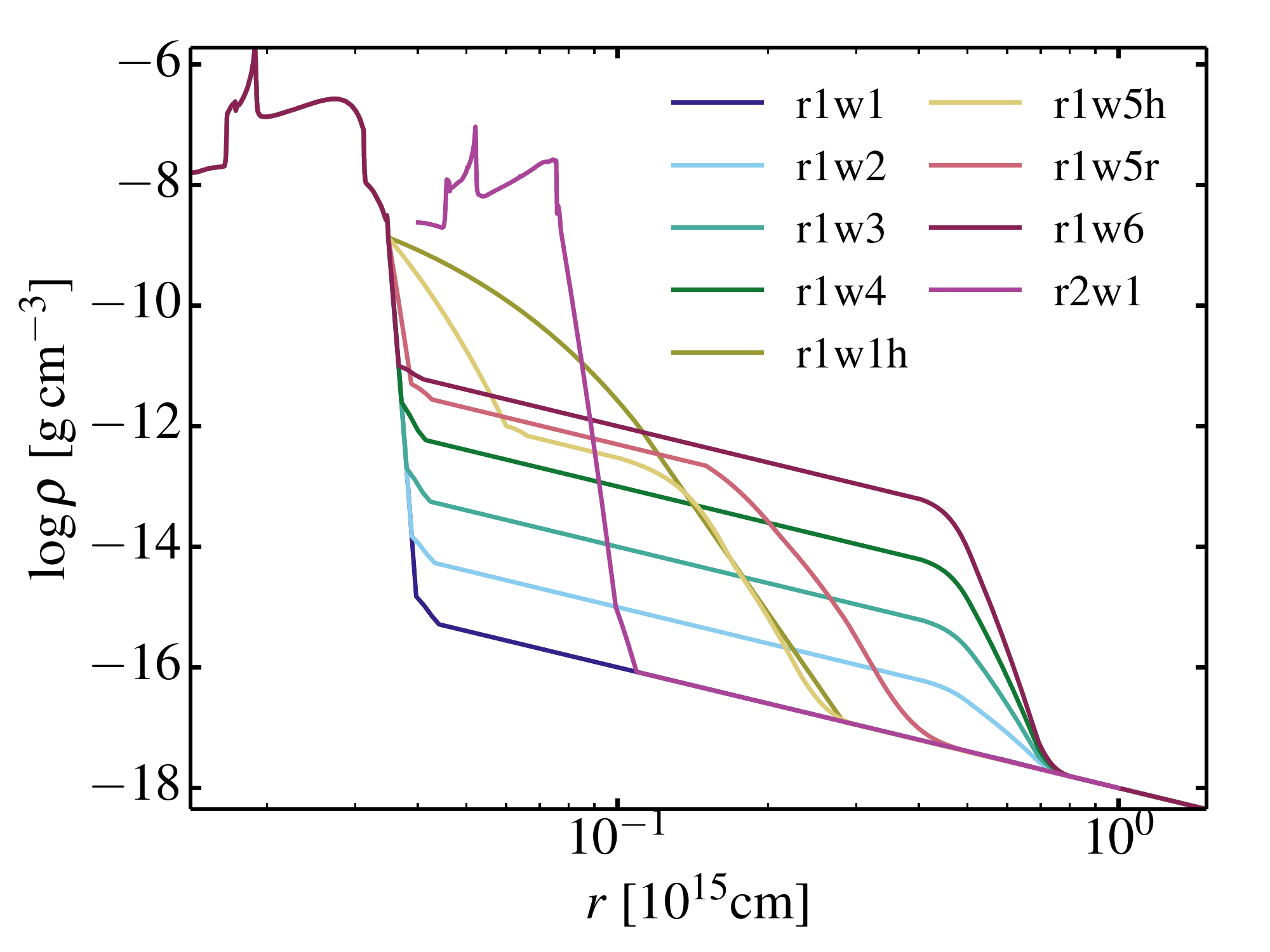,width=8.5cm}
\caption{Initial density structure used in \heracles\ for models
 r1w1, r1w2, r1w3, r1w4, r1w1h, r1w5h, r1w5r, r1w6, and r2w1.
 Model m15mlt3 (r1) is used for the ejecta of model r1w[1-6][h,p,r] and model m15mlt1 (r2)
 for the r2w1, taken at a few hours before shock breakout \citep{d13_sn2p}.
 The atmosphere/wind density is built using a variety of wind mass loss rates and atmospheric
 scale heights. All models transition to a wind mass loss rate of 10$^{-6}$\,\msunyr\ beyond
 2--5$\times$10$^{14}$\,cm ($\equiv$\,5--10\,$R_\star$).
 [See Section~\ref{sect_method} and Table~\ref{tab_sum} for details.]
\label{fig_init_set}
}
\end{figure}

\begin{figure*}
\epsfig{file=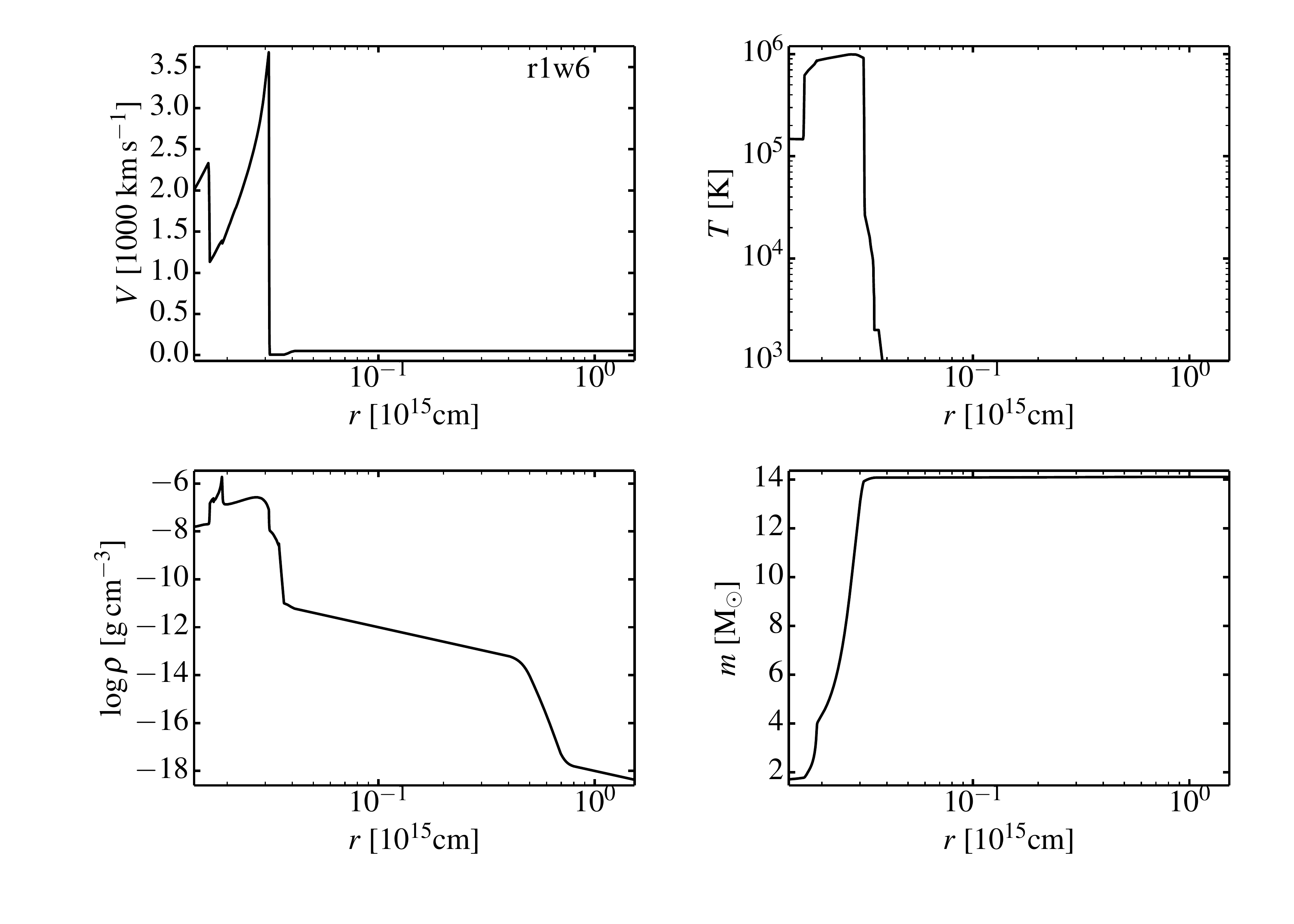,width=17cm}
\vspace{-0.8cm}
\caption{Initial structure used in \heracles\ for model r1w6. The time is a few
hours before shock breakout (the prior evolution of the shock through the stellar
interior was computed with \v1d; \citealt{d13_sn2p}).
\label{fig_init_r1w6}
}
\end{figure*}


\section{The environment of red-supergiant stars}
\label{sect_rsg}

    The complexity of RSG atmospheres prevents a physically consistent description
    of the density profile above the RSG surface radius (i.e., the location corresponding
    to $R_{\star}$ in a stellar evolution model, as produced for example in \mesa).
    RSG atmospheres are very complex, with density inhomogeneities, up-flows and down-flows
    seen out to several stellar radii
    \citep{kervella_betelgeuse_09,kervella_betelgeuse_11,ohnaka_betelgeuse_11,kervella_betelgeuse_16}.
    These observations suggest that there is no well defined atmosphere and that the wind is not
    launched cleanly from $R_{\star}$ but is instead slowly accelerated over an extended region.

    A RSG model computed by a stellar evolution code (here \mesa, but the point
    applies to all such codes) extends out to a minimum density of 10$^{-9}$\,g\,cm$^{-3}$, while
    the base density of a 10$^{-5}$\,\msunyr\ RSG wind is around 10$^{-14}$\,g\,cm$^{-3}$. There
    is therefore a density drop of 5 orders in magnitude between the $R_{\star}$ from \mesa\
    and the actual wind.
    It is in this intermediate region, whose properties are poorly known both observationally
    and theoretically,  that the SN spectrum will form during the first few days
    after explosion (between roughly 0.5 and 2$\times$\,10$^{14}$\,cm).
    In the present study, we assume that the distribution of this surrounding material
    is spherically symmetric around the star.

    In the next section, we present two types of density structures for the near environment
    of the RSG star: an atmosphere with an extended scale height and/or a dense wind.
    Early-time observations of Type II SNe do not allow one to constrain the pre-SN velocity structure
    of this material, hence cannot firmly establish the steady-state nature of a wind. They cannot either
    determine whether the atmospheric material is in hydrostatic equilibrium. This results in part from
    the limitations of observations, which may not have a high enough resolution. But it also stems physically
    from the acceleration produced by the intense radiation associated at shock breakout and after.
    Hence, these early-time SN observations only constrain
    the extent and density/mass of this material. Spatially-resolved observations of RSG clearly show
    that the environment of RSG is neither static nor steady-state
    \citep{kervella_betelgeuse_09,kervella_betelgeuse_11,ohnaka_betelgeuse_11,kervella_betelgeuse_16}.


\section{Initial conditions and methodology}
\label{sect_method}

\subsection{Initial conditions}

   The initial ejecta structure, corresponding to the RSG star explosion, is taken
   from \citet{d13_sn2p}. We choose models m15mlt3 (here renamed r1;
   $R_{\star}=$\,501\,\rsun) and m15mlt1 (here renamed r2; $R_{\star}=$\,1107\,\rsun)
   in order to cover explosions from compact and extended RSG stars.
    In practice, the initial ejecta structure (radius, velocity, density, temperature) is taken
    from the radiation hydrodynamic simulation (carried out with \v1d; see \citealt{d13_sn2p}
    for details) at $\lesssim$\,10$^4$\,s before shock breakout. In other words, the shock
    is within $\lesssim$\,100\,\rsun\ of the progenitor surface.

    When embedding the RSG start initial interior structure (the shocked and un-shocked envelope layers)
    in an atmosphere/wind, we adopt a variety of density structures, varying the wind mass loss rate \mdot,
    the atmospheric scale height  $H_\rho$, or both.
    In a first set of simulations, we add an atmosphere with a density
    scale height of 0.01\,$R_{\star}$ (this corresponds to the surface value
    in the \mesa\ model) until the density reaches down to the wind density
    corresponding to wind mass loss rates between 10$^{-6}$ and 10$^{-2}$\,\msunyr.
    In our nomenclature, model suffixes w1, w2, w3, w4, w5, w6 correspond to
    wind mass loss rates of 10$^{-6}$, 10$^{-5}$, 10$^{-4}$, 10$^{-3}$, 5$\times$\,10$^{-3}$,
    and 10$^{-2}$\,\msunyr.
    To limit the extent of the dense wind, we smoothly switch to a low mass loss rate of
    10$^{-6}$\,\msunyr\ beyond 5$\times$\,10$^{14}$\,cm, except for model r1w5r
    in which the transition to a low wind density is at  2$\times$\,10$^{14}$\,cm.
    For model r2, we only study the case of a low wind mass loss rate (model r2w1)
    because we believe SNe II-P do not stem from such extended RSG progenitors \citep{d13_sn2p} and
    also because the presence of a dense wind is the main ingredient controlling the resulting SN
    radiation (hence such simulations based on model r1 are sufficient).

    In two additional models, we study the impact of increasing the atmospheric density scale height.
    We use $H_{\rho}=$\,0.3\,$R_{\star}$ down to 10$^{-12}$\,g\,cm$^{-3}$, followed by a power-law
    density
    profile with exponent 12 (model r1w1h) or $H_{\rho}=$\,0.1\,$R_{\star}$ followed by a wind mass loss
    rate of 5$\times$\,10$^{-3}$\,\msunyr\ (model r1w5h). In both models, the wind density is forced to be at
    least that for a wind mass loss rate of 10$^{-6}$\,\msunyr.
    Such variations in atmospheric scale height may appear large. However, they correspond
    to variations in density that are not greater than those imposed in models r1w1-r1w6.
    Secondly, they represent different ways of going from a density of 10$^{-9}$\,g\,cm$^{-3}$ at $R_{\star}$
    to a typical RSG wind density at a few $R_{\star}$. RSG atmospheres are observed to be extended,
    much more than given by the atmospheric scale height in \mesa\ models at $R_{\star}$.

\begin{table*}
\caption{
Summary of properties of the \heracles\ simulations and their initial conditions.
It takes $t_{\rm rise}$ for the recorded luminosity to rise from 5$\times$\,10$^8$\,\lsun
to $L_{\rm bol,max}$.
$\Delta t_{\rm max}$ is the time during which $L_{\rm bol}$ remains above a tenth of
$L_{\rm bol,max}$.
$f_{\rm drop}$ is the factor by which $L_{\rm bol}$ decreases in one day after the time of maximum.
$\int L_{\rm bol} \, dt$  gives the time integrated bolometric luminosity from first detection until 15\,d later.
$R_{\star}$ is the progenitor radius, which corresponds to the RSG surface in the \mesa\ model.
$M_{\rm ejecta}$ and $E_{\rm kin}$ are the ejecta mass and kinetic energy (without the
atmosphere/wind material).
$H_{\rho}$ is the atmospheric scale height in the intermediate region between $R_{\star}$
and the base of the wind. $n_\rho$ is the exponent of the power low joining this atmosphere
and the base of the wind. The next column gives the wind mass loss rate within
2--5$\times$\,10$^{14}$\,cm -- beyond this radius the wind mass loss rate is always
set to 10$^{-6}$\,\msunyr.
The subscript `ext' refers to the external material located beyond $R_{\star}$.
Its optical depth is below unity before shock breakout (because of its low original
temperature, typically below the RSG effective temperature of $\sim$\,3000\,K).
Here, we give its optical depth for post shock breakout conditions, i.e.,
when H and He are fully ionized ($\kappa=$\,0.34\,cm$^2$\,g$^{-1}$).
\label{tab_sum}
}
\begin{center}
\begin{tabular}{l@{\hspace{3mm}}c@{\hspace{3mm}}c@{\hspace{3mm}}
c@{\hspace{3mm}}c@{\hspace{3mm}}c@{\hspace{3mm}}
c@{\hspace{3mm}}c@{\hspace{3mm}}c@{\hspace{3mm}}
c@{\hspace{3mm}}c@{\hspace{3mm}}c@{\hspace{3mm}}
c@{\hspace{3mm}}c@{\hspace{3mm}}c@{\hspace{3mm}}}
\hline
Model & $t_{\rm rise}$ & $\Delta t_{\rm max}$ & $L_{\rm bol,max}$ & $f_{\rm drop}$ & $\int L_{\rm bol} \, dt$  &
 $R_{\star}$  &   $M_{\rm ejecta}$  & $E_{\rm kin}$  &
   $H_{\rho}$   & $n_\rho$ &  \mdot\   & $M_{\rm ext}$ &   $\tau_{\rm ext}$ \\
   & [d] & [d]  & [\ergs]  &   & [erg]  & [\rsun]  & [\msun] & [erg]& [$R_{\star}$]  &   &  [\msunyr]  & [\msun] &   \\
   \hline
r1w1        &   0.018  &  0.062 & 7.38(44)   &   146.2  & 6.65(48)   & 501   &  12.52  & 1.35(51) & 0.01  & $\dots$ &   1(-6)  & 2.75(-3)  &  160  \\
r1w2        &   0.019   &  0.068 & 6.71(44)   &   130.4  & 6.68(48)     & 501   &  12.52  & 1.35(51) &  0.01 & $\dots$   &   1(-5) & 2.79(-3)   &  160   \\
r1w3        &   0.025   &  0.15 & 3.30(44)    &  57.4   & 6.66(48)    & 501   &  12.52  & 1.35(51)   & 0.01 & $\dots$   &   1(-4) & 3.05(-3)  &  160   \\
r1w4         &  0.186  &  5.11 &  6.53(43)    &    5.61   & 8.67(48)   & 501   &  12.52  & 1.35(51)    & 0.01  & $\dots$  &   1(-3)& 5.59(-3)  &   169 \\
r1w1h      & 0.30  &  0.755 &  4.16(44) &     11.98 &  2.10(49)     & 501   &  12.52  & 1.35(51) &  0.3    & 12  &   1(-6)& 1.62(-1)  &   4780 \\
r1w5h      & 0.21  & 1.03 & 1.79(44) &    8.72    &   1.10(49)    & 501   &  12.52  & 1.35(51) & 0.1     &   $\dots$  & 3(-3) &  3.57(-2)  & 1600  \\
r1w5r        &  0.53  &  2.36  & 6.25(43)     &    1.74   &  9.48(48) & 501   &  12.52  & 1.35(51)&   0.01 & $\dots$  & 5(-3)&1.02(-2) & 353   \\
r1w6         &   1.94   &  7.00 & 5.24(43)     &   2.02    & 1.99(49) & 501   &  12.52  & 1.35(51)    &  0.01  & $\dots$ & 1(-2) & 3.04(-2) &  246   \\
r2w1      &   0.081   &  0.179 & 8.03(44) &  84.91   & 1.44(49) & 1107 & 12.57 & 1.24(51) & 0.01 & $\dots$  &   1(-6) & 6.14(-2)   &  956 \\
 \hline
\end{tabular}
\end{center}
\end{table*}

    In all cases, the atmosphere/wind is given an initial temperature of 2000\,K
    and the terminal velocity of the wind is $v_\infty$ of 50\,\kms.
    We assume the atmosphere is in hydrostatic equilibrium (zero velocity) and that the wind
    accelerates promptly to $v_\infty$ within a fraction of $R_{\star}$.
    The exact choice of velocity profile is irrelevant for various reasons. At shock breakout, the
    wind base will be strongly accelerated to velocities larger than $v_\infty$. The shock and ejecta velocity
    are also 100 times larger. In this context, it is the density profile that matters for the dynamics.

    Figure~\ref{fig_init_set} shows the density profile versus radius for the whole
    set of \heracles\ simulations. The profile for other quantities is shown for model
    r1w6 in Fig~\ref{fig_init_r1w6}.

    Table~\ref{tab_sum} summarizes the initial model properties.
     We define the external mass $M_{\rm ext}$ as the added mass beyond $R_\star$
     (as given in the stellar evolution model).
     This external material is not necessarily the wind itself, which is probably blown from a larger radius,
     but can be seen
     as some sort of cocoon of material stagnating at the base of the wind. In reality, the boundary between
     the star (interior to $R_\star$), this cocoon of mass, and the external wind is ill defined. Hence, the
     quoted external mass should be interpreted with caution. One conclusion from this work
     is that this material may not stem from a super-wind phase but instead corresponds to the
     cocoon of material that is seen around all spatially resolved RSG atmospheres.

    Table~\ref{tab_sum} also gives the optical depth of the atmosphere/wind assuming full ionization of H and He.
    The external material is cold and optically thin in the RSG progenitor (and its optical depth
    is $\lesssim$\,1, although this greatly depends on whether dust or molecules are present)
    but it becomes fully ionized (at least for
    H and He) as soon as the shock breaks out. Hence, it is the optical depth of the ionized atmosphere/wind
    that matters for the evolution during and after shock breakout.
    Even in the case of a weak wind mass loss rate of 10$^{-6}$\,\msunyr, the optical depth
    of the atmosphere above $R_{\star}$ is $>$\,100. This implies that shock breakout will take place
    outside of the region covered by the \mesa\ model. In fact, the SN photosphere stays in this external
    region for about a week after shock breakout.

    While we give a velocity profile to the material in the environment of the RSG star (which we call a
    wind or an atmosphere), the early-time SN radiation is primarily influenced by the extent and density/mass
    of this material --- the influence of the initial velocity is negligible hence unconstrained. So, for all simulations,
    the results would be unchanged if the velocity of the CSE was set to zero.

\subsection{Multi-group radiation-hydrodynamic simulations with \heracles}

   The interaction configurations described above are used as initial conditions for
   the 1-D multi-group radiation-hydrodynamics simulations with the code \heracles\
   \citep{gonzalez_heracles_07,vaytet_mg_11}.
   The approach is identical to that used and described in \citet{D15_2n,D16_2n}.
   We use 8 groups that cover from the UV to the near-IR: one group for the entire Lyman
   continuum (including the X-ray range), two groups for the Balmer continuum,
   two for the Paschen continuum, and three groups for the Brackett continuum and
   beyond.\footnote{Using more groups
    would make the simulation more costly and would not alter the fundamental results discussed here.
    At present, most simulations in the community are limited to grey transport. Our 8 groups are
    positioned to capture the relevant photoionization edges -- the absorptive opacity varies very smoothly
    between these edges.}
    A multi-group approach is superior to a grey approach because it better describes the opacity of the
    material to the radiation when they have widely different temperatures
    (e.g., high energy radiation crossing a cold gas; see \citealt{D15_2n} for discussion).
    Because the thermal energy of the gas is a negligible fraction of the total radiation energy
    in SNe, we adopt a simple equation of state that treats the gas as ideal with $\gamma=$\,5/3.
    We have done tests using a mean-atomic-weight $\mu=$\,0.67 (corresponding to full
    ionization) and 1.35 (neutral) and these yields the same results at early times.\footnote{The major
    drawback here is not the neglect of the very small ionization/excitation energy. The main limitation
    in radiation hydrodynamics simulations is the treatment of the gas in LTE at $\tau\lesssim$\,20 since
    the temperature/ionization is in reality influenced by both
    time dependence and non-LTE effects \citep{UC05,D08_time}, while the color temperature
    is influenced by electron scattering and line blanketing \citep{E96_epm,D05_epm}.}

    We concentrate on the early-time properties so we limit the simulations to
    times prior to $\sim$\,15\,d. For the shock speeds relevant to the present simulations,
    this requires placing the outer boundary at $R_{\rm max}=$\,1.5\,$\times$\,10$^{15}$\,cm.
    We cover the ejecta down to a radius of a few 100\,\rsun, which is deep enough
    to cover the reverse shock progressing into the He core.

    Eulerian coordinates are not
    ideal for SN studies in 1D (although much better than Lagrangian coordinates when we
    extend such simulations to multiple dimensions, as in \citealt{vlasis_2n_16}) because the structure
    expands by orders of magnitude through the evolution. Resolving the ejecta at all times requires
    high resolution. We use 1-D spherical polar coordinates with a total of 10,000 radial points.
    To better resolve the ejecta at smaller radii (earlier times), we use a grid with a constant
    spacing from the minimum radius at $\sim$\,10$^{13}$\,cm up to $R_{\rm t} = 5\times$10$^{14}$\,cm,
    and then switch to a grid with a constant spacing in the log up to $R_{\rm max}$.
    The grid is designed to have no sharp jump in spacing at $R_{\rm t}$.

    Because we focus
    on early times, we adopt a fixed composition corresponding to the outer ejecta
    of models m15mlt3 and m15mlt1 ($X_{\rm H}=$\,0.65, $X_{\rm He}=$\,0.33, and solar
    metallicity). This composition is only relevant for the computation of the opacities.

    The bolometric light curves extracted from the \heracles\ simulations are computed using
    the total radiative flux at the outer boundary.
    The bolometric luminosity should not be affected by the LTE assumption
    for the gas properties. However, multi-band light curves are sensitive to non-LTE effects,
    the wavelength dependent albedo, line blanketing, and are therefore computed
    with the 1-D non-LTE code \cmfgen. Post-processing an LTE calculation with a non-LTE code
    is not optimal, but it is a good start.
    For example, for the same hydrodynamical snapshot, assuming LTE in \cmfgen\
    produces redder/cooler spectra than the corresponding non-LTE calculation.

\subsection{non-LTE radiation-transfer simulations with \cmfgen}

     At selected epochs in the \heracles\ simulations, the radius/velocity/density/temperature structure
     is remapped into \cmfgen\ \citep{HD12,D15_2n} for the calculation of the emergent flux,
     from which the multi-band light curves and the spectral evolution are obtained.
     The code treats the non-monotonic velocity field produced by \heracles. Hence, emission, absorption,
     and scattering from the fast ejecta regions and the slower atmosphere/wind regions is taken into account.
     We also allow for the non-coherent frequency redistribution by thermal electrons.
     We fix the temperature during the \cmfgen\ calculation because it results in a large part from
     the dynamics (shock deposited energy).

     For a given snapshot, we only extract the region of the \heracles\ simulation between
     the electron-scattering optical depths of 10$^{-5}$ and 50 (the inner boundary is optically
     thick at all wavelengths).
     The grid in \cmfgen\ uses 100 points with the prescriptions that any two consecutive
     points should at most have a jump of 5\% in radius, of 10\% in temperature, of 10\% in electron density,
     of 10\% in density, and of 10\% in the log of the optical depth. If two consecutive points in the \heracles\
     simulation violate any of these criteria, additional points are inserted in the \cmfgen\ grid.
     A finer grid is also used at the inner and outer boundaries.

     For our \cmfgen\ calculations, we include H, He, C, N, O, and Fe, with the
     mass fractions 0.63972, 0.349, 0.00142, 0.0031, 0.00541, and 0.00135, respectively.
     No radioactive decay is accounted for here (all species/isotopes treated are stable) since
     the early-time evolution of SNe II is unaffected by decay heating.
     The minor abundance offset from the \heracles\ simulation is irrelevant.
     Our study focuses primarily on dynamical issues and line profile morphology --- slight variations
     in abundance that may occur through different levels of mixing, overshoot, etc. are not a concern
     at this stage. The study of their influence on early-time spectroscopic properties is deferred.
     For example, the early-time observations of SNe 2013cu and 2013fs suggest that their respective
     progenitors may have different CNO
     surface abundances (although this may stem partly from different ionization conditions):
     SN\,2013cu shows one strong line of
     nitrogen \citep{galyam_13cu_14}, while SN\,2013fs shows primarily lines
     of oxygen \citep{yaron_13fs_17}.

     Depending on the post-shock-breakout phase, we adjust the model atoms in order to treat only
     the ions that contribute to the spectrum formation. At most, we include H\one, He\one--\two,
     C\two--\five, N\two--\six, O\two--\six, Fe\two--\eight.


\begin{figure*}
\epsfig{file=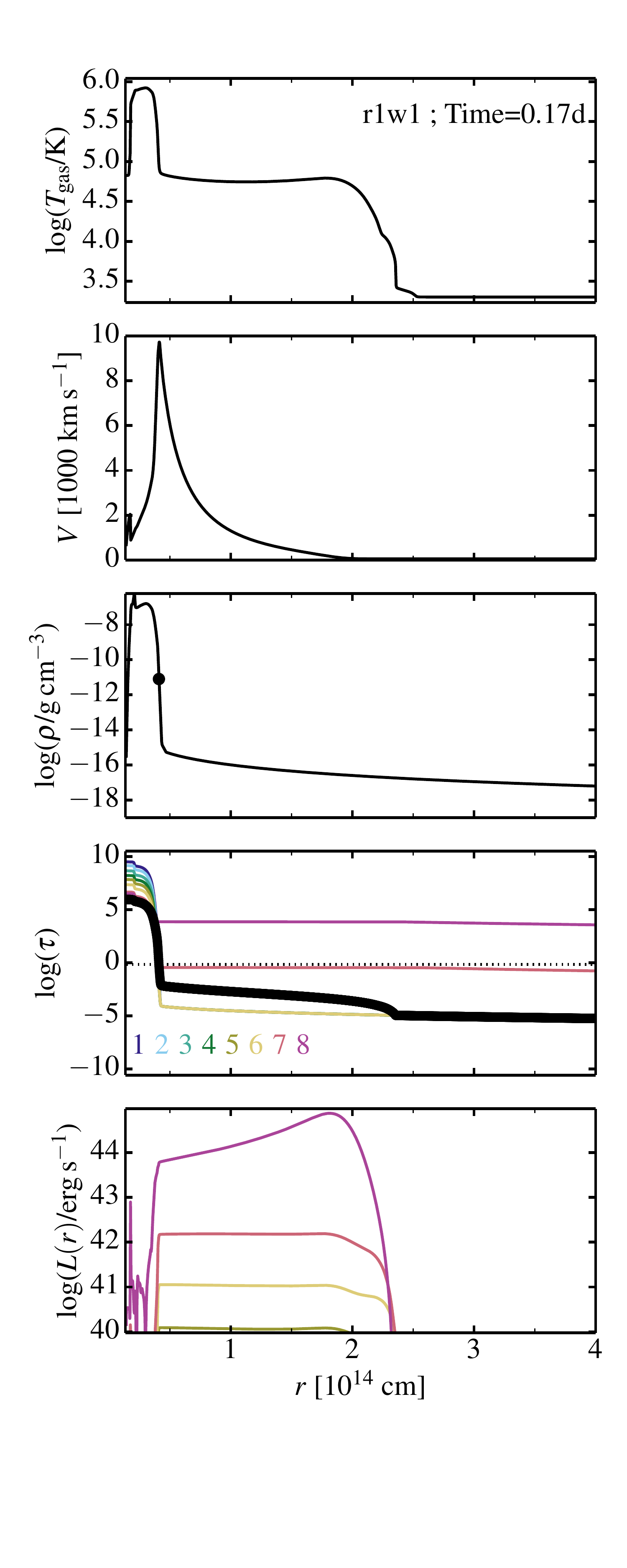,width=6.15cm}
\epsfig{file=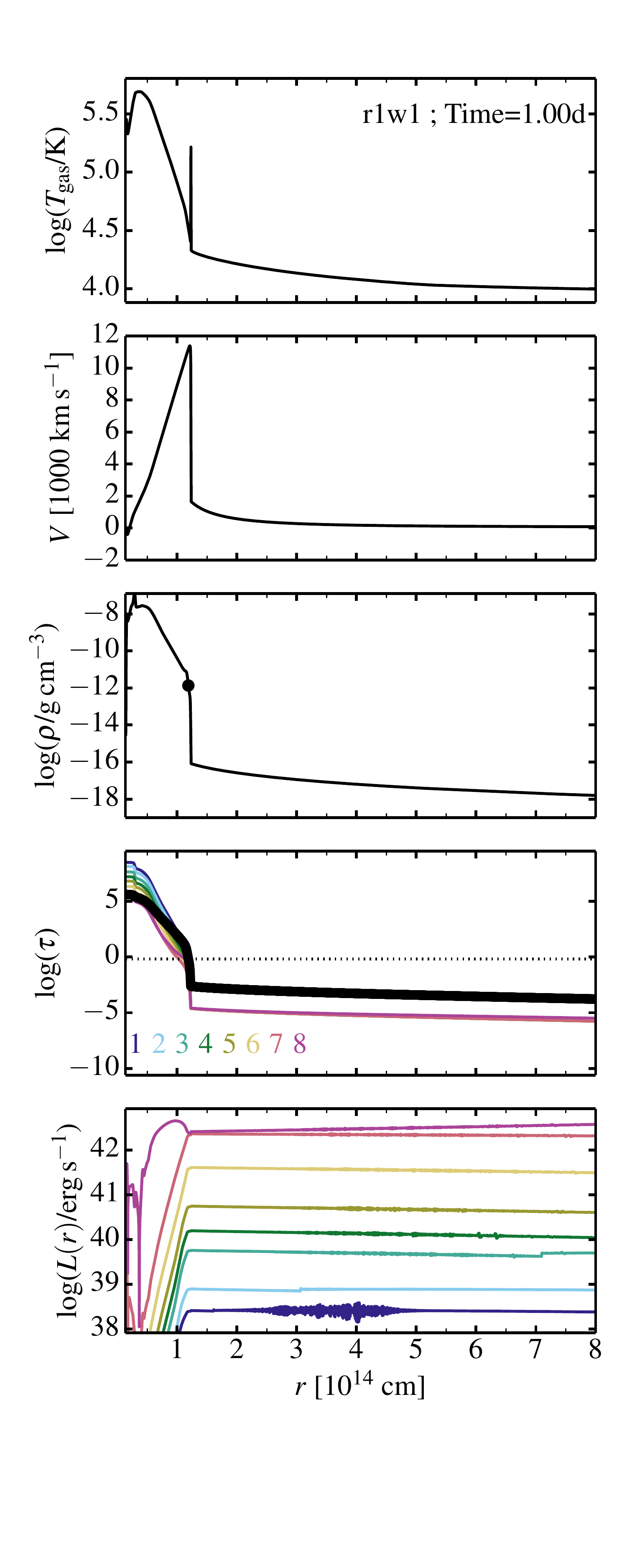,width=6.15cm}
\epsfig{file=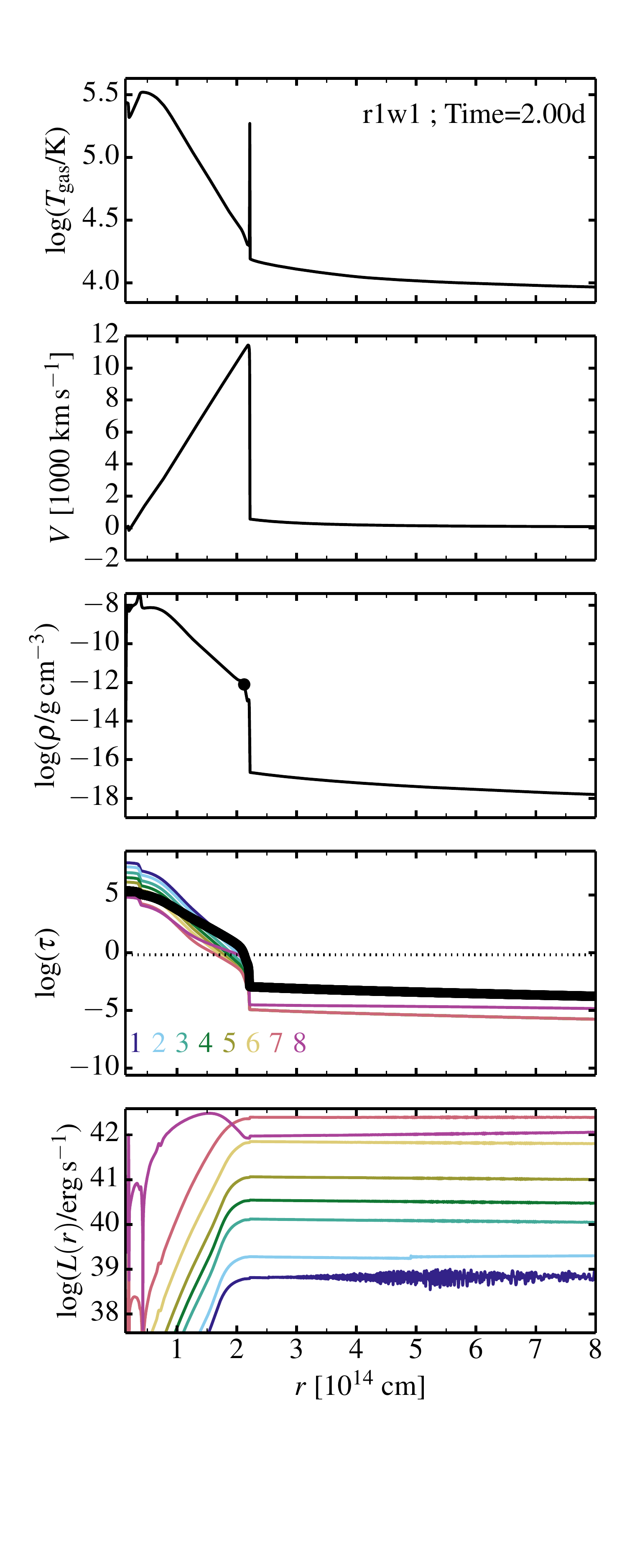,width=6.15cm}
\vspace{-1.5cm}
\caption{Ejecta properties computed by \heracles\ for model r1w1 (weak wind
mass loss rate of 10$^{-6}$\,\msunyr) at 0.17, 1, and 2\,d after the start of the \heracles\ simulation.
The black dot in each of the middle panels corresponds to the electron-scattering photosphere.
In the optical-depth panel, the colored numbers correspond to each energy group, ordered from
low frequency (group 1 covers the far-IR) to high frequency (group 8 covers the Lyman continuum).
The thick black line corresponds to the electron-scattering optical depth.
Through this early evolution, we can see the ejecta accelerate, the wind acceleration by the
SN radiation, the propagation of the radiation burst that coincided with shock breakout (at $\sim$\,0.1\,d),
the heating and rapid ionization of the wind, the establishment of radiative equilibrium in the optically-thin
regions at 2\,d.
\label{fig_snap_r1w1}
}
\end{figure*}

\begin{figure*}
\epsfig{file=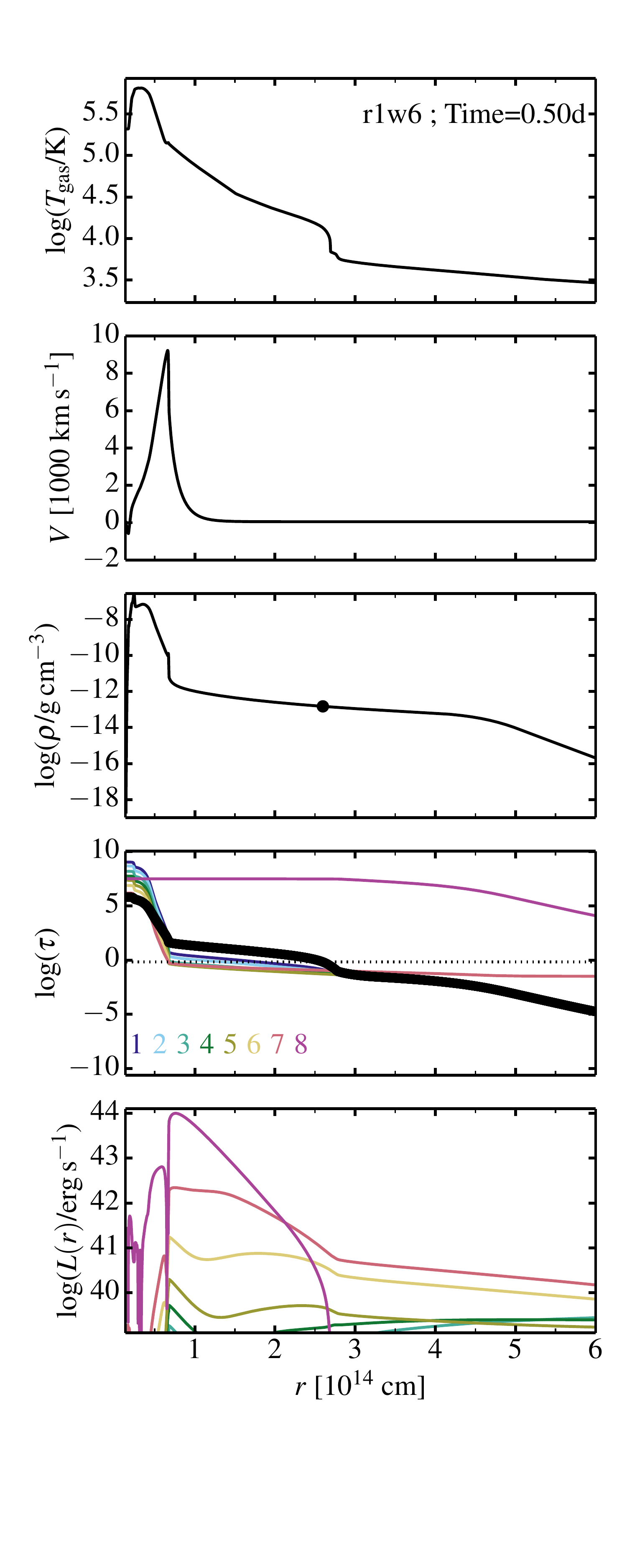,width=6.15cm}
\epsfig{file=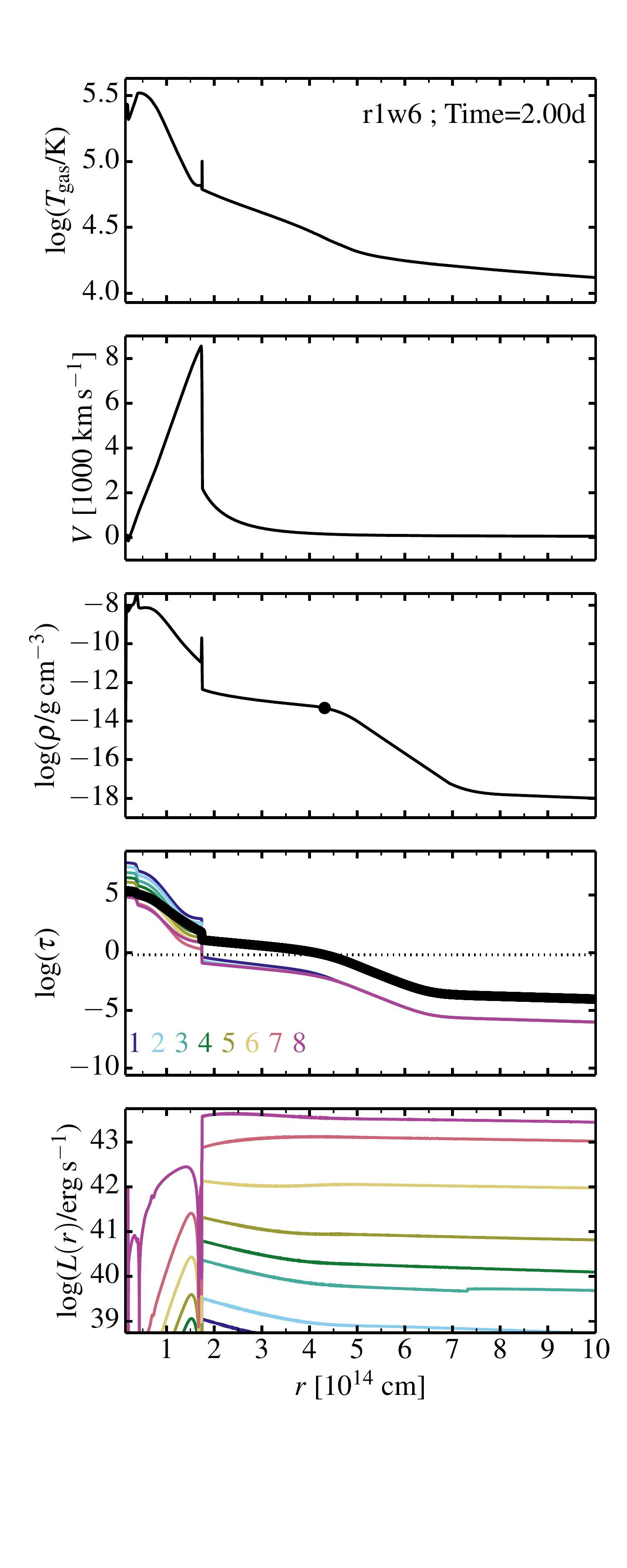,width=6.15cm}
\epsfig{file=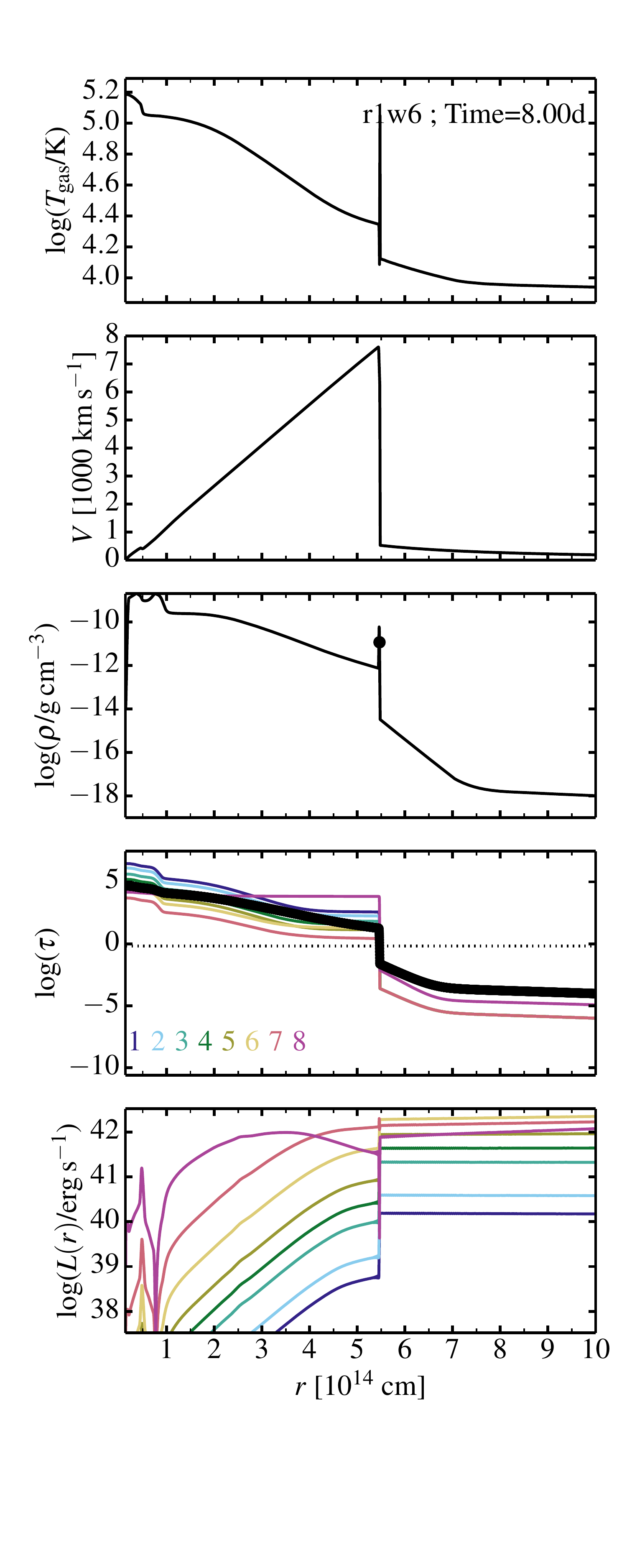,width=6.15cm}
\vspace{-1.5cm}
\caption{Ejecta properties computed by \heracles\ for model r1w6 (strong wind
mass loss rate of 10$^{-2}$\,\msunyr) at 0.5, 2, and 8\,d after the start of the \heracles\ simulation.
Compared to model r1w1, the ejecta is more decelerated by the dense wind (the maximum ejecta velocity
levels off at $\sim$\,7000\,\kms\ compared to 11000\,\kms), more radiation energy
gets trapped within the optically-thick wind leading to a delayed but longer-lived shock break-out signal,
a much more massive cold-dense-shell forms (containing the swept up wind material). All these properties
produce unambiguous radiative signatures.
\label{fig_snap_r1w6}
}
\end{figure*}

\begin{figure*}
\epsfig{file=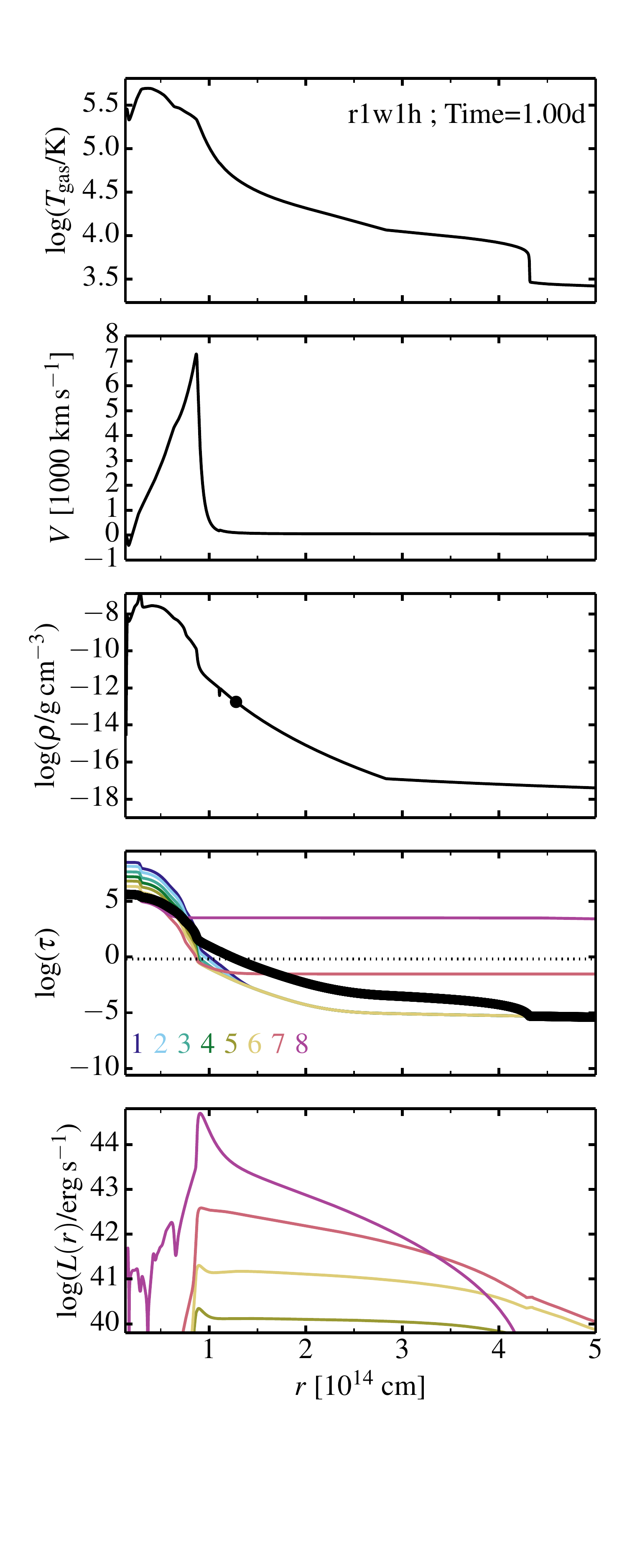,width=6.15cm}
\epsfig{file=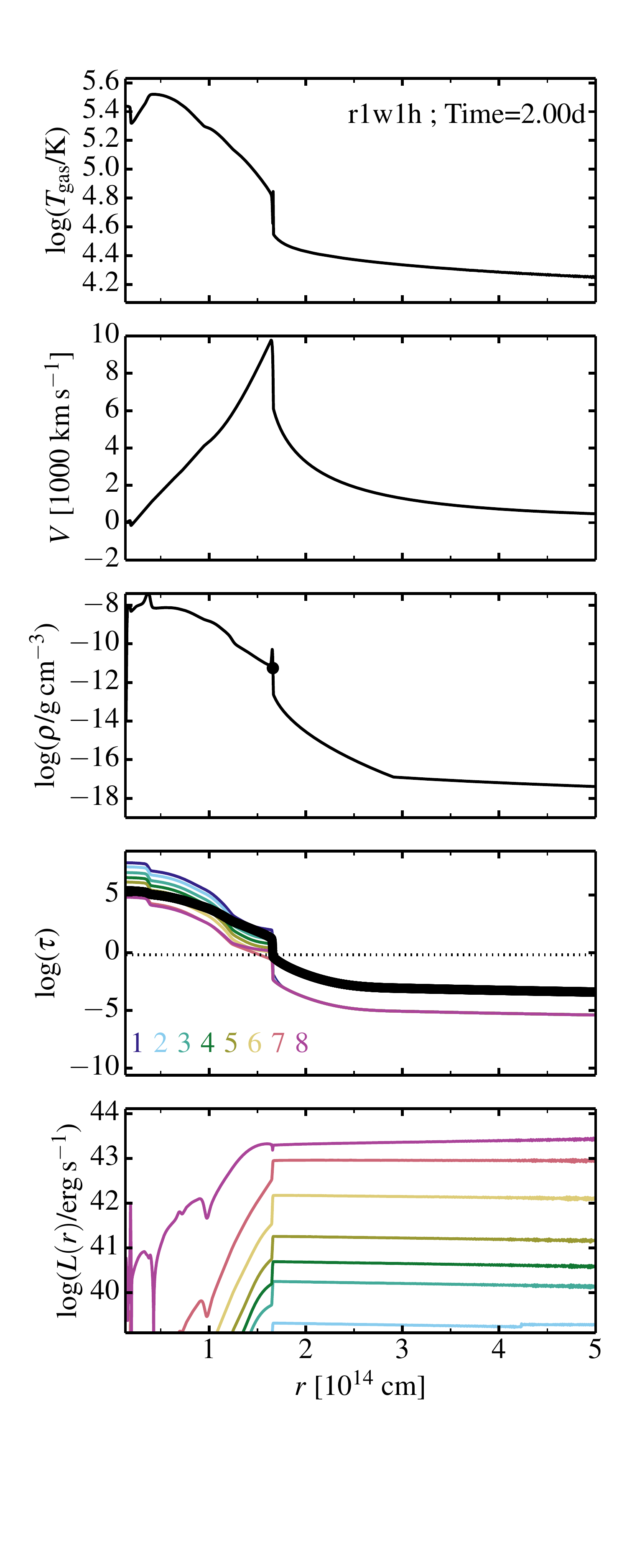,width=6.15cm}
\epsfig{file=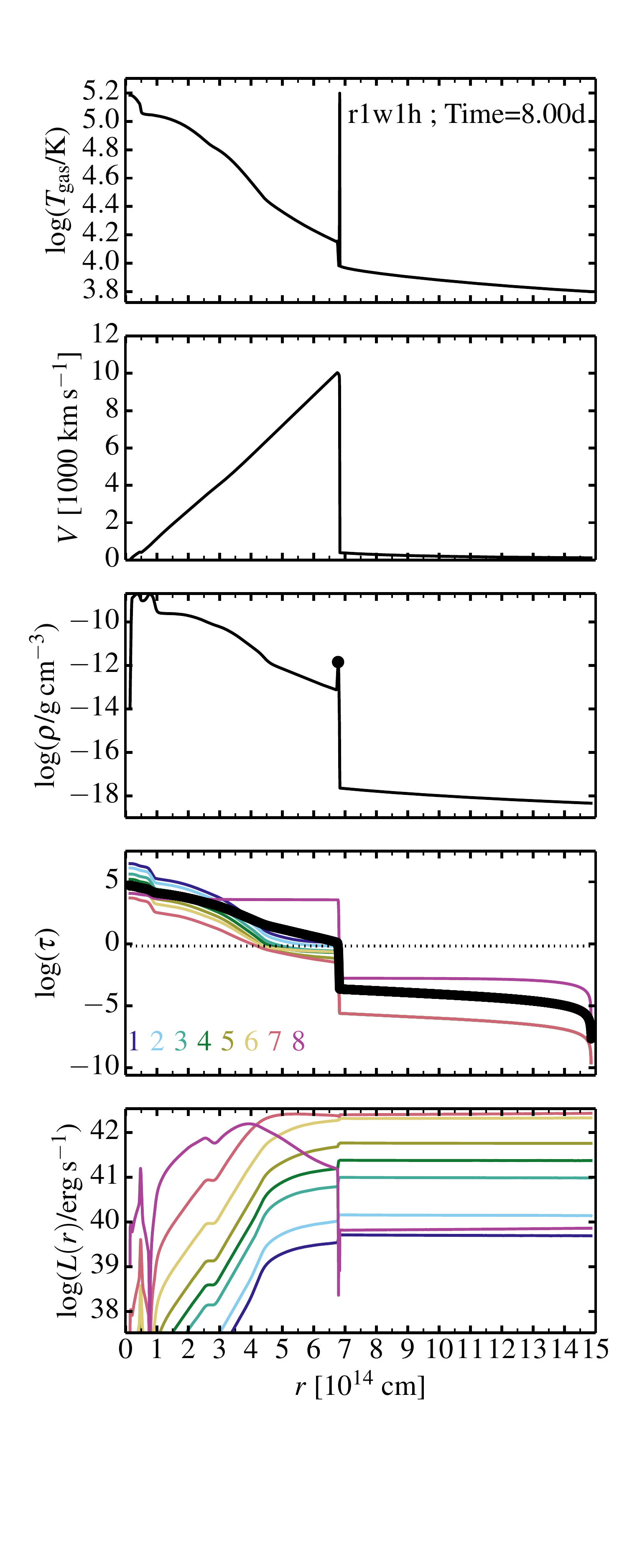,width=6.15cm}
\vspace{-1.5cm}
\caption{Ejecta properties computed by \heracles\ for model r1w1h (model with a
weak wind but an extended atmospheric scale height) at 1, 2, and 8\,d after the start
of the \heracles\ simulation.
Compared to model r1w6, the shock is not as strongly decelerated by the atmosphere (which has
a steep density profile), and the photosphere remains close to the fast-moving ejecta, inhibiting
the formation of an long-lived shock-breakout signal.
\label{fig_snap_r1w1h}
}
\end{figure*}

\section{Results from radiation-hydrodynamic simulations with \heracles}
\label{sect_heracles}

\subsection{Dynamical evolution for representative models of the sample}

   The dynamical evolution obtained in all \heracles\ simulations is qualitatively similar, but variations
in wind mass loss rate and atmospheric scale height introduce important quantitative differences.
We describe in some detail below the cases of a weak wind (model r1w1;  Fig.~\ref{fig_snap_r1w1}),
a strong wind (model r1w6; Fig.~\ref{fig_snap_r1w6}), and an extended atmospheric scale height
(model r1w1h; Fig.~\ref{fig_snap_r1w1h}).
In this section, times are given with respect to the start of the \heracles\ simulation (about 10$^4$\,s
before shock breakout), unless stated otherwise.

    Figure~\ref{fig_snap_r1w1} shows snapshots at 0.17, 1, and 2\,d after the start of the \heracles\
simulation (the shock crosses $R_{\star}$ at $\sim$\,0.1\,d).
As the shock approaches $R_{\star}$, a radiative
precursor forms that ionizes the atmosphere and the base of wind. This shifts the photosphere
outwards, beyond the $R_{\star}$ of the \mesa\ model, at a density of about 10$^{-12}$\,g\,cm$^{-3}$.
This implies that shock breakout actually occurs in regions of the star not covered by the \mesa\ model.
This arises because the \mesa\ model has a cold optically-thin atmosphere.
The radiative precursor heats, ionizes, and accelerates the material
that it crosses on a light travel time. The strong radiative flux
and energy accelerate both the ejecta and the wind. The ejecta reaches its asymptotic kinetic energy
of 1.34$\times$\,10$^{51}$\,erg at 10--15\,d. The acceleration time scale depends on ejecta depth.
The fastest material reaches 11500\,\kms\ at 0.5\,d only (bottom row panels of Fig.~\ref{fig_var}),
while most of the H-rich ejecta regions
are within a few per cent of their asymptotic velocity at 5\,d. It is the deep ejecta layers that take
much longer, because of the reverse shock and the slower expansion (the initial radius of the corresponding
shells remains a sizable fraction of the current shell radius for longer).
Because of the interaction with the low-density wind, the outer ejecta velocity (where it is maximum)
is reduced by only $\sim$\,2\% between 0.5\,d and 15\,d later.

    The dynamical evolution is much different if the wind mass loss rate is large.
Figure~\ref{fig_snap_r1w6} shows snapshots at 0.5, 2, and 8\,d after the start of the \heracles\
simulation. As the shock overtakes $R_{\star}$, the atmosphere and the wind get progressively
ionized. The opacity in the Lyman continuum is initially large (cold wind), so the heating occurs
partially by absorption of lower energy photons because they have a longer mean free path.
The photosphere progresses outwards in the dense wind until $\sim$\,1d and settles
at $\sim$\,4$\times$\,10$^{14}$\,cm (four times further out at 1\,d than in model r1w1).
It takes nearly a week for
the shock to reach this radius. During that time, energy from the shock is deposited in the
wind (which is now optically-thick)  and escapes on a diffusion time scale.
As previously, the wind is accelerated by the radiation that crosses it, while the ejecta is both
accelerated by radiation-pressure gradients and
decelerated by the wind. The maximum ejecta velocity
is in this case only 7200\,\kms\ at 15\,d, hence 30\% less than in model r1w1 at the same time.
The asymptotic kinetic energy is 1.32$\times$\,10$^{51}$\,erg, hence 2$\times$\,10$^{49}$\,erg
lower than in the weak-wind model r1w1. This excess energy boosts the SN luminosity over the first week
of evolution in the same proportion (see below).
The drop in kinetic energy therefore arises because the wind is of sufficiently low-density
that the shock energy can be sapped by radiative losses.

Varying the wind mass loss rate between 10$^{-6}$ and 10$^{-2}$\,\msunyr\ or varying the extent
of the high wind density region (model r1w5r) produce similar
properties that are intermediate between those of models r1w1 and r1w6 (Fig.~\ref{fig_var}).

Figure~\ref{fig_snap_r1w1h} shows snapshots at 1, 2, and 8\,d after the start of the \heracles\
simulation for model r1w1h. Increasing the atmospheric scale height enhances the mass exterior
to $R_{\star}$  but yields properties that are quantitatively different from enhancing the wind density.
Indeed, a greater scale height extends the star but the bulk of the atmosphere mass resides in high density
layers close to $R_{\star}$, so the diffusion time through the atmosphere remains small.
Consequently, there is no extended shock-breakout signal like in the high-mass loss rate model r1w6
and the photosphere after shock breakout remains closer to the fast-moving ejecta layers (rather than
within the slowly moving wind like in model r1w6). The additional mass in the atmosphere leads
to maximum velocities 1000\,\kms\ smaller than in model r1w1. The interaction of the ejecta
with the atmosphere/wind (whose mass outside of $R_{\star}$ is $\sim$\,0.16\,\msun)
leads to the extraction of 2$\times$\,10$^{49}$\,erg compared to 7$\times$\,10$^{48}$\,erg
in model r1w1 (Table~\ref{tab_sum}). As for model r1w6, this energy is primarily radiated, but with a
very different power profile, as we discuss in the next section.

\begin{figure*}
\epsfig{file=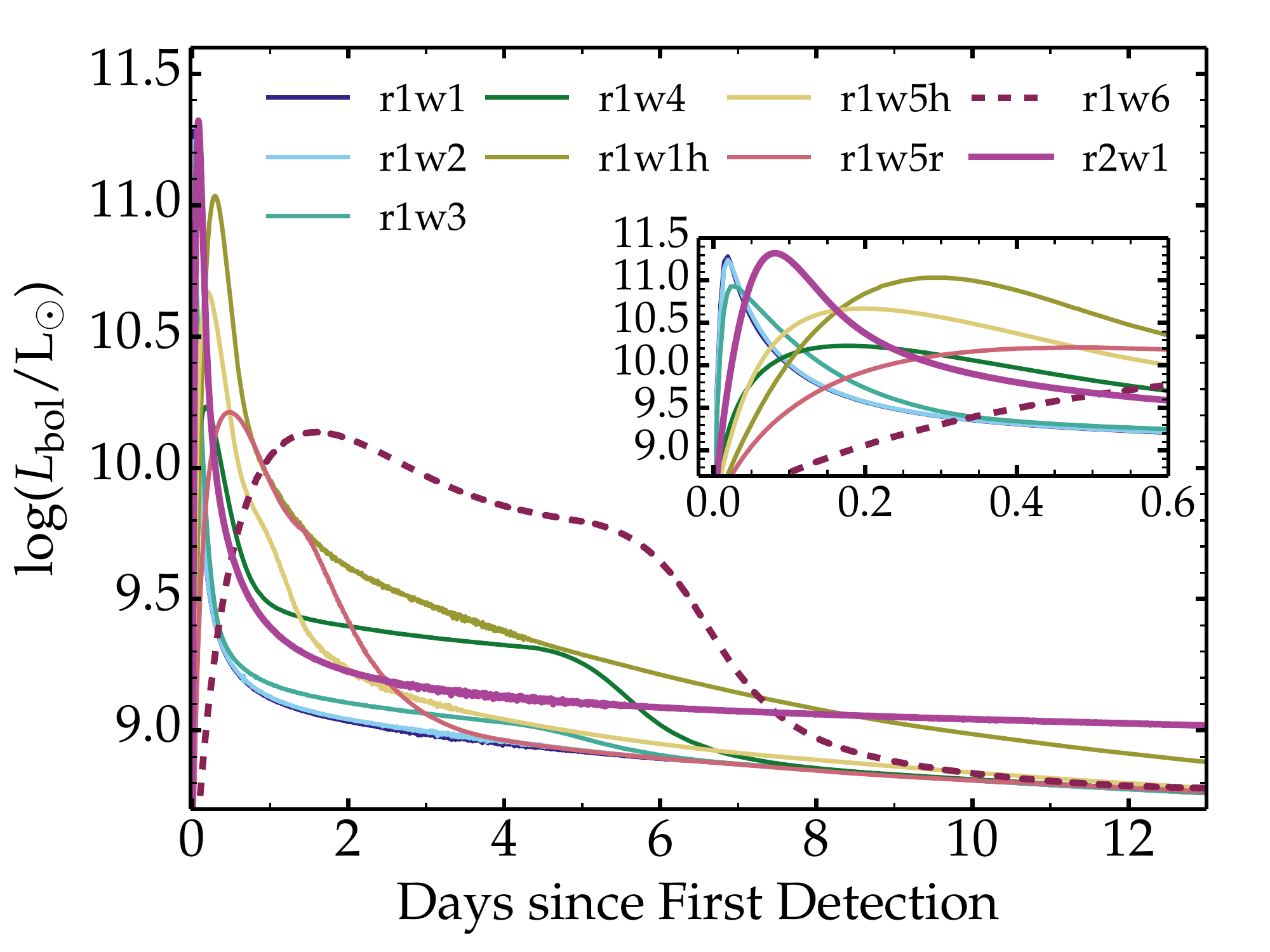,width=17cm}
\caption{
Bolometric light curves computed with \heracles\ for our full set of models.
Times are given since first signal detection at the outer boundary.
The time resolution in the \heracles\ simulation is always less than a minute (and it is less than a
second around the phase of shock breakout).
The insets show the same quantity but zooming on the earliest times, concomitant with shock breakout.
Progenitors with a larger radius or with a denser atmosphere/wind (here confined to within 5--10\,$R_{\star}$)
have a longer shock-breakout signal and an enhanced luminosity at early times powered by the interaction
with the atmosphere/wind.
\label{fig_lbol_heracles}
}
\end{figure*}

\subsection{Bolometric, Far-UV, UV, and optical light curves}

   Figure~\ref{fig_lbol_heracles} shows the bolometric light curves for the models computed
with \heracles\ (see previous section, Fig.~\ref{fig_init_set}, and
Table~\ref{tab_sum} for a summary of initial properties and results).
The bolometric luminosity is evaluated at the outer boundary from the sum of the fluxes in each group.
Because of the time delay for the radiation to reach the outer boundary at
$R_{\rm max}=$\,1.5$\times$\,10$^{15}$\,cm, we shift all light curves in time so that the time
origin corresponds to the first rise (at $R_{\rm max}$) to a luminosity of
5$\times$\,10$^8$\,\lsun. Because of optical depth effects, this time differs between models.

\subsubsection{Sensivity to the wind mass loss rate}

Models with increasing wind mass loss rates and the same atmospheric scale height
(r1w[1-6] and r1w5r) have a longer rise time (from 0.03 to 1.72\,d), a longer shock-breakout
signal duration
(from about an hour in model r1w1 up to a week in model r1w6, although in the latter the high-brightness
phase after breakout includes both the precursor radiation and the leakage of shock-deposited energy in
the wind), and a smaller bolometric maximum (from a maximum of 7.4$\times$10$^{44}$\,\ergs\
in model r1w1 down to 5.2$\times$10$^{43}$\,\ergs\ in model r1w6).
Light-travel time effects, which spread the shock breakout signal over $R_{\star}/c$, only
matter for the weak-wind models. For dense winds, the intrinsic duration of the shock breakout
signal exceeds $R_{\star}/c$. This lengthening of the shock-breakout signal
is related to the increasing photon diffusion time
through the optically-thick atmosphere/wind region. If this was merely a spreading of the
breakout signal, the bolometric
maximum would decrease as the shock breakout signal duration increases (and it does), but we would obtain
the same time-integrated bolometric luminosity.
Instead, we find that the time-integrated bolometric luminosity over the first 15 days of
evolution increases from 7.26$\times$\,10$^{48}$\,erg in the weak wind model r1w1
to 2.02$\times$\,10$^{49}$\,erg  in the strong wind model r1w6.
The time integrated bolometric luminosity is in all cases equal within a few percent to the change
in the total energy (the sum of the kinetic, radiative, and thermal contributions) on the grid.
Variations in $\int L_{\rm bol} \, dt$ thus reflect energy extraction from the ejecta as it interacts
with the wind. Hence, if a dense wind is present, there is both a spread of the shock breakout
signal by diffusion through the wind and a boost to the radiative flux from the interaction.

\begin{figure*}
\epsfig{file=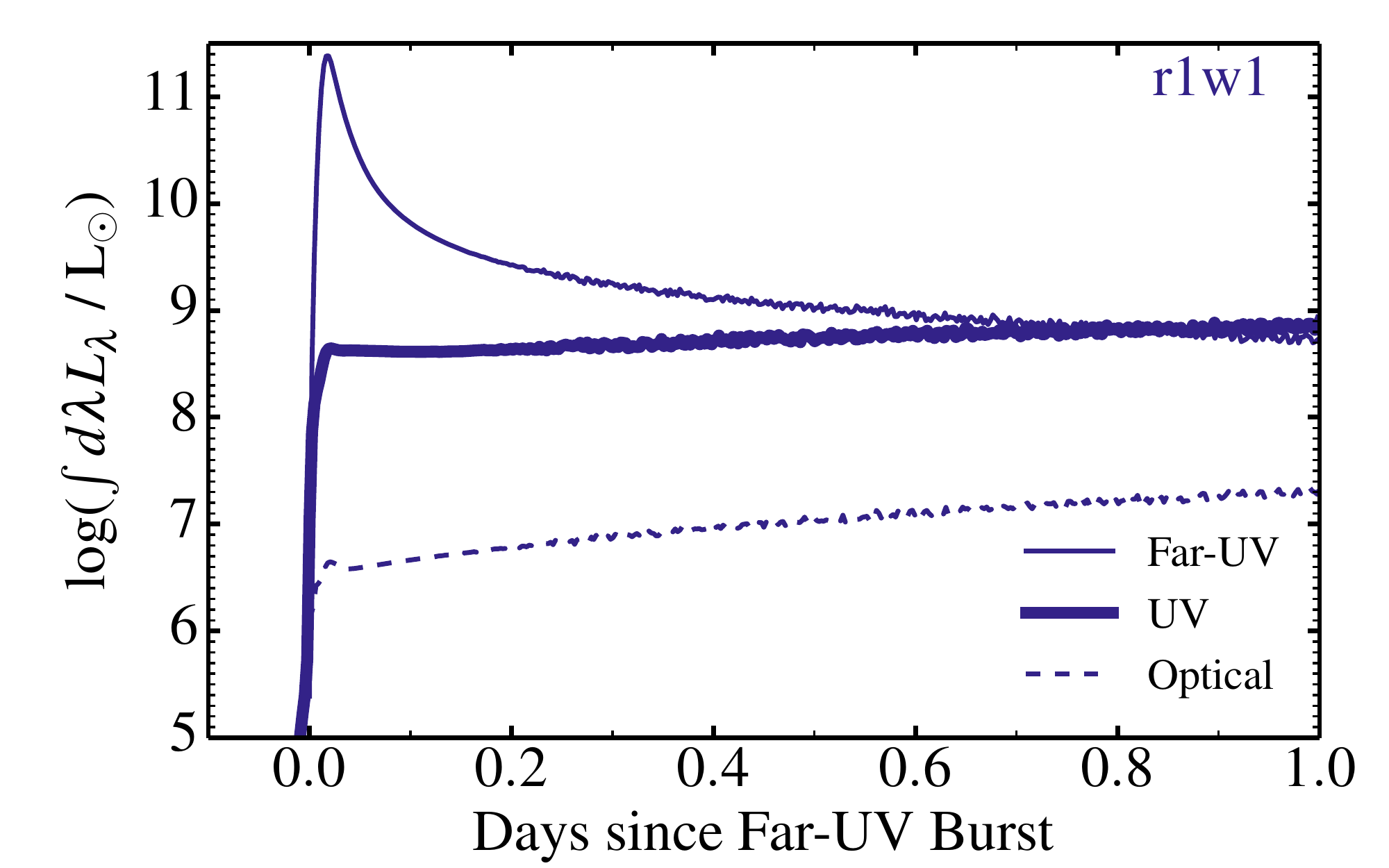,width=9.2cm}
\epsfig{file=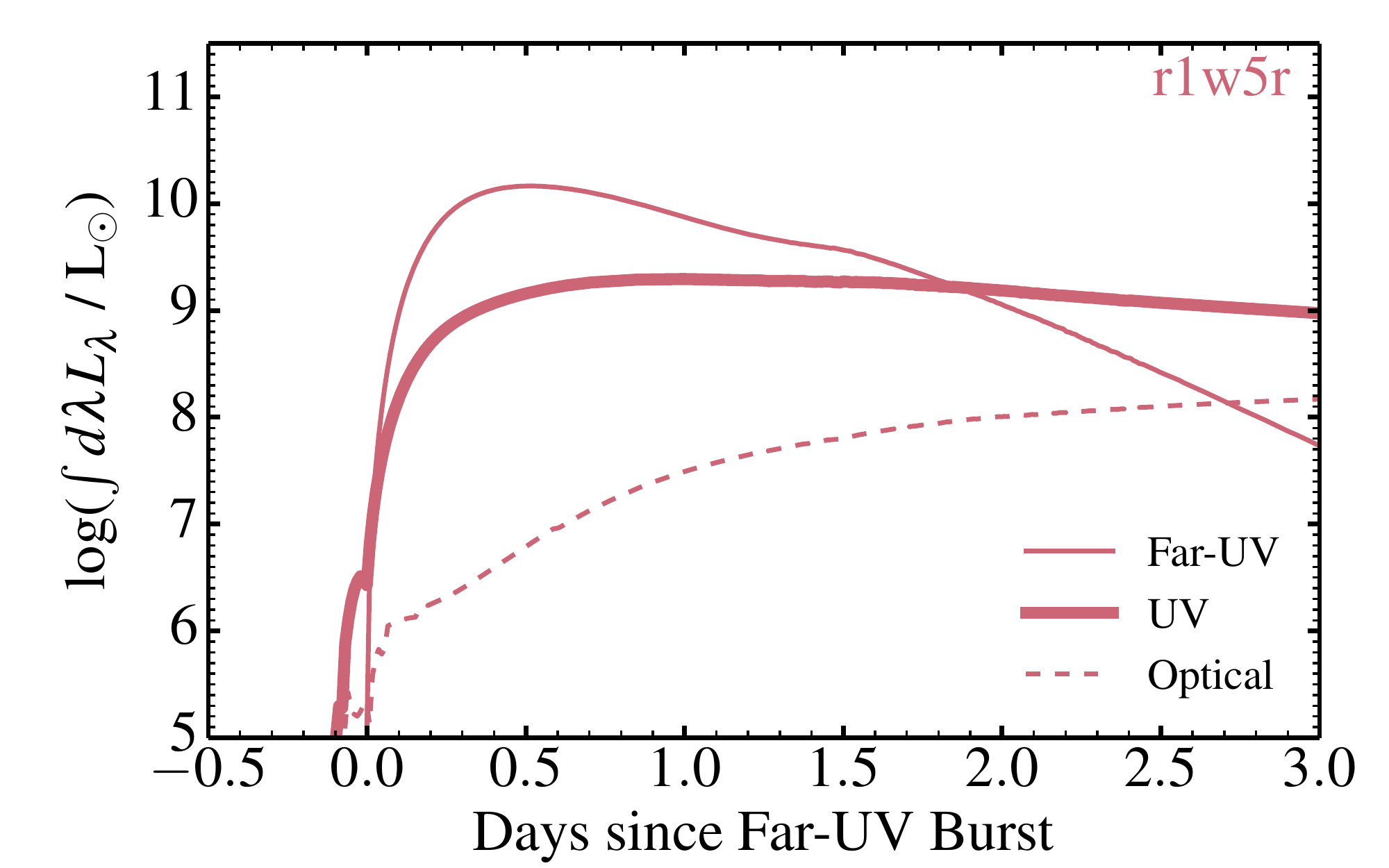,width=9.2cm}
\epsfig{file=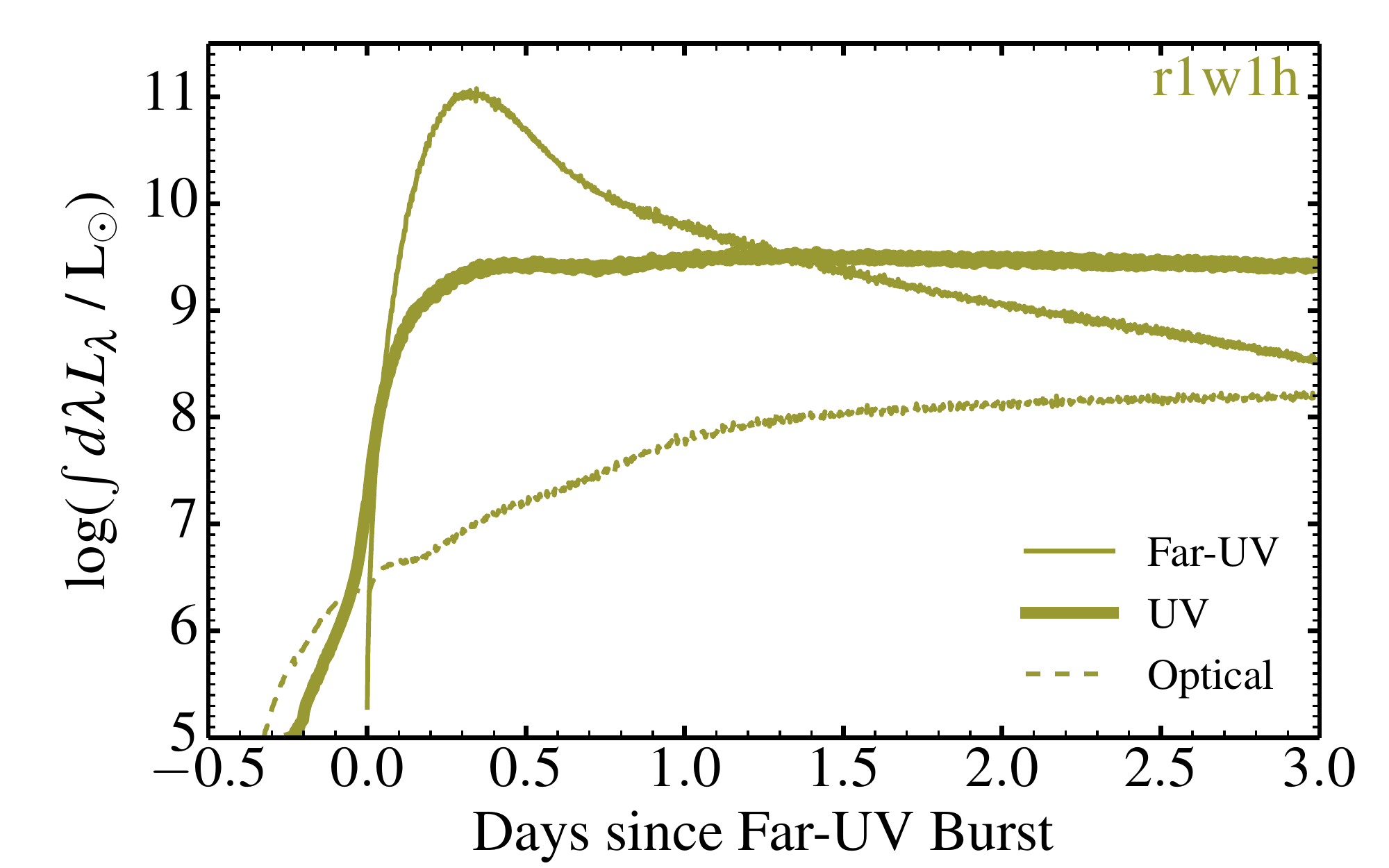,width=9.2cm}
\epsfig{file=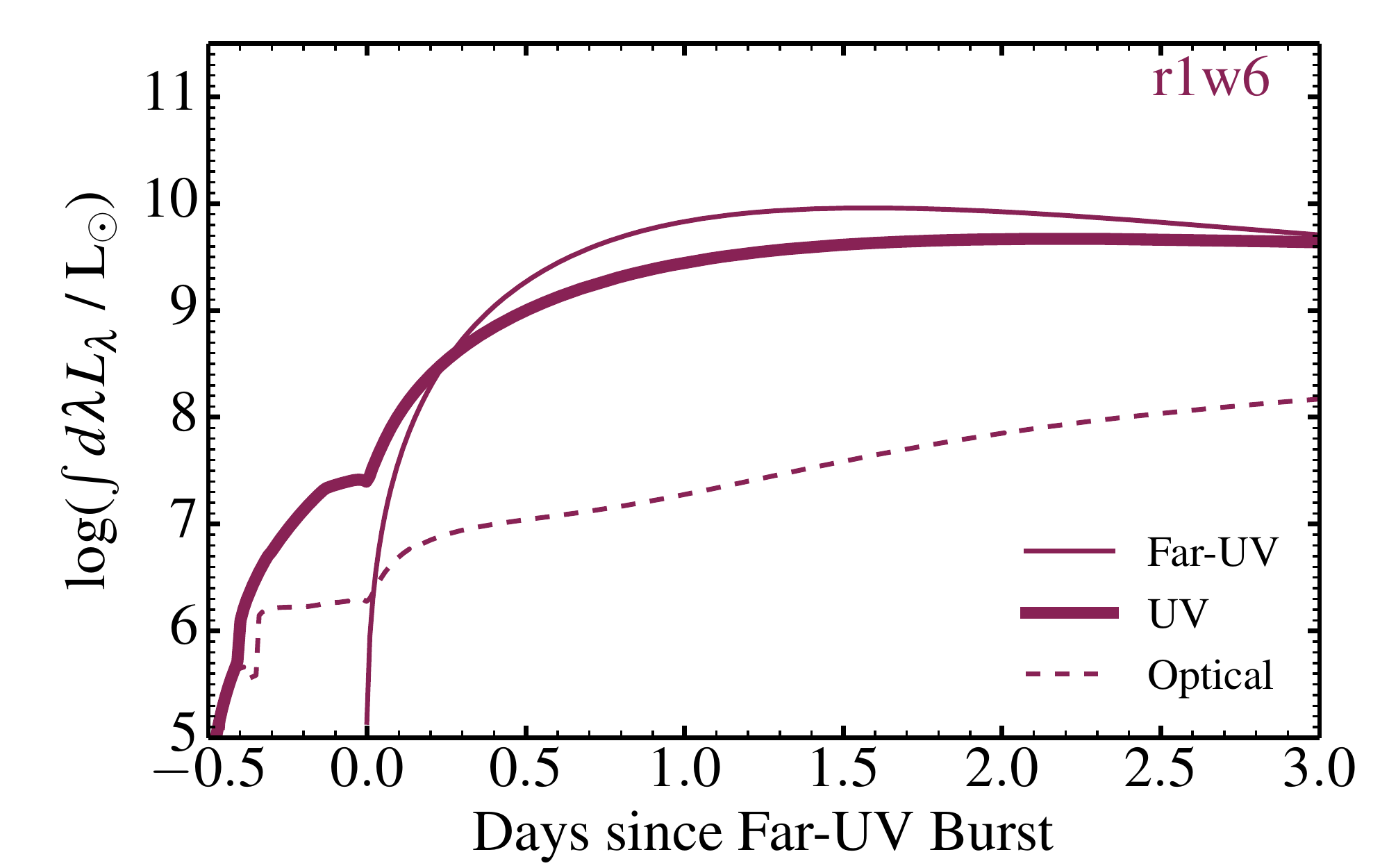,width=9.2cm}
\caption{Far-UV, UV, and optical light curve computed with \heracles\ for models r1w1, r1w1h, r1w5r,
and r1w6.
In practice, we show the evolution of the luminosity in the energy group number 8 (Far-UV),
the luminosity from energy groups 6 and 7 between the Lyman and the Balmer edges (UV), and
the luminosity from energy groups 4 and 5 between the Balmer and the Paschen edges (optical).
Up to one day after the Far-UV burst, the bulk of the energy is emitted blueward of the Lyman
edge but progressively shifts to the UV until it falls mostly in the optical after about 10\,d.
Models with a higher mass loss rate (right column) have a broader and redder shock-breakout signal.
\label{fig_lbol_grp}
}
\end{figure*}

The presence of a dense wind leaves unambiguous signatures on the bolometric light curve.
In model r1w6 there is a sudden drop in bolometric luminosity at one week, which coincides with
the epoch when the interaction with the dense parts of the wind ceases.
When the interaction with the dense wind is over
in r1w3/r1w4/r1w5r/r1w6 at 5--10\,d, the light curve agrees very closely with
the weak-wind cases r1w1-r1w2.
A short-lived super-wind phase with a mass loss rate greater
than a few 10$^{-4}$\,\msunyr\ prior to core-collapse should therefore leave an imprint on
the bolometric light curve at early times (see also \citealt{moriya_rsg_csm_11}).
If RSG stars have a super-wind phase before core collapse, one should expect a diversity of early-time light
curve properties in a large sample of SNe II-P, with such breaks occurring at different epochs
and with different magnitudes.

\subsubsection{Sensivity to the progenitor radius}

In model r2w1, the progenitor star at death is more than twice as large.
The rise time to bolometric maximum and the shock breakout signal are three times
longer than in model r1w1, while the bolometric maximum is about the same (Table~\ref{tab_sum}).
The differences are caused by the increased light travel time across the progenitor star.

\subsubsection{Sensivity to atmospheric scale height}

Models with an extended atmospheric scale height (r1w1h and r1w5h) show a very different behavior
from the weak-wind model r1w1.
The rise time to maximum is short, the duration of the maximum phase is short, but the
bolometric luminosity is significantly higher for 10--15\,d. The higher density in the atmosphere
(but close to $R_{\star}$) leads to a strong interaction. Energy is deposited in this outer region
of moderate optical depth (i.e., $\sim$\,5000; see Table~\ref{tab_sum}), and released in the form
of radiation for 15\,d. The time-integrated bolometric luminosity is 2.06--2.15$\times$\,10$^{49}$\,erg and
comparable to that of model r1w6, but here the distribution of material in the atmosphere
is much more confined. In this case, there is no break in the light curve.
Compared to model r1w1h, the model r1w5h has a smaller density scale height but a greater
density at large distances, producing a longer precursor and a smaller luminosity boost to later times.

\subsubsection{Wavelength dependence of the shock breakout signal and observability}

The complex morphology of bolometric light curves shown in Fig.~\ref{fig_lbol_heracles}
appears differently in selected spectral bands.
Figure~\ref{fig_lbol_grp} shows the Far-UV, UV, and optical light curves from the \heracles\
simulations of models r1w1, r1w5r, r1w1h, and r1w6.
If the atmosphere/wind has a negligible mass/extent,
a clear burst of radiation is seen simultaneously in the Lyman, Balmer, and Paschen continua (top row
of Fig.~\ref{fig_lbol_grp}). However, in the presence of a dense/extended atmosphere/wind,
optical photons emerge earlier and more progressively than in the Balmer/Lyman continuum.
While a clear burst may occur in the
Far-UV, the optical signal will show a much slower rise with a smaller discontinuity (low-cadence
observations may still capture a burst, but because of a lack of resolution).
The multi-group approach in \heracles\ captures this effect, which arises because
the opacity for Lyman-continuum photons is orders of magnitude larger than for optical photons.

\begin{figure*}
\epsfig{file=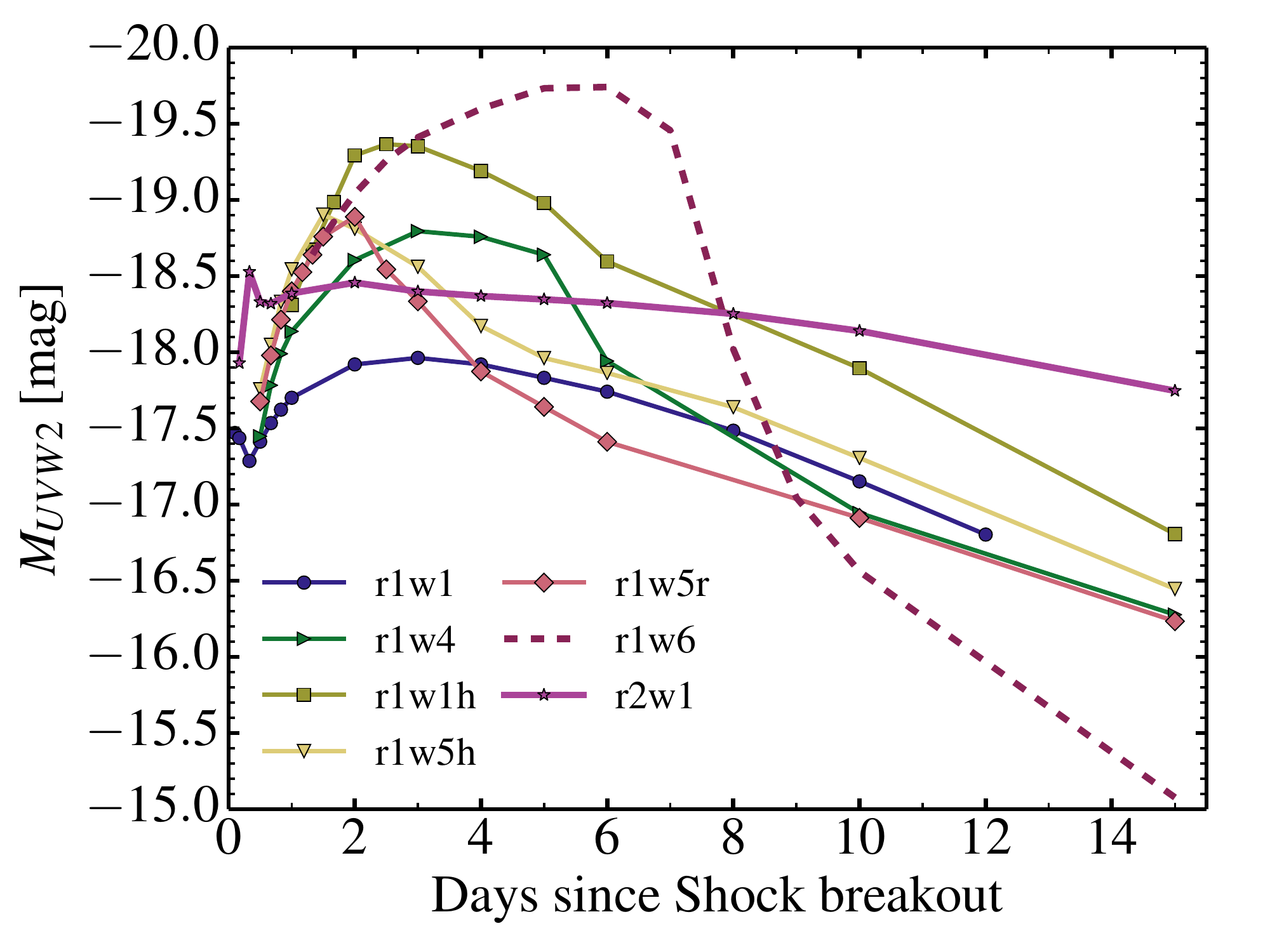,width=6.15cm}
\epsfig{file=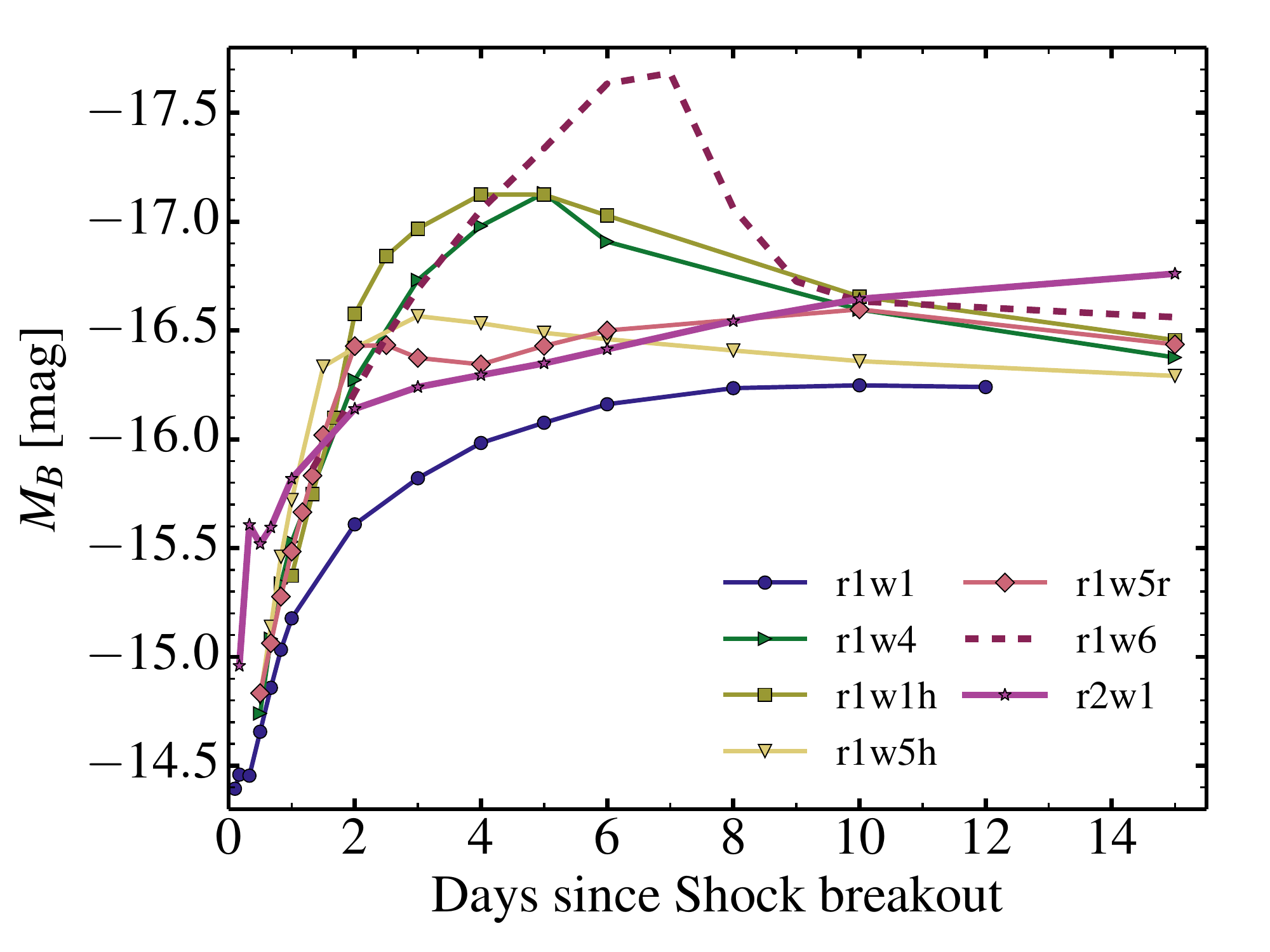,width=6.15cm}
\epsfig{file=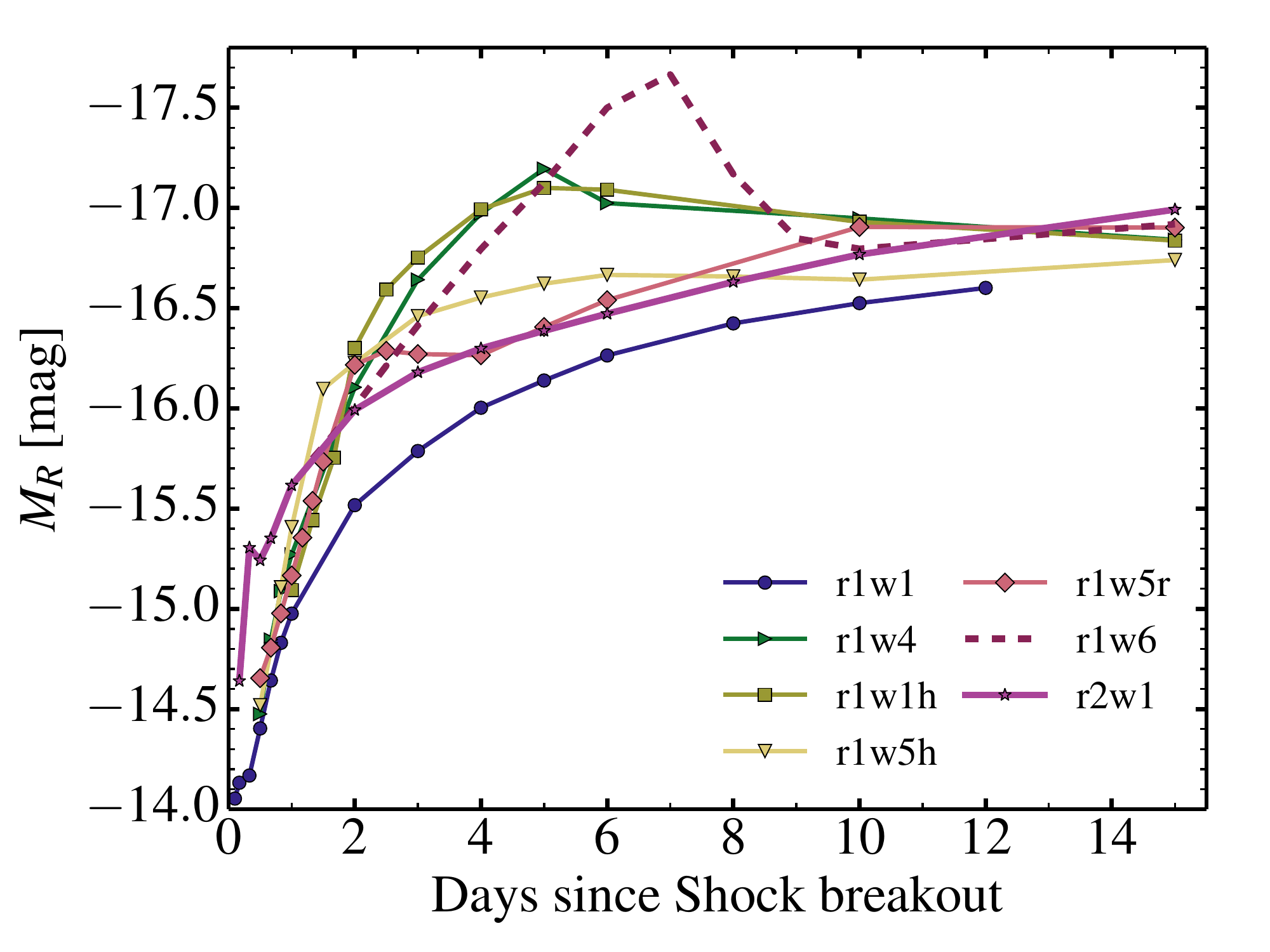,width=6.15cm}
\caption{
$UVW2$, $B$, and $R$ band light curves obtained with \cmfgen\ for models
r1w1, r1w1h, r1w4, r1w5h, r1w5r, r1w6, and r2w1.
The time origin corresponds to shock breakout (taken to be
when the luminosity at the photosphere rises to 10$^{42}$\,\ergs).
The variations in atmosphere/wind density have considerable impact on the early-time light curves,
generally reducing the rise times and potentially introducing breaks in the light curves (e.g., model r1w6).
\label{fig_lc_band}
}
\end{figure*}

\begin{figure*}
\epsfig{file=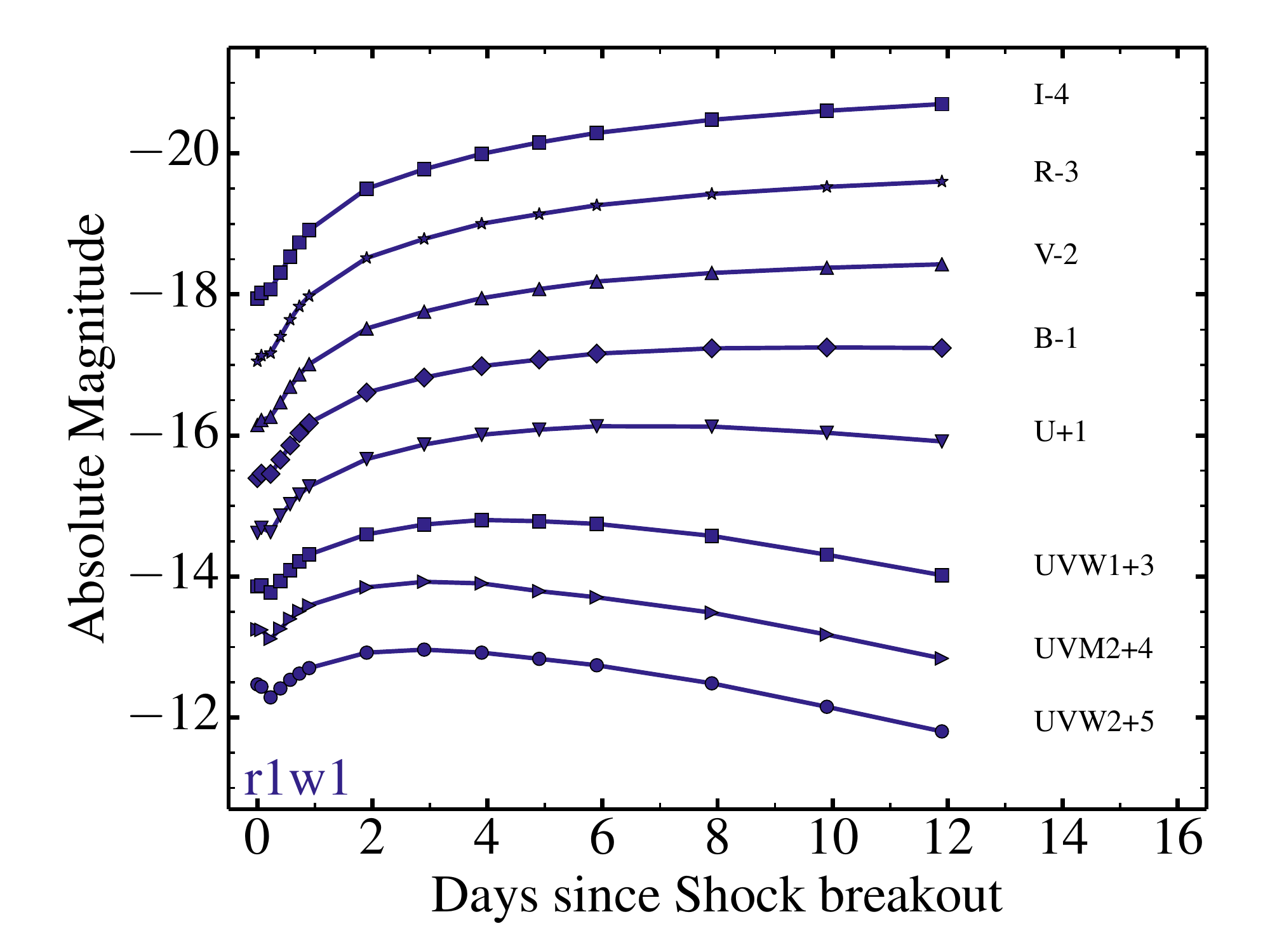,width=6.15cm}
\epsfig{file=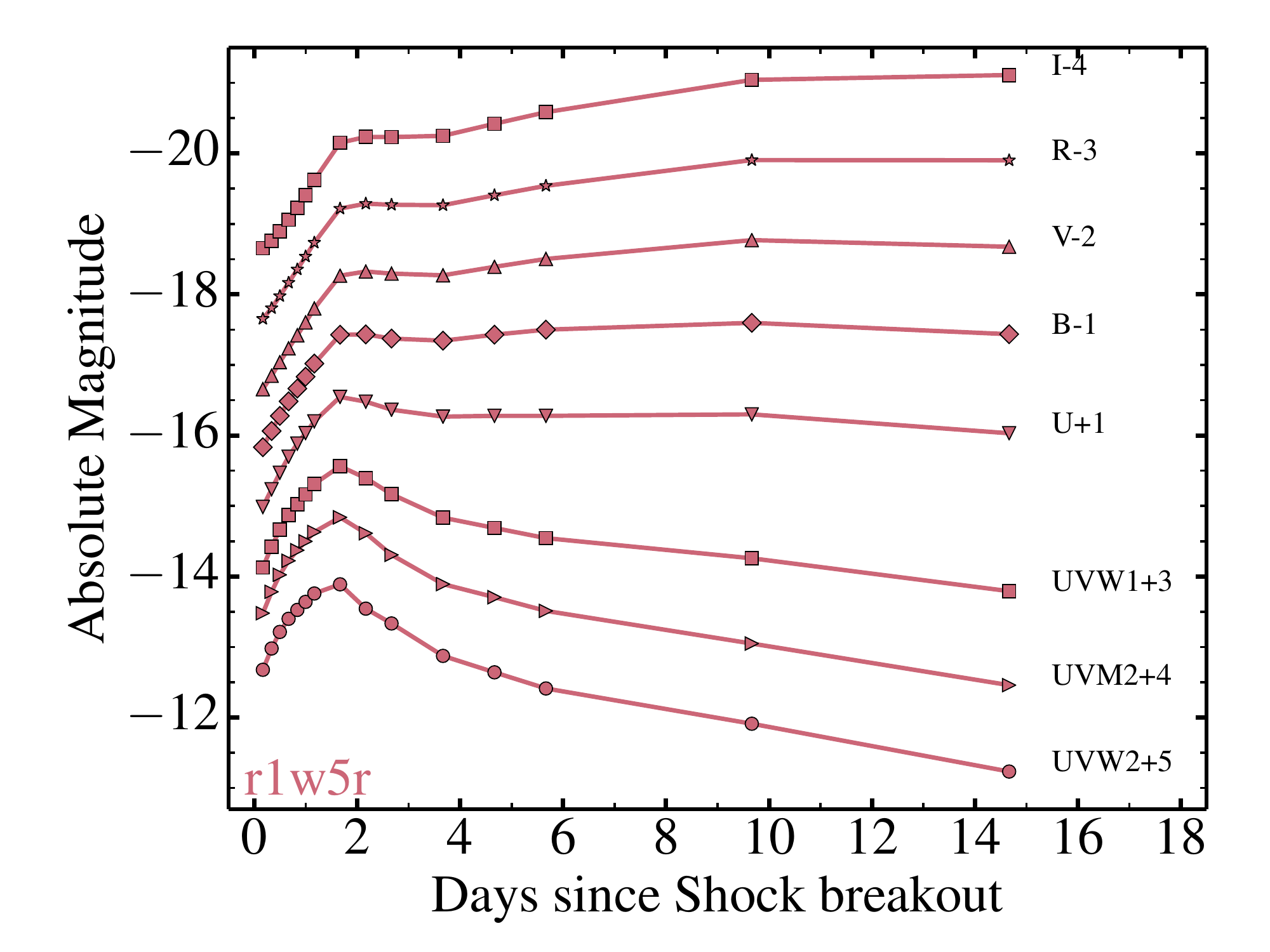,width=6.15cm}
\epsfig{file=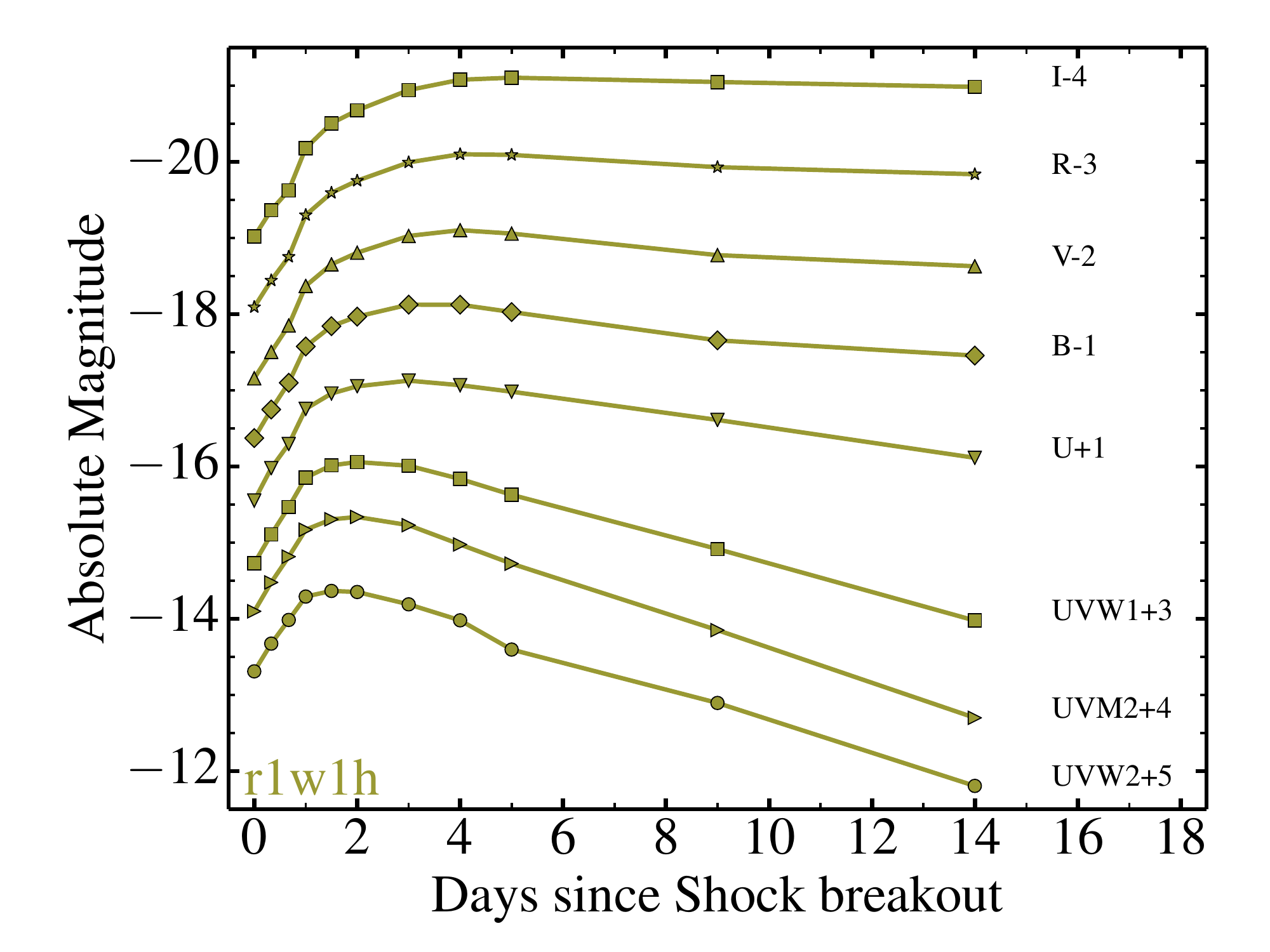,width=6.15cm}
\caption{
Multi-band UV and optical light curves for the weak-wind model r1w1 (left),
the strong wind model r1w5r (center), and the model with an extended
scale height r1w1h (right).
The time origin corresponds to shock breakout (taken to be when the luminosity at the photosphere
rises to 10$^{42}$\,\ergs). Models r1w5r and r1w1h both show much shorter rise times in most/all
bands than the weak-wind model r1w1.
\label{fig_mb_lc}
}
\end{figure*}

For a blackbody at a temperature $T$ the peak of the spectral energy distribution occurs
at $\sim 2900/T_4$\,\AA, where $T_4=T/10^4$\,K.
Because of the high temperatures in the spectrum formation region (roughly between
optical depth 0.1 and 10;
see Fig.~\ref{fig_var}), the bulk of the flux falls in the Lyman continuum for a few days,
then in the Balmer continuum up to about 10\,d, and in the optical at later times
(this evolution is model dependent). The diversity
of atmosphere/wind properties causes a large scatter in this evolution in our set of models.
For example, in the strong wind case (model r1w6), the radiation energy from shock breakout
$E_{\rm sbo}$ does not immediately escape but instead gets trapped within a large volume
$dV$ bounded by (roughly) $R_{\star}$ and the photospheric radius at about
5$\times$10$^{14}$\,cm.
The larger this volume $dV$, the lower the representative radiation temperature $T_{\rm rad}$
because $E_{\rm sbo} \sim a_{\rm R} T_{\rm rad}^4 dV$ ($a_{\rm R}$ is the radiation constant).
In model r1w6, the Far-UV flux is much weaker and the UV flux is boosted relative to model r1w1.

It is clear from Fig.~\ref{fig_lbol_grp} that the detection of shock breakout is very difficult in the optical,
although not impossible \citep{garnavich_sbo_16}. The challenge is both the brightness of the burst in
the corresponding spectral range, but also the way the brightness changes with time.
\citet{garnavich_sbo_16} report the very weak and short lived bump ($<$1\,hr)
in the optical light curve of KSN\,2011d,
 something that we do not actually obtain in our \heracles\ simulations, perhaps because of
 numerical diffusion and accuracy (the optical flux is one part in 10$^5$ of the Far-UV/UV flux).
 Hunting for shock breakout in the UV or the far-UV, where the jump in flux in orders of magnitude,
  is obviously more suitable \citep{cenko_cutie_17}.

\begin{figure*}
\epsfig{file=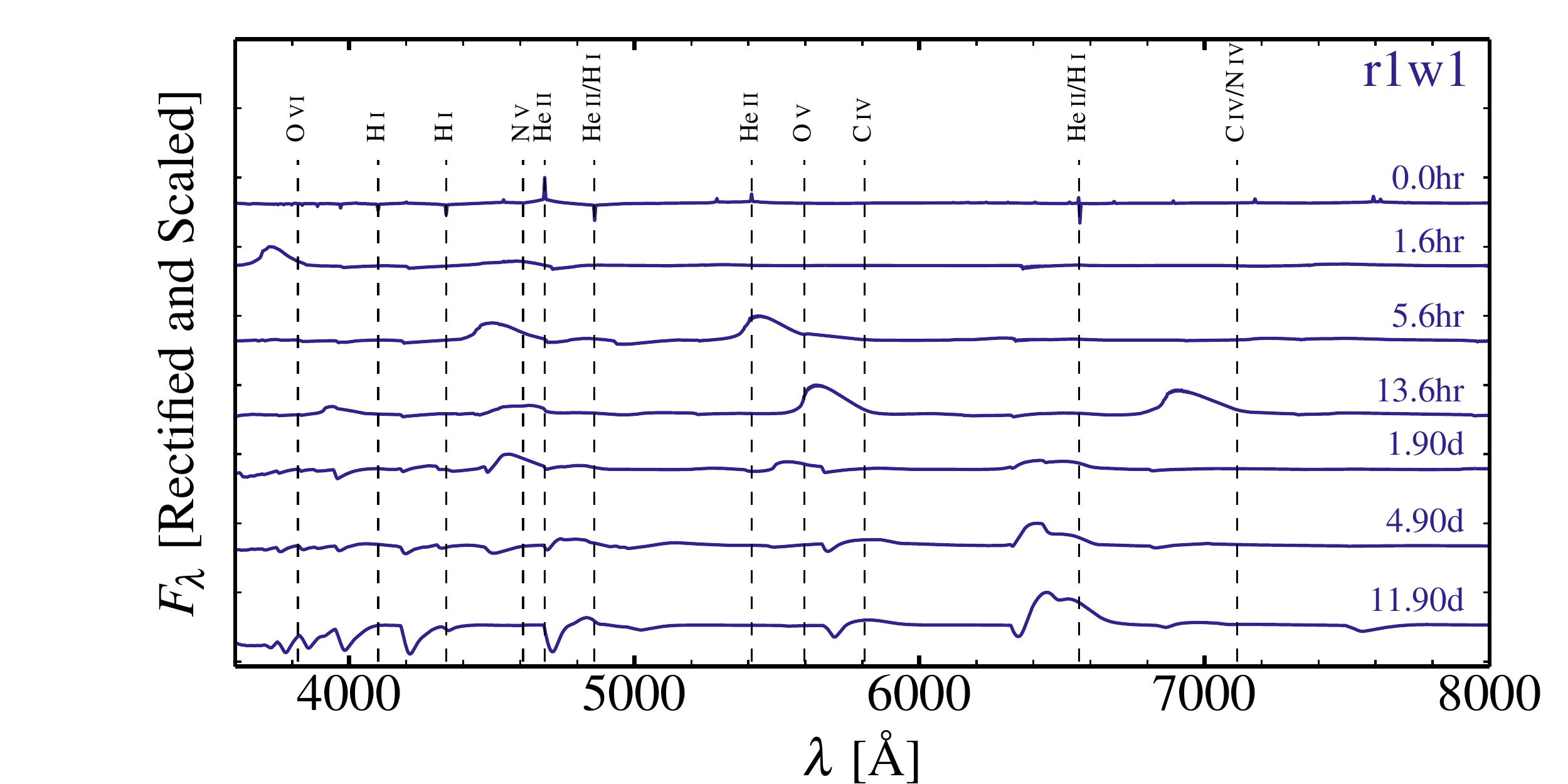,width=15.cm}
\epsfig{file=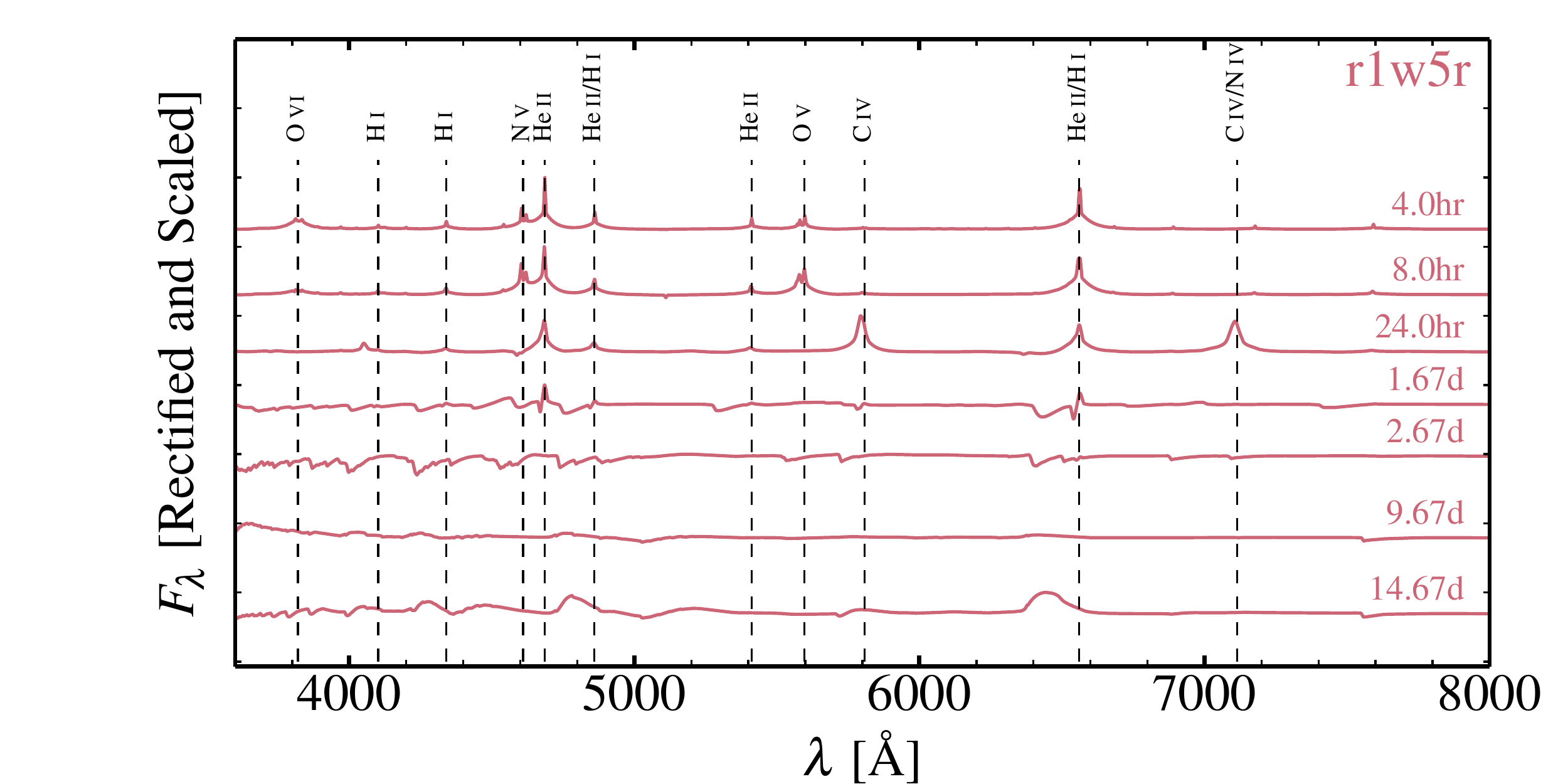,width=15.cm}
\epsfig{file=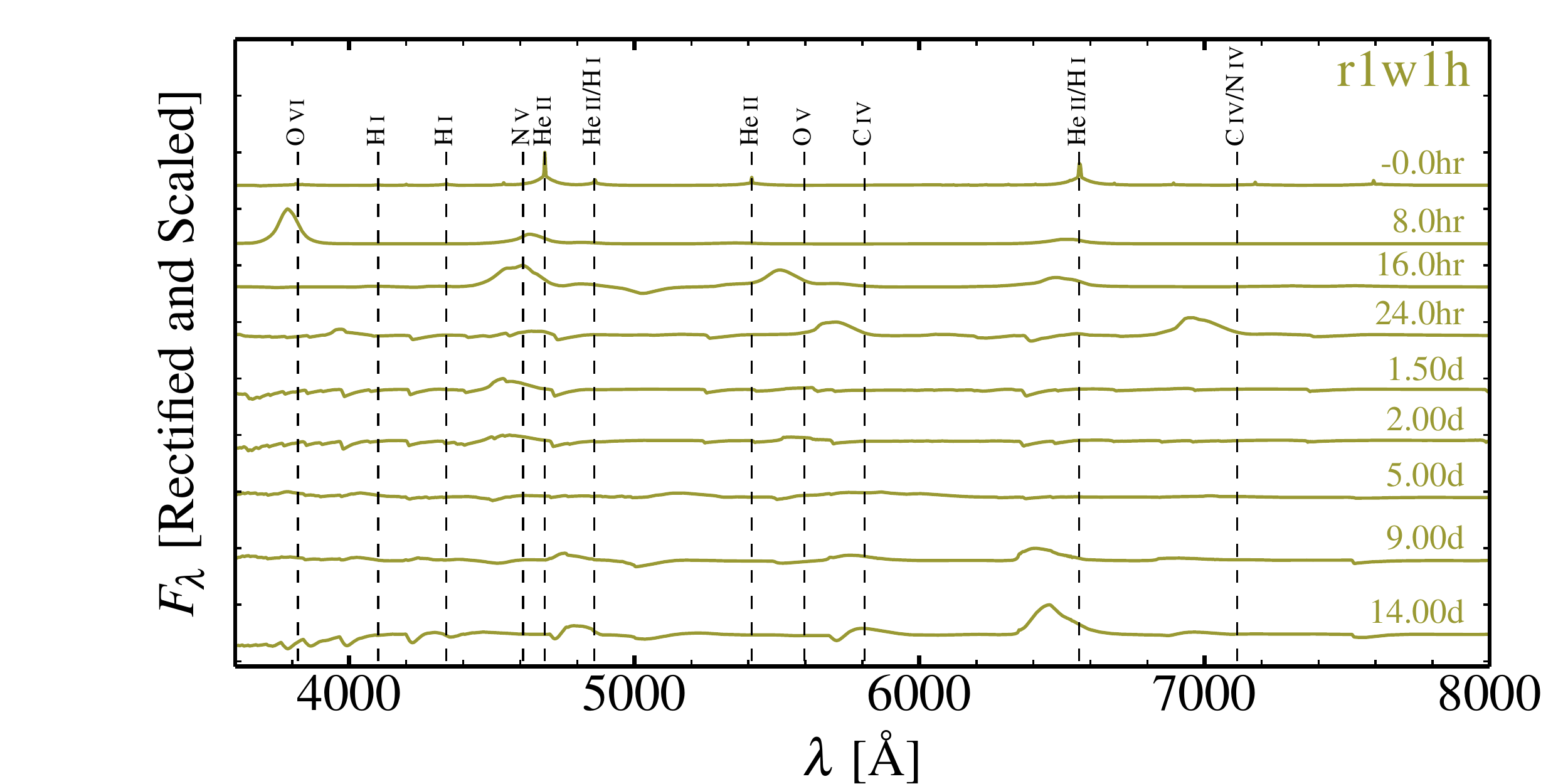,width=15.cm}
\caption{Same as Fig.~\ref{fig_mb_lc}, but now showing
multi-epoch optical spectra. We show the quantity
[$(F_\lambda/F_{\rm c} -1) \times \alpha + 1]$, where $F_{\rm c}$ is the continuum flux.
To better show the weak lines, we take $\alpha=$\,3.
The time origin corresponds to shock breakout (taken to be
when the luminosity at the photosphere rises to 10$^{42}$\,\ergs.).
All three models show narrow line profiles at the earliest times, but only the
models with a dense atmosphere/wind retain this property, for a duration that depends
on the density and extent of this external material, as well as the SN shock speed.
When spectral lines are primarily electron-scattering broadened, the peak emission
is centered on the line rest wavelength. When spectral lines are primarily Doppler broadened,
the peak emission is blueshifted, which can compromise a robust identification or lead
to an erroneous measure of the redshift if the blueshift is ignored, as in \citet{gezari_08es_09}.
\label{fig_spec}
}
\end{figure*}

\section{Results from non-LTE steady-state radiative transfer simulations with \cmfgen}
\label{sect_cmfgen}

For a selection of \heracles\ simulations (models r1w2 and r1w3 are omitted because they
are intermediate cases of adjacent models r1w1 and r1w4), we compute \cmfgen\ models
that solve for the non-LTE properties of the gas as well as the radiation field.
We assume steady state in \cmfgen\ but time dependence is accounted for in \heracles.
The flux computed in \cmfgen\ at a given time corresponds to a later time
in the \heracles\ simulation, by at most 0.5\,d. This inconsistency does not affect the relative spectral evolution
computed by \cmfgen.
The bolometric luminosity computed by \cmfgen\ (see Fig.~\ref{fig_lbol_cmfgen} in the appendix)
agrees to within 10-20\% with that computed by \heracles\ (modulo the $\lesssim$\,0.5\,d shift mentioned
above). For each snapshot treated, we also find that the \cmfgen\ results for the electron density
and optical depth are within a few per cent of those computed by \heracles.

\subsection{Multi-band light curves}

Figure~\ref{fig_lc_band} shows the light curves in $UVW2$ (Swift filter), $B$, and $R$
(absolute magnitudes) for our \cmfgen\ simulations.
A qualitatively similar evolution is obtained for other UV and optical bands.
We find that the brightness increases initially to reach a peak within a maximum of 6\,d in $UVW2$ ,
from 4--8\,d in $U$, 5--10\,d in $B$, from 6 to $>$15\,d in $V$, and from 5 to $>$15\,d in $R$.

In the weak-wind models r1w1 and r2w1, there is a discontinuity at the earliest times
in all three filters. This discontinuity corresponds to the shock breakout signal.
The optical brightness jump in model r2w1 is greater than in the corresponding \heracles\ model
because in this study \cmfgen\ assumes
steady state and therefore ignores light-travel time effects, while \heracles\ is time dependent
and accounts for time delays (spreading the breakout signal over $R_\star/c$).
In other models, the extended/dense atmosphere/wind spreads that signal (computed by \heracles\
or by \cmfgen) and no sharp discontinuity is seen.

In the case of a weak wind, the photometric evolution reflects the combined effects of expansion
(which increases the surface area of the emitting surface) and cooling (which shifts the peak
of the spectral energy distribution to longer wavelengths).
For a given model, the peak brightness occurs earlier in bluer bands.
In a given band, the peak brightness occurs earlier for a more compact progenitor star (compare
models r1w1 and r2w1).
Models with increasing wind mass loss rates exhibit a brightness boost in all bands from the stronger
interaction, but with an obvious break in models r1w4, r1w5r, and r1w6 when that interaction ceases
(or when the atmosphere/wind has been completely swept-up so that it no longer introduces
an optical-depth effect).

Apart from model r2w1, all models have the same explosion energy and ejecta mass,
so the presence of a dense atmosphere/wind alone can introduce significant UV-optical light
curve variations at early times,
up to 1--2\,mag for this set of models and displace the times of maximum by more than a week.
Once interaction is weak in all models ($\gtrsim$\,10\,d), the magnitude scatter in a given optical band
is $\lesssim$\,0.2\,mag, but remains very large in the UV (if we exclude the strong wind model r1w6
and the model from a bigger star r2w1, the scatter is only 0.6\,mag at 15\,d).

Models with strong wind mass loss exhibit a break in all UV/optical bands (as we obtain
in the \heracles\ light curves; Fig.~\ref{fig_lbol_heracles}).
This is not a generic property of SNe II light curves,
but this jump could be turned into a more progressive decline by using a slowly declining mass loss rate
(i.e., from $>$10$^{-4}$\,\msunyr down to 10$^{-6}$\,\msunyr; see Fig.~\ref{fig_init_set}).
In contrast, no break is seen in the models with an extended atmospheric scale height, and the
rise time to maximum in optical bands is brought to earlier times, in better agreement with observations
\citep{gonzalez_gaitan_2p_15}.

\begin{figure*}[h]
\epsfig{file=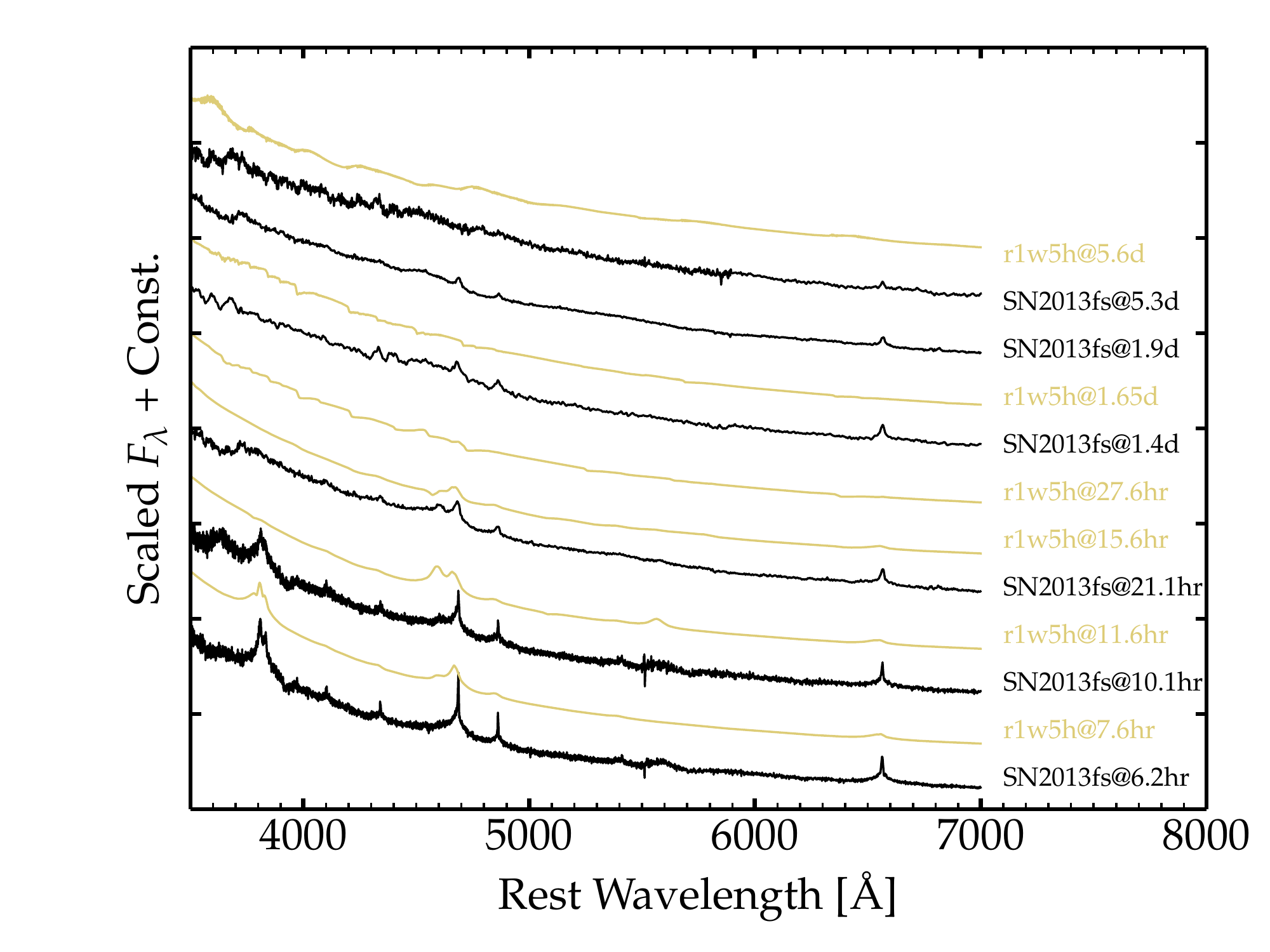,width=15.5cm}
\epsfig{file=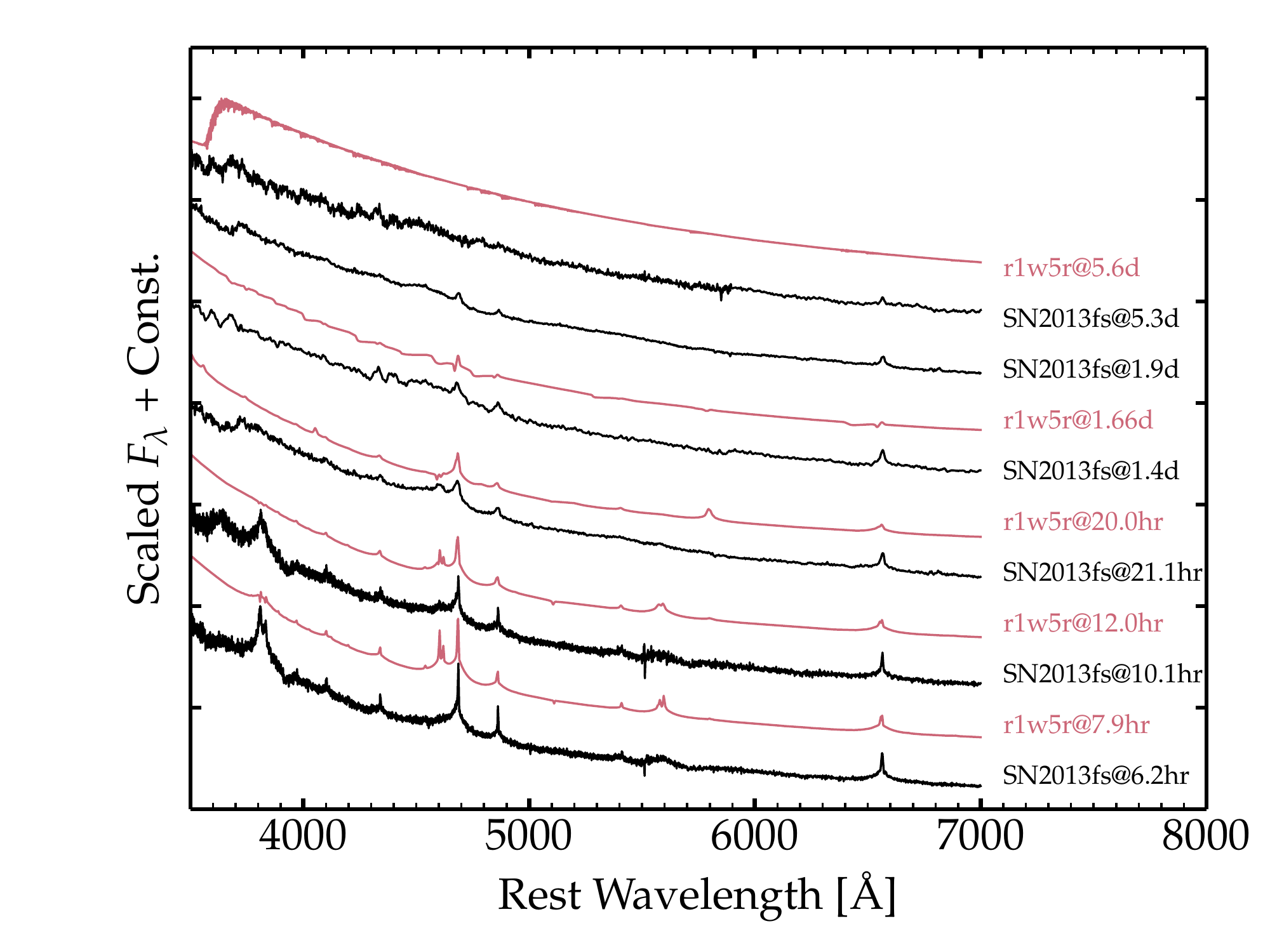,width=15.5cm}
\caption{Comparison of the multi-epoch spectra of SN\,2013fs at early times
with the spectra for models r1w5h (top) and r1w5r (bottom).
All spectra have been scaled and shifted.
The observations have been corrected for redshift but not for reddening.
These two models roughly encompass the properties of SN\,2013fs,
with small offsets in the persistence of some lines (e.g., O\six\,3811--3834\,\AA).
[See Section~\ref{sect_spec} for discussion.]
\label{fig_13fs}
}
\end{figure*}

\subsection{Models of the spectral evolution}
\label{sect_spec}

In this section, we discuss the spectral evolution up to 15\,d after shock breakout
for three representative models of atmosphere/wind configurations: 1) weak wind (model r1w1;
top panel of Fig.~\ref{fig_spec}); 2) strong wind (model r1w5r; middle panel of Fig.~\ref{fig_spec});
3) extended atmospheric scale height (model r1w1h; bottom panel of Fig.~\ref{fig_spec}).
The corresponding multi-band light curves are shown for completeness in Fig.~\ref{fig_mb_lc}.
In the appendix, we show the complete set of spectral calculations for all models
(Figs.~\ref{fig_spec_r1w1_all}--\ref{fig_spec_r2w1_all}).

In all cases, the evolution of the spectrum over that timespan reflects the change in ejecta
properties in the spectrum formation region.
The gas temperature drops from $\sim$\,10$^5$ down to $\sim$\,10$^4$\,K
(which affects the ionization, and hence the sources of opacity/emissivity,
themselves impacting the colors as well as the lines produced).
The ejecta accelerates, as well as some of the wind material,
while some ejecta deceleration takes place if the wind is very dense
(this modulates the Doppler broadening of line profiles).
The frequency redistribution by electron scattering ebbs as the atmosphere/wind optical depth drops.
In nearly all cases, the spectra show at least once the lines associated with
H\,\one\,4340\,\AA, 4862\,\AA, 6562\,\AA,
He\one\,5875\,\AA, 6678\,\AA, He\two\,4686\,\AA, 4860\,\AA, 5411\,\AA, 6562\,\AA,
C\four\,5801--5812\,\AA, 7110\,\AA,
N\four\,4057\,\AA, 7122\,\AA, N\five\,4610\,\AA, O\five\,5597\,\AA,
and O\six\,3811--3834\,\AA.
The spectral evolution is extremely rapid in our models (as observed, e.g., in SN\,2013fs; \citealt{yaron_13fs_17}), so that some lines/ions may only be present for a few hours.
What differs between models is the duration over which these lines are seen and their
morphologies. The critical element is whether line broadening is caused by non-coherent
scattering with thermal electrons (which produces symmetric lines centered at the rest wavelength)
or by large bulk motions (which produces asymmetric lines with a blueshifted peak emission;
see Section~6.2 and Fig.\,15 of \citealt{d09_94w}).

For the weak wind case (top panel of Fig~\ref{fig_spec}; the figure shows rectified spectra that
have been scaled to better reveal the weak lines), the first spectrum, at the time of shock breakout,
show narrow lines with electron-scattering wings (He\two\,4686\,\AA).
At 1.6\,hr, the spectrum then shows Doppler-broadened lines of O\six, N\five, and He\two,
all with a marked blueshift in peak emission.
At 5.6\,hr, O\six\ has disappeared and we see lines of N\five, He\two, and O\five.
At 13.6\,hr, O\five\ has disappeared and we see lines of N\four, N\five, He\two, and C\four.
As the temperature continues to drop, the ionization decreases, the spectrum
is less and less blue, and we eventually see the typical Type II-P SN  spectrum with lines
of H\one\ and He\one.
Because of optical depth and occultation effects, all lines have a blue-shifted peak emission
\citep{DH05a,anderson_blueshift_14}, except at the first epoch.
A blue-shifted line emission peak is unambiguous evidence
that the line is not primarily broadened by non-coherent electron scattering, which instead
tends to cause a symmetric (narrow) profile.

In contrast, for a stronger wind case (model r1w5r; middle panel of Fig~\ref{fig_spec}; the strong
wind case model r1w6 is shown in Fig.~\ref{fig_spec_r1w6_all}),
the spectrum exhibits narrow and symmetric emission profiles typical of Type IIn SNe for up to 1\,d.
At 4.0\,hr, we see the same lines as in the weak wind case but the lines, which show no absorption
component, are now centered on the rest wavelength.
Lines form in the slow moving (or weakly accelerating wind) and are affected by frequency
redistribution by thermal electrons.
At 12.0\,hr, the ionization has dropped (O\six\ has disappeared and O\five\ is present)
but the narrow emission features are still present.
After 1\,d, there is only weak evidence for the reprocessing of photons by the atmosphere/wind.
Its optical depth is too small to produce
strong electron-scattering wings, although we do see a very narrow emission in He\two\ for up to 2\,d.
The spectrum starts to show P-Cygni profiles with a broad blueshifted
absorption and a very weak emission. At this time, the spectrum forms in the dense shell of swept-up
atmosphere/wind material. The lack of an emission component in line profiles is caused by
the steeply declining density profile within that dense shell.
Such properties and evolution are reminiscent of SN\,1998S \citep{D16_2n}.

\begin{figure*}
\epsfig{file=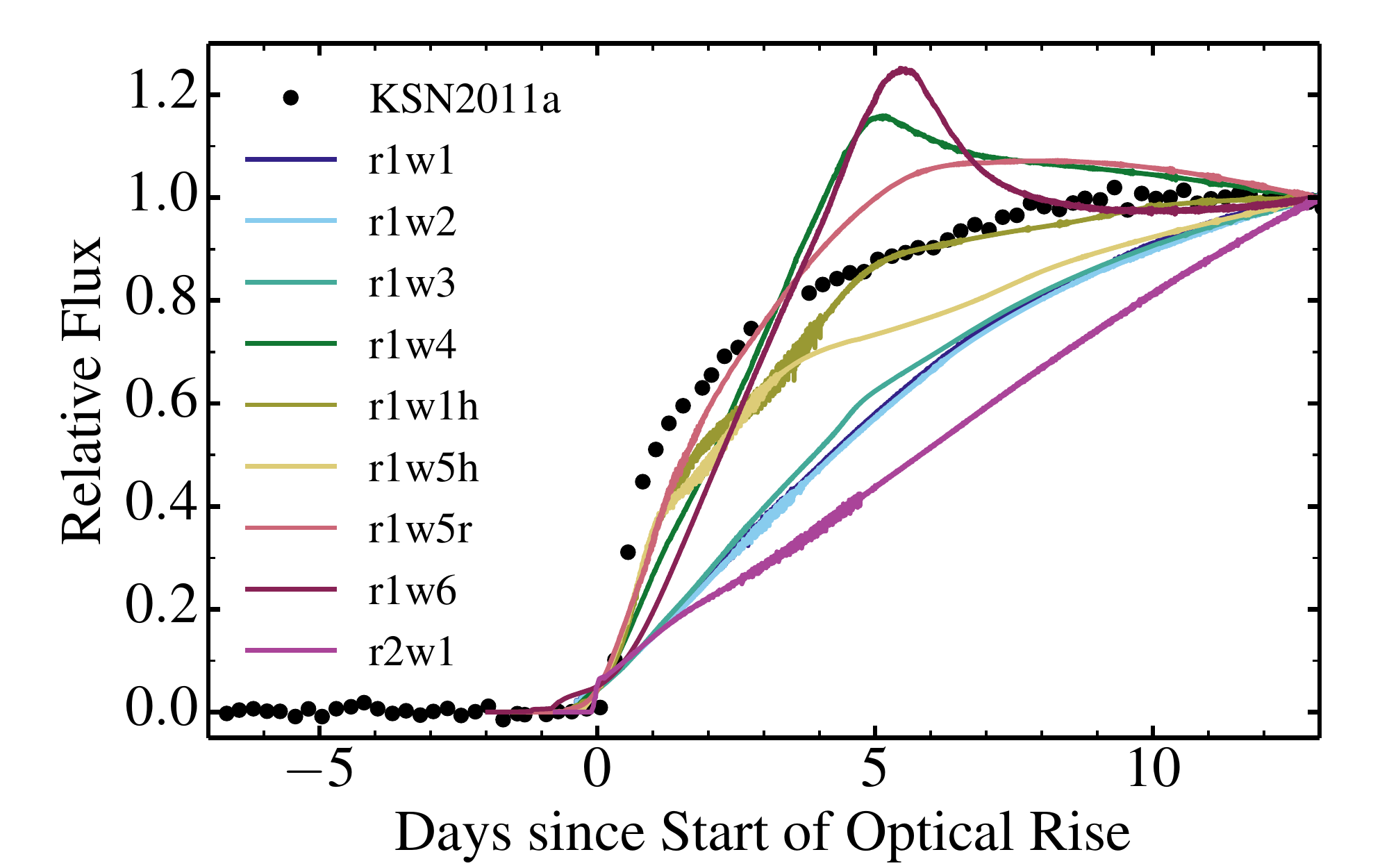,width=6.15cm}
\epsfig{file=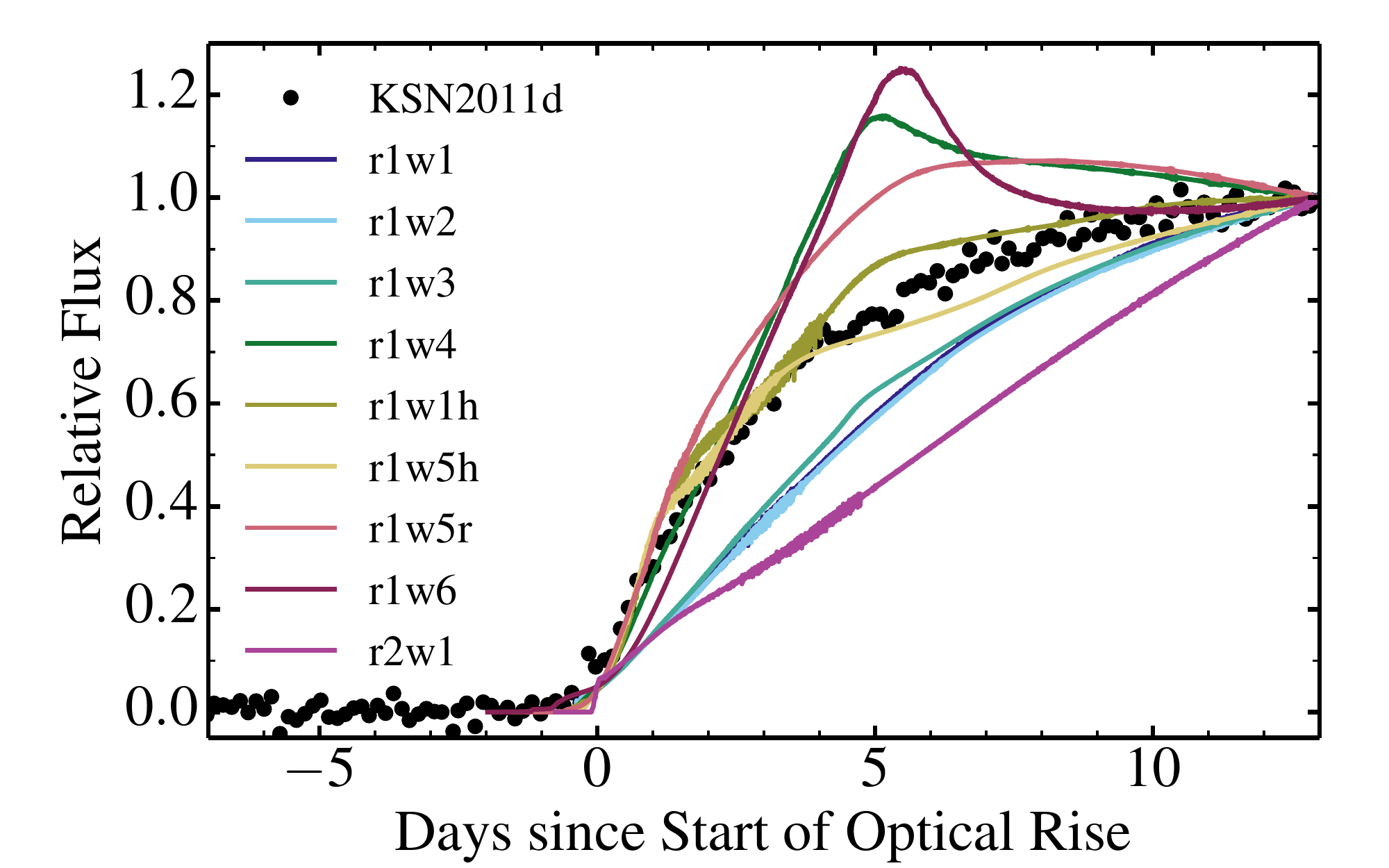,width=6.15cm}
\epsfig{file=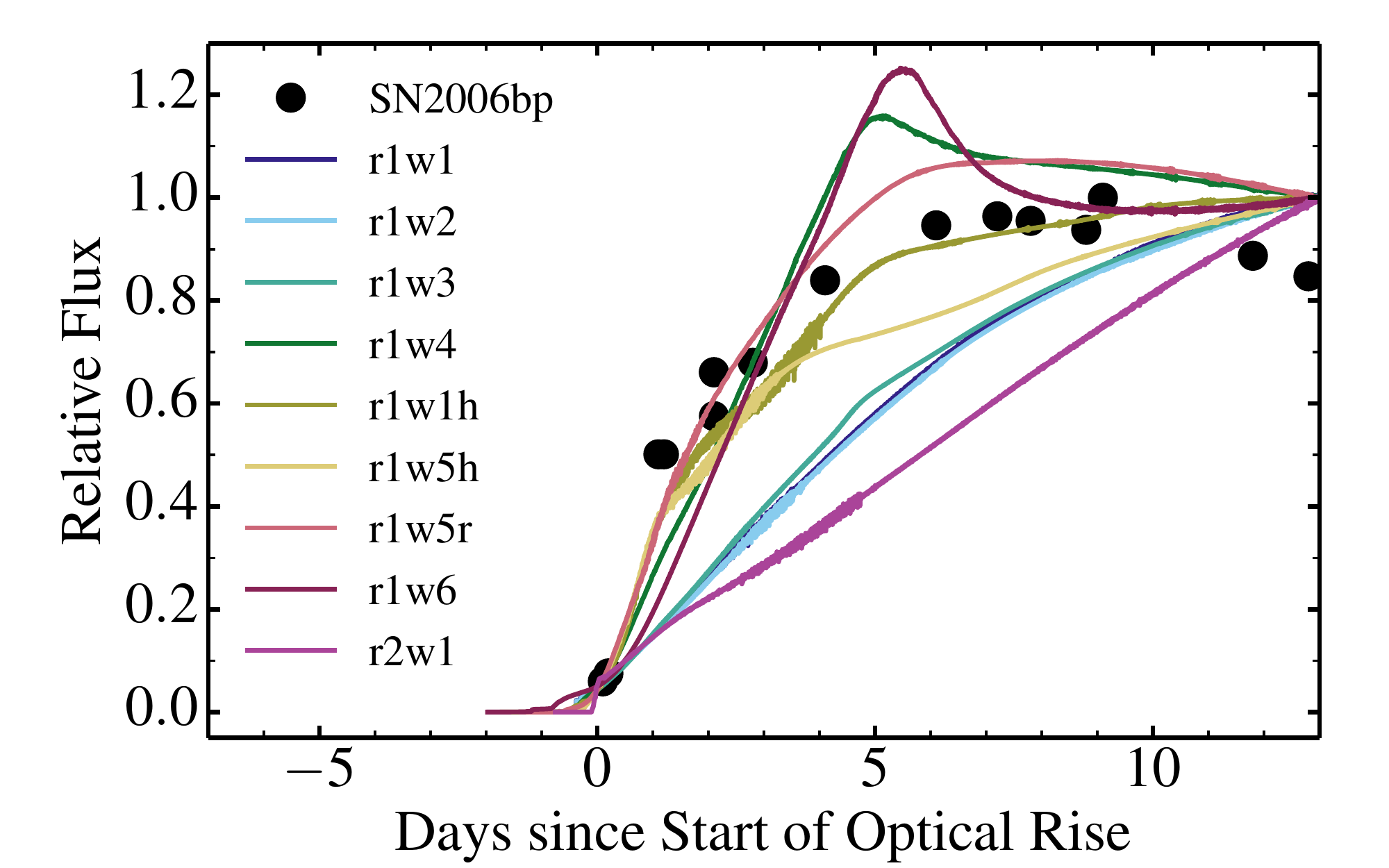,width=6.15cm}
\caption{
Comparison between the optical flux (relative to the flux at 13\,d) from our \heracles\ simulations
(we account for the flux between the Balmer and Paschen edges) and the observations of
KSN\,2011a (left), KSN\,2011d (middle; \citealt{garnavich_sbo_16})
and SN\,2006bp (right; \citealt{quimby_06bp_07}).
The time origin is conveniently adjusted to correspond to the rise of the optical flux.
A better match to the light curves is obtained for models with a dense/confined atmosphere/wind.
None of our models match the fast rise of KSN\,2011a.
\label{fig_kepler}
}
\end{figure*}

For the model r1w1h with an extended atmospheric scale height
(bottom panel of Fig~\ref{fig_spec}), despite the large amount of
atmospheric material
(more than ten times as much as in model r1w5r), the signatures of interaction
(i.e., narrow symmetric lines) are present only in the first spectrum,
which coincides with the time of shock breakout. Only He\two\,4686\,\AA\
shows this morphology. The spectral evolution proceeds as in model r1w1, although emission
profiles are much stronger because the density does not fall as steeply as in model r1w1.
The biggest difference between model r1w1h and model r1w1 is not spectroscopic but
bolometric and photometric, with model r1w1h showing a much more  extended phase of
high luminosity and an earlier time of maximum in all UV/optical bands
(Figs.~\ref{fig_lbol_heracles}--\ref{fig_lbol_grp}--\ref{fig_lc_band}).

All models are shown in the appendix, with properties that vary somewhat from the above.
For example, in the strong wind cases r1w4 and r1w6, the ionization seen in the spectra at all epochs
is much lower than in other models with a more confined and tenuous atmosphere/wind.
One signature of this is the absence of O\five\ and O\six\ lines in model r1w6.
In model r1w4, O\six\ is also absent but O\five\ is present. In model r1w5r, the progenitor
environment is dense but
less extended. Consequently, the atmosphere/wind temperature reaches much higher temperatures
and the spectral lines identified are similar to those in the weak wind model r1w1.
In other words, the extent/mass/density of the atmosphere/wind has a strong influence on the ionization
in the spectrum formation region, which complements what can be inferred from the
line profile morphology.

While we do not have a model that fits all multi-epoch spectra of SN\,2013fs, our set
of models show a good correspondence at selected epochs for a sample of features.
For example, model r1w5h follows closely the ionization seen in SN\,2013fs (top
panel of Fig.~\ref{fig_13fs}), while
model r1w5r shows a line profile morphology that evolves closely to that seen in SN\,2013fs
(bottom panel of Fig.~\ref{fig_13fs}).
Building a full progenitor/explosion/interaction model for any observed SN  is obviously a challenge.
The model presented in \citet{yaron_13fs_17} fits better the observation, but it does so at one
epoch only, has not hydrodynamical consistency, and imposes radiative equilibrium.
So, our approach lacks the flexibility needed to achieve a good fit (this would require
hundreds of simulations), but it has physical consistency so that a grid of models,
as presented here, can cover the broad parameter space in which a given SN resides
and identify the basic trends.


\section{Comparison to other observations}
\label{sect_obs}

Figure~\ref{fig_kepler} shows a comparison of optical light curves from our \heracles\ simulations
(we account for the flux between the Balmer and Paschen edges) with the observations
of KSN\,2011a and KSN\,2011d \citep{garnavich_sbo_16}, and SN\,2006bp \citep{quimby_06bp_07}.
Our explosion models from a big star (r2w1) and/or weak winds rise too slowly in the
optical.  Models accounting for an extended dense wind produce a flux excess in the optical
light curve that is not observed. The best match to the KSN\,2011d and  SN\,2006bp is for models with an
extended scale height or with a dense wind confined to the surface of the star (model
r1w5r).\footnote{These \heracles\ optical light curves reflect closely the $R$-band light curve properties obtained
with \cmfgen\ (see Fig.~\ref{fig_lc_band}).} None of our models, even with interaction, match the fast rise
of KSN\,2011a.

Our results are somewhat in tension with \citet{garnavich_sbo_16}, who find a
good match for KSN\,2011d for a RSG with a weak or no wind and a radius of 490\,\rsun.
For KSN\,2011a, they argue for wind interaction and/or a compact RSG progenitor with
a radius of 280\,\rsun. Their inferences are based on the semi-analytical modeling
of \citet{rw11}, who assume that the progenitor density profile goes as $1/r^{1.5}$.
This density profile corresponds to the deep convective envelope, not the outer layers
in which the SN spectrum forms for a week, so this may be the origin of the difference
in our results. It still remains very surprising that a RSG could radiate
in the optical at 1\,d as much as $\sim$\,50\% of the flux that it radiates at 10--15\,d.
This is the case for SN\,1987A, but KSN\,2011a does not have a Type II-pec light curve morphology.
Optical spectra for this SN would probably have helped understanding this very fast optical rise.

The KSN\,2011d observation of \citet{garnavich_sbo_16} also reveals an initial short and weak optical burst
of $<$\,1\,hr. As discussed earlier, our \heracles\ simulation do not show such a sharp feature in
the optical light curve, probably because of an accuracy issue (numerical diffusion). The \cmfgen\
simulations for model r2w1 show a sharp and small optical burst but this is most likely because
of the neglect of time delays. However, in the case of KSN\,2011d,
the observation of this bump suggests that the RSG progenitor was not surrounded by an extended
and dense wind.

In addition, the earliest time spectra of SN\,2006bp do not exhibit narrow line profiles
with extended symmetric wings. Instead, SN\,2006bp shows a P-Cygni profiles with
blue-shifted emission peak (e.g., He\two\,4686\,\AA, C\four\,5808\,\AA, and H$\alpha$),
together with a narrow component at the corresponding rest wavelength \citep{quimby_06bp_07}.
The strong blueshifted emission is clear evidence that the atmosphere/wind material,
if present at the corresponding epoch, is already optically thin (otherwise line emission
would peak at the rest wavelength and electron-scattering wings would be seen).


\section{Conclusion}
\label{sect_conc}

   We have used 1-D Eulerian multi-group radiation-hydrodynamics and 1-D non-LTE radiative transfer modeling
   to characterize the bolometric, photometric, and spectroscopic signatures of RSG explosions embedded
   in an atmosphere/wind of modest extent
   (within $\sim$\,10\,$R_{\star}$) and mass ($\lesssim$\,10$^{-1}$\,\msun).
   Our work is conceptually analogous to the studies of \citet{moriya_rsg_csm_11} and \citet{morozova_2l_2p_17}
   but includes the computation of multi-epoch spectra as a post-treatment of multi-group radiation hydrodynamics.
   Our initial ejecta/atmosphere/wind structures also extend to low density (and large radii),
   which ensures the proper computation of the shock breakout and the associated burst of radiation.
   For the spectral calculations, we work from physical models of the explosion/interaction and make allowance
   for the non-monotonicity of the velocity field, both not treated in \citet{yaron_13fs_17}. The numerical
   approach is similar to a recent study for SN\,1998S \citep{D16_2n}.
   Our simulations are 1D and therefore do not predict the disruption of the cold-dense-shell that forms
   from the swept-up atmosphere/wind material.
   We also adopt a spherically-symmetric density distribution for this external material initially, although there
   is ample evidence that RSG environments are structured and complex.

    In the case of a weak progenitor wind (model r1w1), the model spectra show a blue nearly featureless spectrum
    for about a week, with lines of He\two, C\four, N\five, O\five, O\six\ that may be present for no more than
    a few hours. As time passes, the temperature and ionization decrease and eventually H\one\ and He\one\
    lines appear, producing a spectrum analogous to those of standard SNe II-P.
    In the absence of interaction, all lines form in the fast expanding ejecta and exhibit a strong peak blueshift
    at all times. Photometrically, this model shows a rise time in the $B$ band of about 10\,d.

    In the case of a strong progenitor wind (up to a maximum of 10$^{-2}$\,\msunyr\ in model r1w6),
    the spectral and photometric evolution during the first 15\,d
    are very different. As long as the spectrum forms in the slow (un-shocked) and optically thick wind,
    the lines exhibit a narrow emission peak with symmetric wings in place of a blueshifted
    Doppler-broadened emission peak. The dense wind material is eventually swept up into a dense shell.
    When the spectrum forms in that shell, the profiles start exhibiting a blueshifted and broad
    absorption with weak or
    no emission. Eventually, the photosphere recedes through this shell and the spectrum then
    resembles a more typical SN II spectrum with P-Cygni profile associated with H\one\ and He\one\ lines.
    Photometrically, the shock breakout in the progenitor environment is fainter and longer, except if the wind material
    is massive enough to allow the extraction of ejecta kinetic energy.
    In our simulations, we find that the time-integrated luminosity over the first 10--15\,d can be boosted
    by a factor of a few.

    The greater the luminosity boost from the ejecta/wind interaction, the greater
    the ejecta deceleration. In our strong-wind model r1w6, the maximum velocity is
    only $\sim$\,7000\,\kms\ at 10\,d, compared
    to $\sim$\,11000\,\kms\ in our weak-wind model r1w1. This implies narrower (Doppler-broadened) line profiles
    for the SN. In model r1w6, no line shows absorption/emission at $\gtrsim$\,7000\kms. The greater
    the wind mass loss rate, the greater the reduction. Because this deceleration does not affect the inner ejecta,
    the ejecta interaction with a dense wind should produce SNe that have narrower lines at early time,
    and normal line widths at later times. \citet{faran_sn2l_14} report that
    faster-declining Type II SNe indeed have narrower
    line profiles early on, but also broader ones at late times, which suggests ejecta/wind interaction alone may not
    explain fast-declining Type II. This needs to be considered when assessing the suitability of
    ejecta/wind interaction to explain SNe II-L \citep{morozova_2l_2p_17}.

    Early interaction with a dense wind
    reduces the rise time in optical bands and may help
    reduce the discrepancies with the observations \citep{gonzalez_gaitan_2p_15}.
    The early time light curve depends, however, on the properties of the atmosphere/wind
    (i.e., the atmospheric scale height and the wind density/extrent).
    We find that when the interaction with a dense wind (with \mdot$<$10$^{-2}$\,\msunyr)
    ebbs, the multi-band light curves show a break that is not obviously
    seen in Type II SN observations. A second issue with a strong wind is that they should come with a variety of
    strength in Nature, and therefore the observations should reveal a wide range of early time light curve
    properties, with breaks occurring at different epochs, function of the extent of the high density wind.

    We have also investigated configurations in which the RSG atmosphere has a larger scale height.
    The corresponding change in density profile brings important differences. The bolometric
    and multi-band light curves no longer show a break, the rise times are very short, the bolometric
    luminosity may be boosted for 15\,d, all of which seem to be in agreement with the observations
    \citep{gonzalez_gaitan_2p_15}. This stems from the strong interaction with the dense material at the
    base of the atmosphere. Its optical depth is large but not as large as the underlying ejecta so that the
    extra shock deposited energy can be released progressively over 15\,d.
    The spectral signatures of atmosphere/wind interaction
    are short lived. The extra energy in this configuration produces blue colors for longer.

\begin{figure*}
\epsfig{file=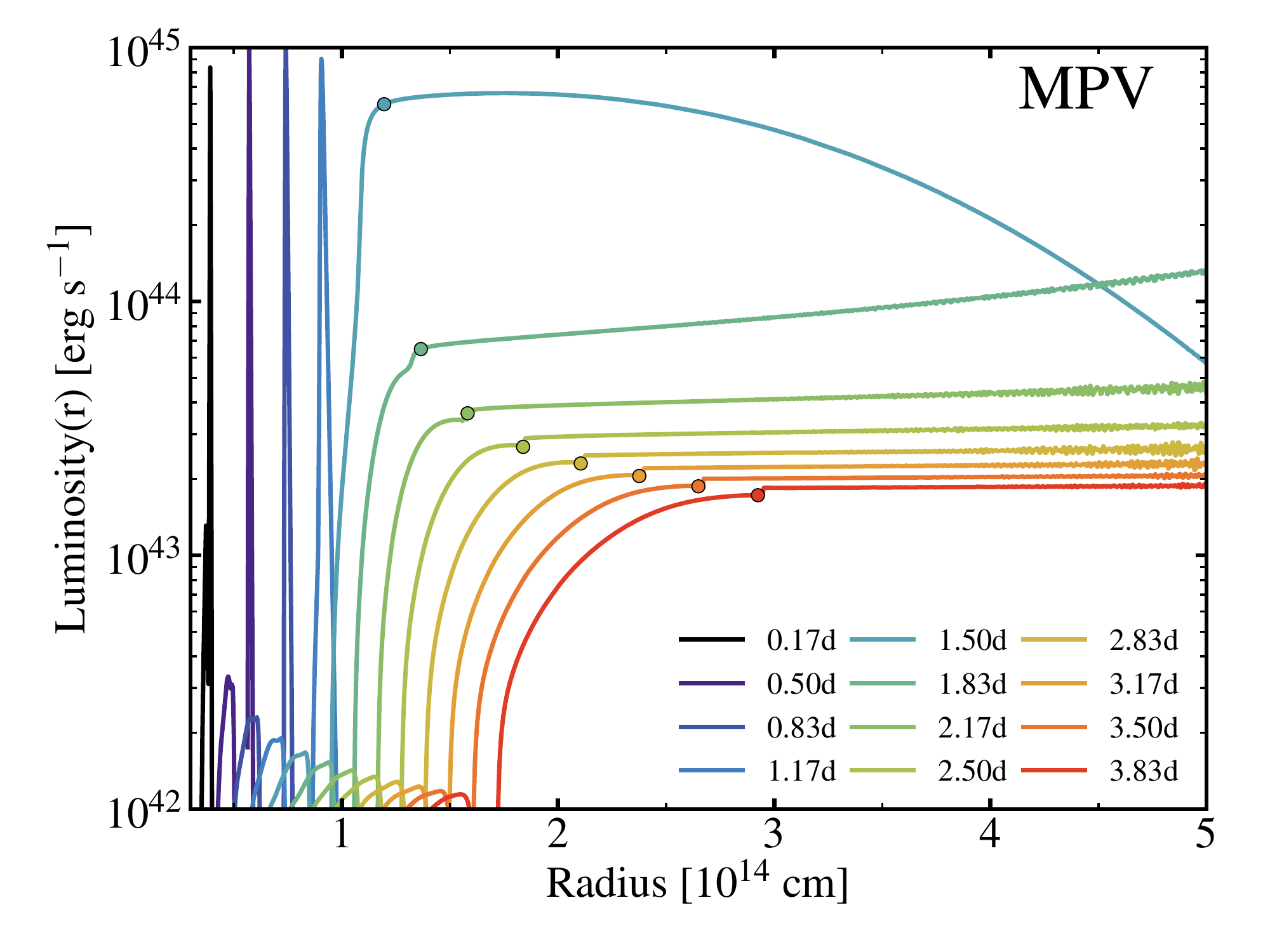,width=9.2cm}
\epsfig{file=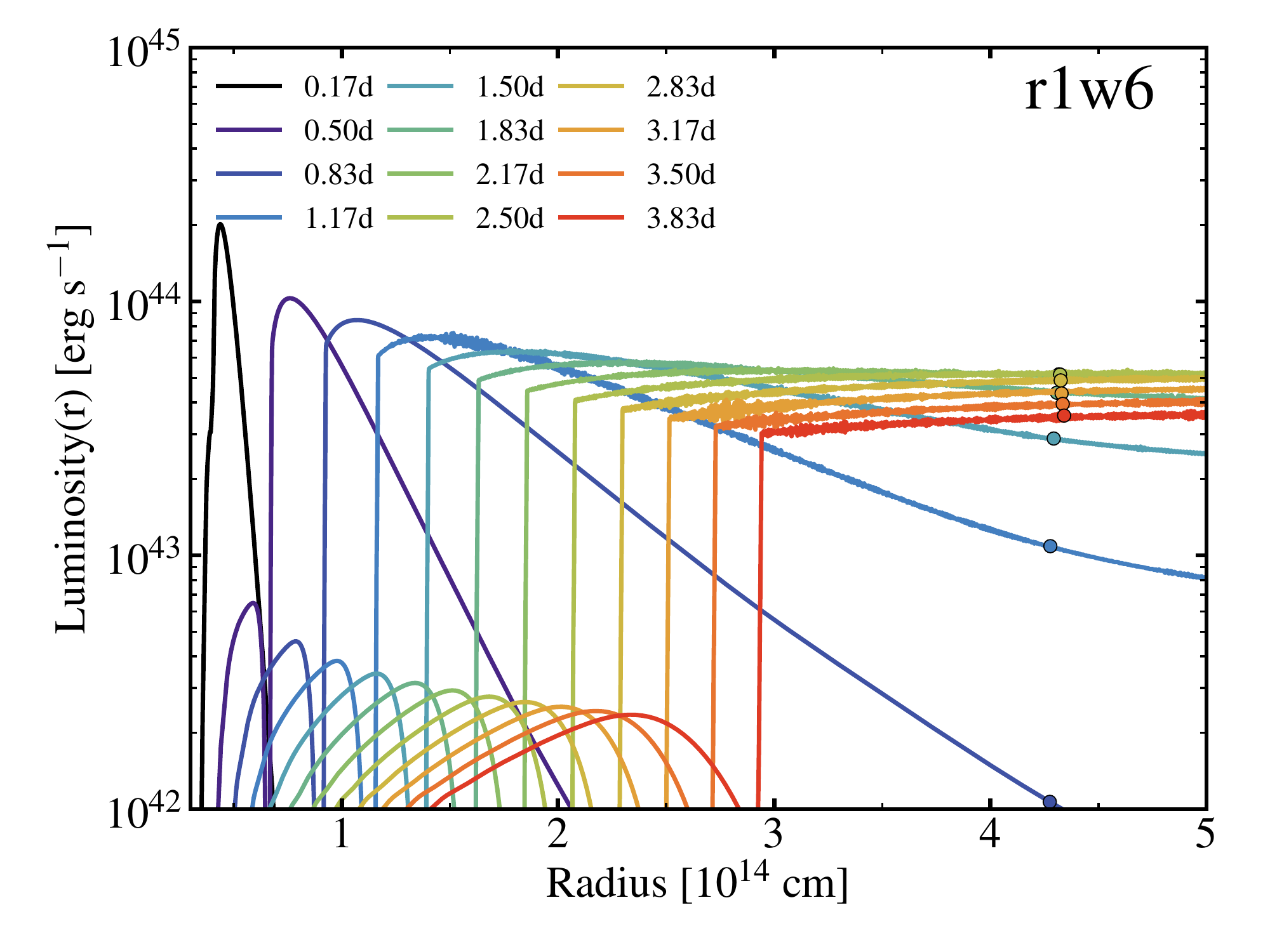,width=9.2cm}
\epsfig{file=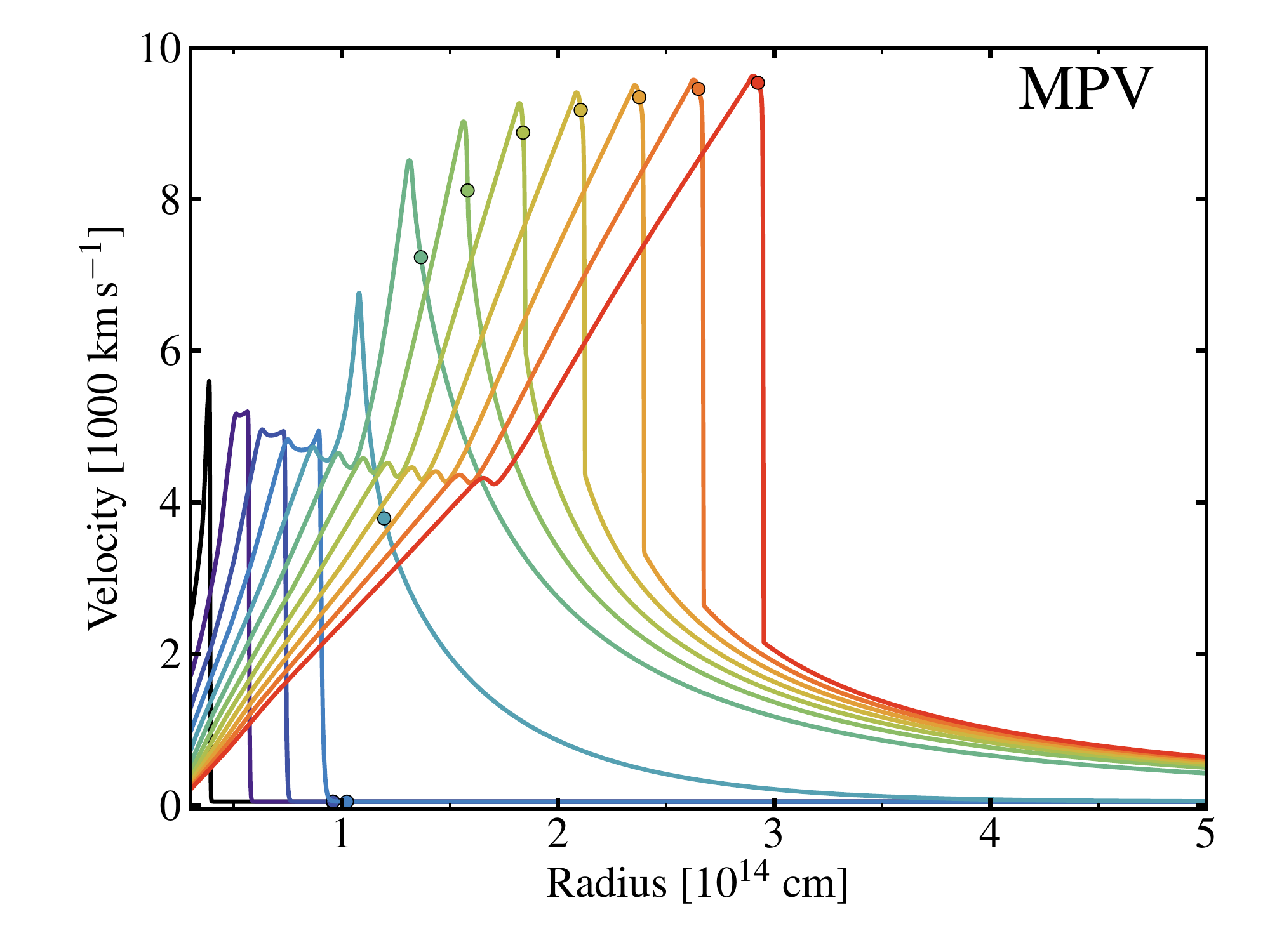,width=9.2cm}
\epsfig{file=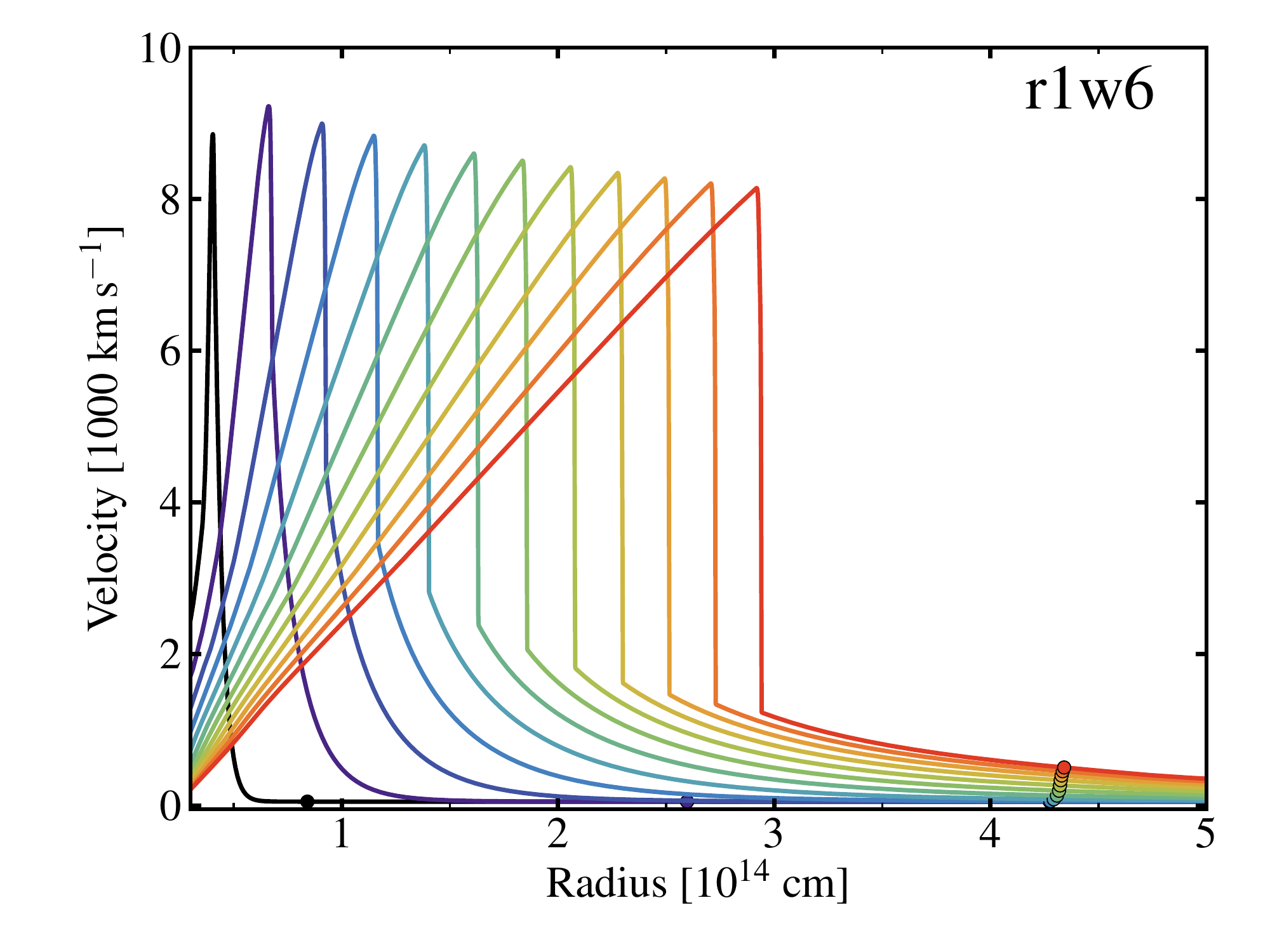,width=9.2cm}
\caption{Left:
Evolution of the local radiative luminosity (top) and velocity (bottom) for
the SN\,2013fs model of \citet[MPV]{morozova_2l_2p_17}.
Right: Same as left, but now for model r1w6 characterized by a dense wind
(but about 50 times less dense).
The filled circle is the location of the photosphere at each epoch. In all panels,
the same set of epochs is shown (from 0.17 until 3.83\,d after the start of the simulation).
In model r1w6, the luminosity maximum occurs well below the photosphere, and
the slow material in between is at the origin of the narrow line profile cores with extended wings.
In the configuration of MPV, this buffer of slow material is absent, there is no radiative precursor,
and the event will show broad lines (with blueshifted emission peaks) immediately after shock breakout.
The model of MPV, despite the interaction, will therefore not look like a SN IIn and will not match the
earliest spectra of SN\,2013fs.
\label{fig_mpv}
}
\end{figure*}

    We find that all our simulations, irrespective of wind density, exhibit
    narrow line profiles at the earliest times. These narrow spectral signatures last as long the
    shock is embedded within some optically-thick slow-moving material. If the wind is tenuous,
    this region is the atmosphere of the RSG. If the wind is dense, this region is the wind itself.
    Ignoring material acceleration, these narrow line profiles last for the duration of the shock breakout
    (i.e., until the shock overtakes the photosphere).

    Our results reproduce some of the characteristic properties observed in SN \,2013fs \citep{yaron_13fs_17}.
    For about a week, SN \,2013fs shows a blue and nearly featureless spectrum.
    Up until  $<$2\,d, the SN exhibits lines from He\two, N\five, O\five, O\six, with a
    narrow and symmetric profile. Subsequently, the SN exhibits a blue nearly featureless spectrum,
    with narrow and weak emission line peaks that lack broad wings.
    This property implies a low electron scattering optical depth at 5\,d (the dense external material
    is overtaken by $\sim$\,2\,d).
    Our model r1w5r is reminiscent of these properties, with a pre-SN mass loss rate of
    5$\times$10$^{-3}$\,\msunyr\ extending out to about 2$\times$10$^{14}$\,cm.
    These values are consistent with those inferred by \citealt{yaron_13fs_17}, although observations
    strongly suggest that the dense wind does not extend to a radius of 10$^{15}$\,cm --- it should
    be no more than about 2$\times$10$^{14}$\,cm, hence a few stellar radii only. Our interpretation
    of the origin of this external material is, however, completely different.

    The term flash-ionization used to explain the early-time properties \citep{khazov_flash_16}
    is somewhat misleading. The shock breakout
    radiation is indeed characterized by very high temperatures and a huge ionizing flux, which promptly
    ionizes all the H and He in the RSG environment. However, in RSG explosions,
    the photospheric temperature and the ionizing flux remain large for at least a week. The recombination
    time scale at the photospheric densities relevant here are of the order of minutes.
    Hence, the large ionization inferred from spectra stems from the sustained UV flux
    and large temperatures in the spectrum formation region, and not exclusively
    from a flash associated with shock breakout.

    The mechanism at the origin of this external material, which is dense and contiguous to the stellar surface,
    is unclear. Drawing an analogy from the
    very massive star eruptions at the origin of super-luminous SNe of Type IIn
    (see, e.g., \citealt{smith_06gy_07}), one may wonder whether RSG may undergo a super-wind phase
    in the final stages of their evolution. Various mechanisms involving fluid instabilities, nuclear burning,
    or pulsations have been proposed
    \citep{heger_rsg_97,shiode_wave_14,smith_arnett_14,yoon_cantiello_rsg_14,WH15}.

   \citet{morozova_2l_2p_17} proposed a model with a strong pre-SN wind mass loss rate
   of 0.15--1.5\,\msunyr\ out to a radius of 1900\,\rsun\ to explain the observations of SN\,2013fs.
   Such a mass loss rate implies densities of the order of 10$^{-10}$\,g\,cm$^{-3}$. When the
   SN shock arrives in this wind, there is in fact no precursor because the photon mean free
   path is too small. In that case, the shock breaks out when it reaches 1900\,\rsun\
   (i.e., their adopted wind merely extends the star) and there
   is no lengthening of the shock-breakout signal. Because of this, this model cannot produce the IIn signatures
   seen for 2\,d in SN\,2013fs. The IIn signatures can only form if there is radiation leakage from
   the shock into the slow-moving atmosphere/wind above, and this cannot occur if its density is
   too high\footnote{If the densities are of the order of 10$^{-10}$--10$^{-9}$\,g\,cm$^{-3}$
   at 10$^{14}$\,cm, the photon mean free path 1/$\kappa \rho$
   is of the order of 10$^{9}$--10$^{10}$\,cm and no extended radiative precursor takes place.}.
   We have done an \heracles\ simulation using the initial conditions for SN\,2013fs
   proposed by Morozova et al. and we illustrate the results together with those of model
   r1w6 in Fig.~\ref{fig_mpv}.
    In the MPV simulation, the photosphere (filled dot) always resides in the fast moving ejecta
    after shock breakout -- there is no reprocessing by a slow moving dense wind.
    Hence, the model of MPV is unable to reproduce the early-time spectra of SN\,2013fs.
    In contrast, in model r1w6, the photosphere is well above the embedded shock for a few days
    after breakout (from within the star) so that photons emitted at the shock are reprocessed
    by the slow-moving optically thick wind.

    In reality, the observations of SN\,2013fs only provide a constraint on the mass and
    extent of the external material ($\sim$\,0.01\,\msun\ spread over $\sim$\,2$\times$10$^{14}$\,cm) .
    There is no observational constraint on the velocity profile and in particular, whether this
    material is static or if it is part of an outflow. These are speculations.
    There may, however, be a simpler explanation for the observations of SN\,2013fs, which
    has to do with the standard properties of RSG atmospheres, and may therefore apply
    with different magnitudes in all Type II SNe.
     Stellar evolution models of RSG stars set the photosphere at a density of about 10$^{-9}$\,g\,cm$^{-3}$.
     This photospheric
     density is very high for such a large radius because of the low surface temperature, making the contribution
     of electron scattering to the opacity very small. In contrast, the base of a typical RSG wind is
     at a density of 10$^{-14}$\,g\,cm$^{-3}$ (e.g., Betelgeuse has an inferred wind mass loss rate of
     about 3$\times$\,10$^{-6}$\,\msunyr; \citealt{harper_betelgeuse_01}). This represents a drop
     in density by 5 orders of magnitude.
     At the surface of the \mesa\ model, the density scale height is about 0.01\,$R_{\star}$. Extrapolated
     down to 10$^{-14}$\,g\,cm$^{-3}$, the base of the wind would be $\sim$\,60\,\rsun\ above $R_{\star}$,
     hence at about 1.1\,$R_{\star}$.
     Observations of RSG atmospheres suggest that the wind is actually launched from further out.
     The direct environment of Betelgeuse has been spatially resolved with adaptive optics or
     interferometry. \citet{kervella_betelgeuse_11} report on the detection of emission in the NIR out
     to large distances (up to 10$R_{\star}$), with the presence of inhomogeneities (surface brightness
     variations) or clumps.
     \citet{ohnaka_betelgeuse_11} discussed the dynamics of this inhomogeneous atmosphere
     and reports on the presence of upflows and downflows at the 10\,\kms\ level out to several $R_{\star}$.
     Similar conclusions are made by \citet{josselin_rsg_07}.
     These observations suggest that the wind is driven from larger radii, and implies that
     there is an extended intermediate region where material moves up and down on a year timescale,
     and fed somehow from $R_{\star}$, perhaps by pulsations or convective motions.
     The observations of SN\,2013fs may suggest that this buffer region of a few $R_{\star}$
     may actually contain 0.01--0.1\,\msun\ of material.
     This material is optically-thin to electron-scattering in the RSG, but becomes optically thick
     when the shock breaks out. Such a small mass is sufficiently optically thick to give rise
     to the narrow symmetric line profiles observed in SN\,2013fs at $\lesssim$\,2\,d.

     Hence, rather than a super wind, the environment inferred from the observations
     of some Type II SNe at early times
     may simply correspond to the cocoon of material that is observed around RSG stars like Betelgeuse.
     The shock is simply breaking out into the RSG atmosphere.

\begin{acknowledgements}

    LD thanks ESO-Vitacura for their hospitality.
    This work utilised computing resources of the mesocentre SIGAMM,
    hosted by the Observatoire de la C\^ote d'Azur, Nice, France.

\end{acknowledgements}

\appendix

\section{Dynamical evolution from the \heracles\ simulations}

\begin{figure*}
\epsfig{file=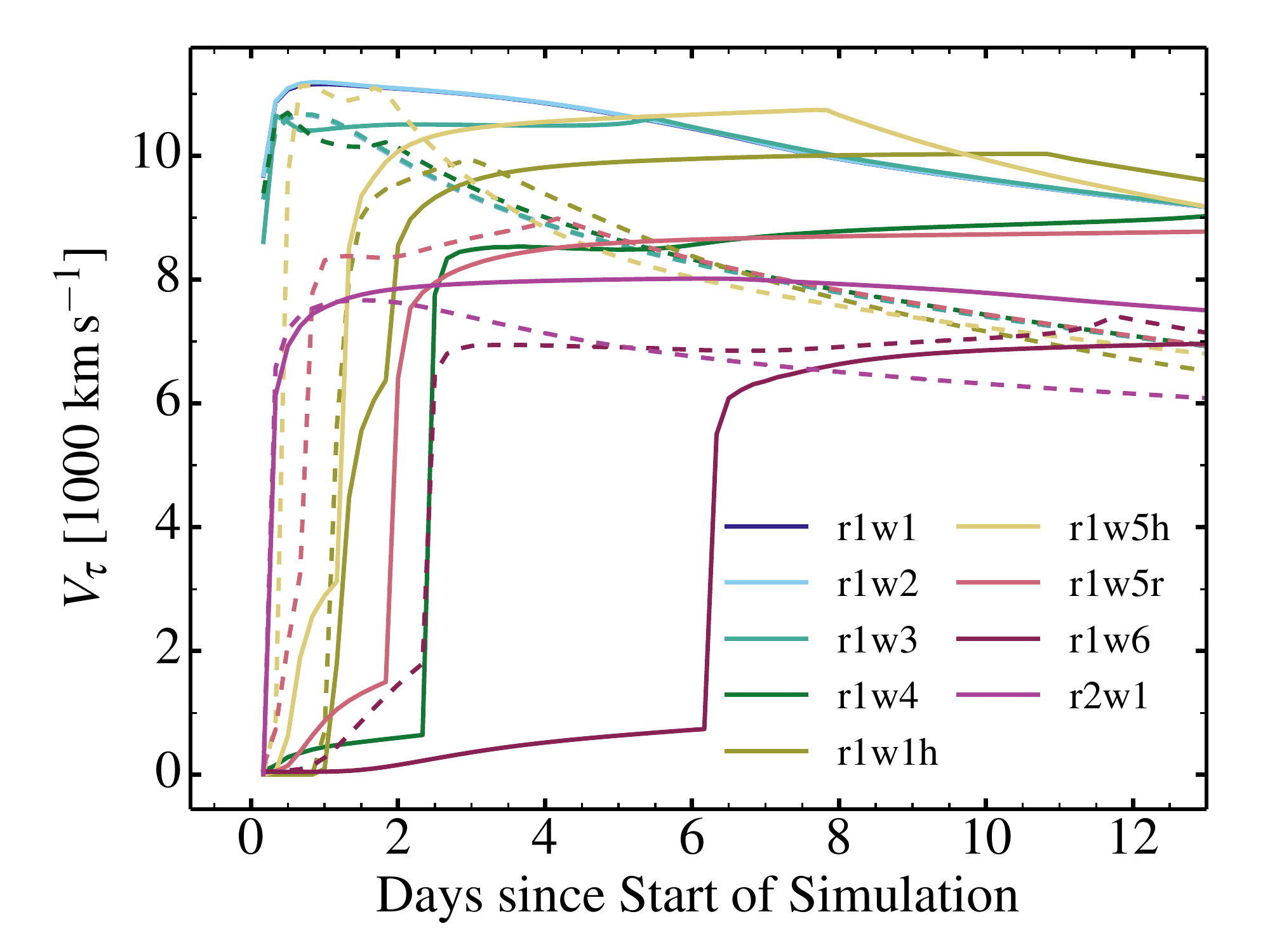,width=9.3cm}
\epsfig{file=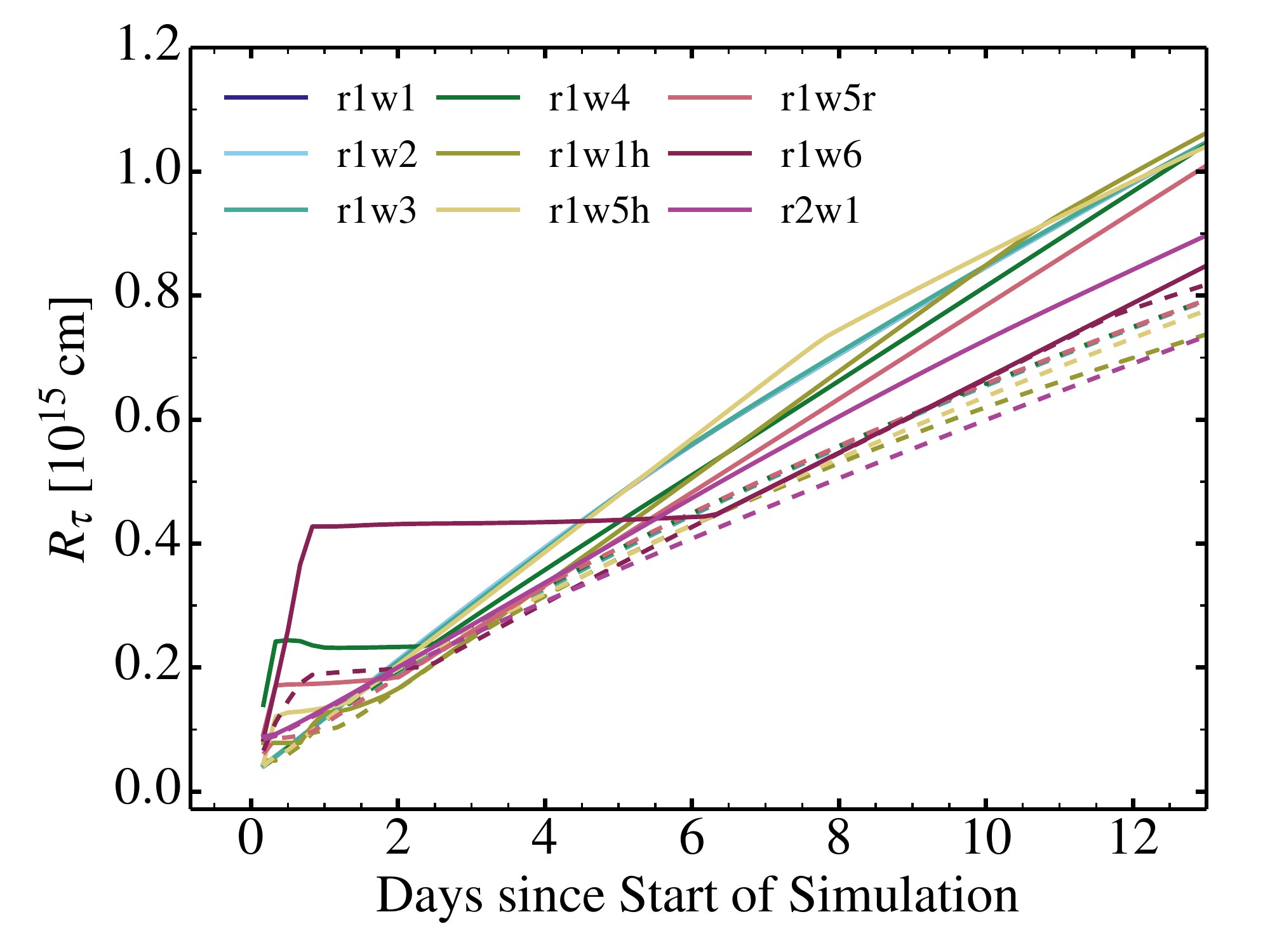,width=9.3cm}
\epsfig{file=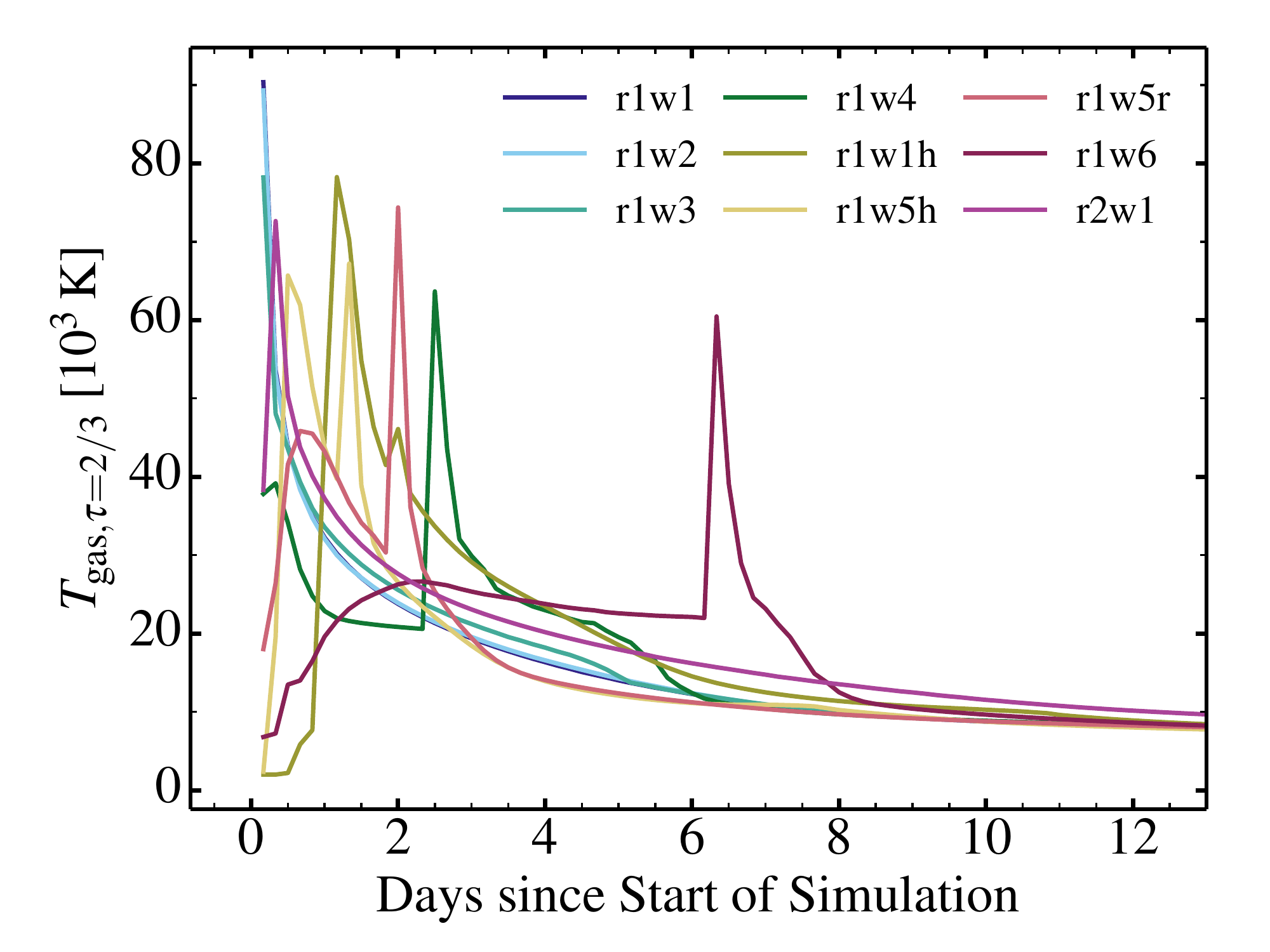,width=9.3cm}
\epsfig{file=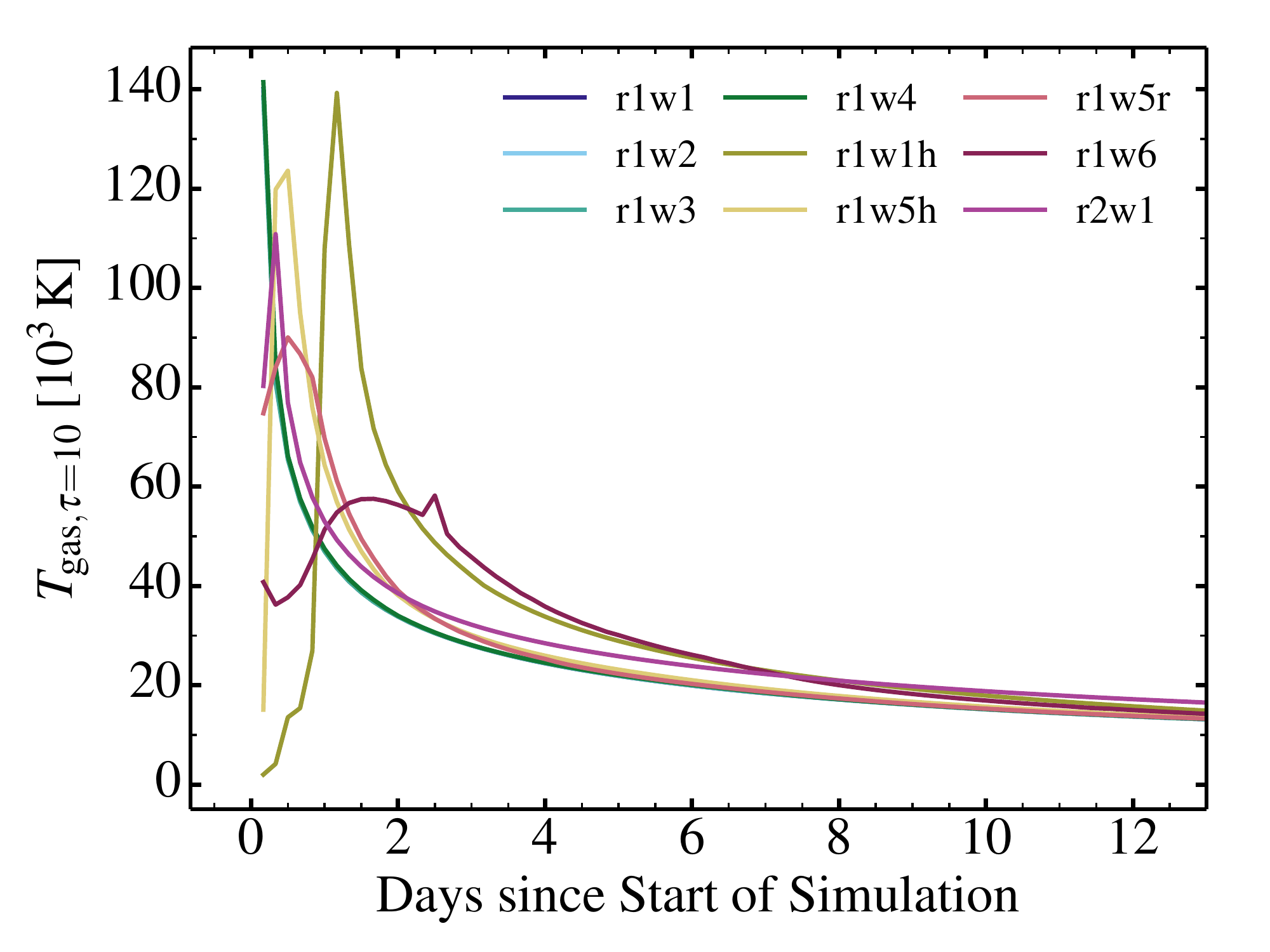,width=9.3cm}
\epsfig{file=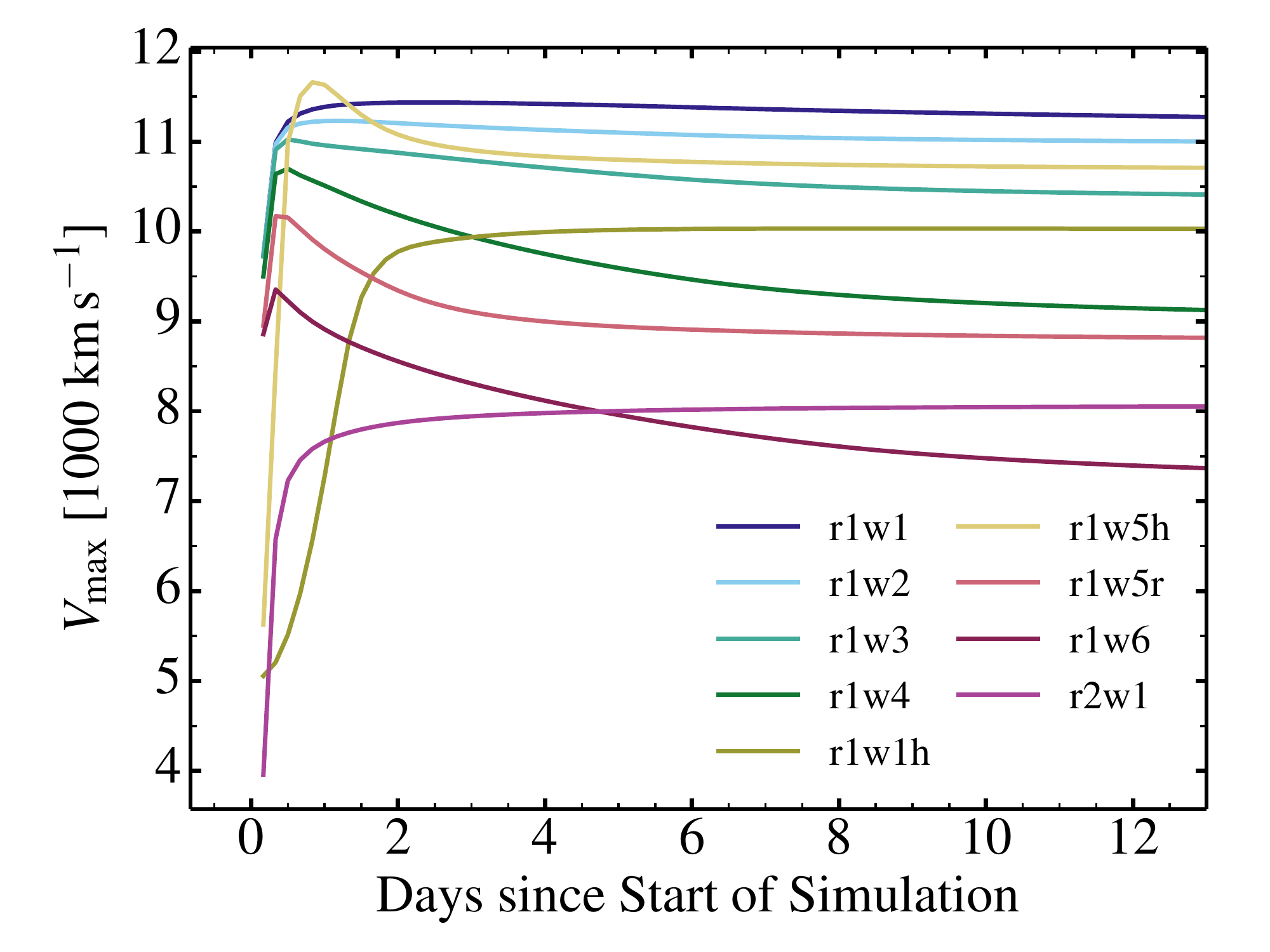,width=9.3cm}
\epsfig{file=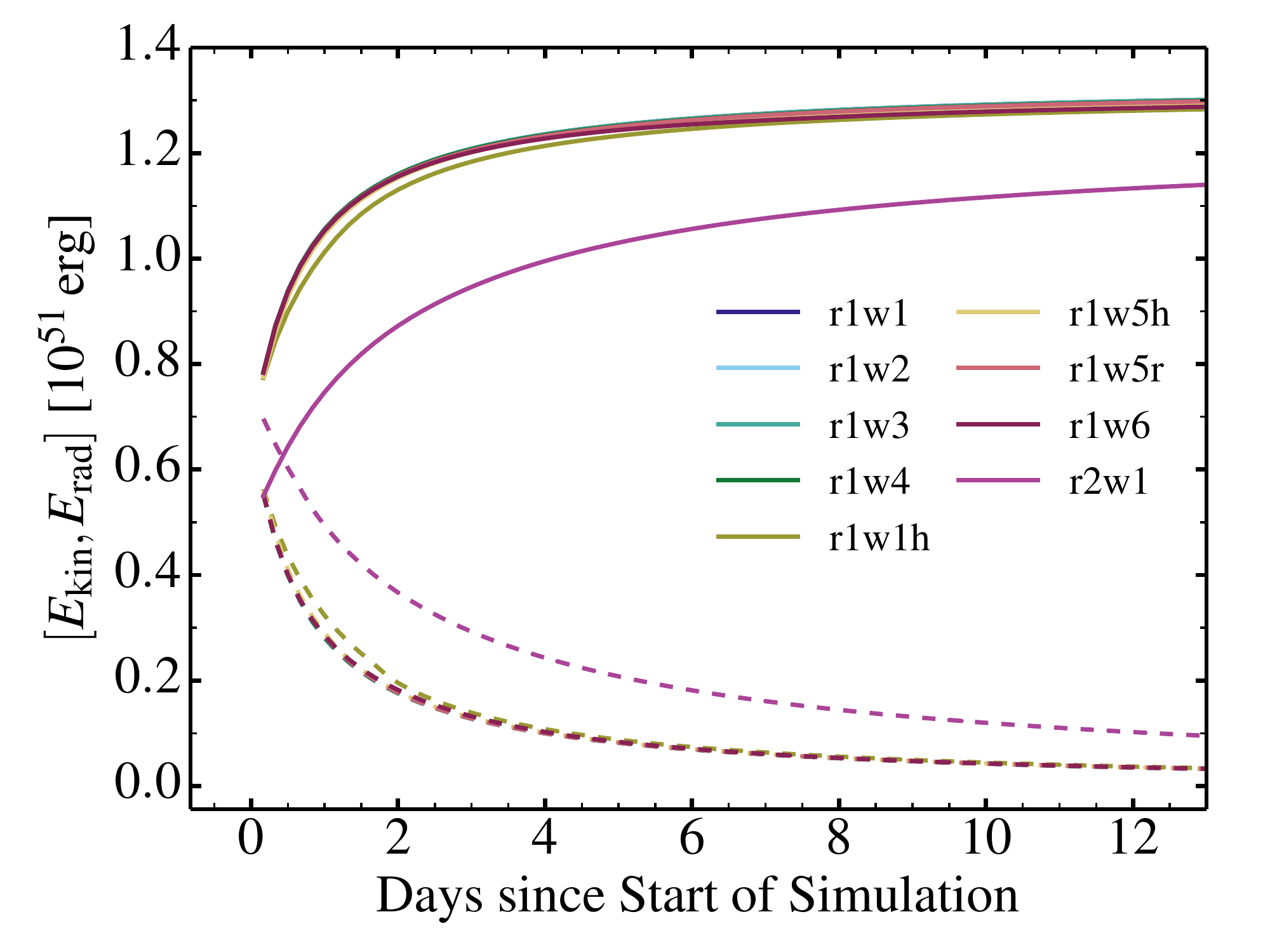,width=9.3cm}
\caption{Evolution of some ejecta properties computed by \heracles\ for our set of models.
In the top row, we show the velocity (left) and the radius (right) at an optical depth of 2/3
(solid line) and 10 (dashed line).
In the middle row, we show the gas temperature at an optical depth of 2/3 (left) and 10 (right).
In the bottom row, the show the evolution of the maximum velocity on the grid (left) and of
the total energy on the grid (kinetic: solid line; radiation: dashed line).
\label{fig_var}
}
\end{figure*}

\section{Bolometric light curves from \cmfgen}

\begin{figure*}
\epsfig{file=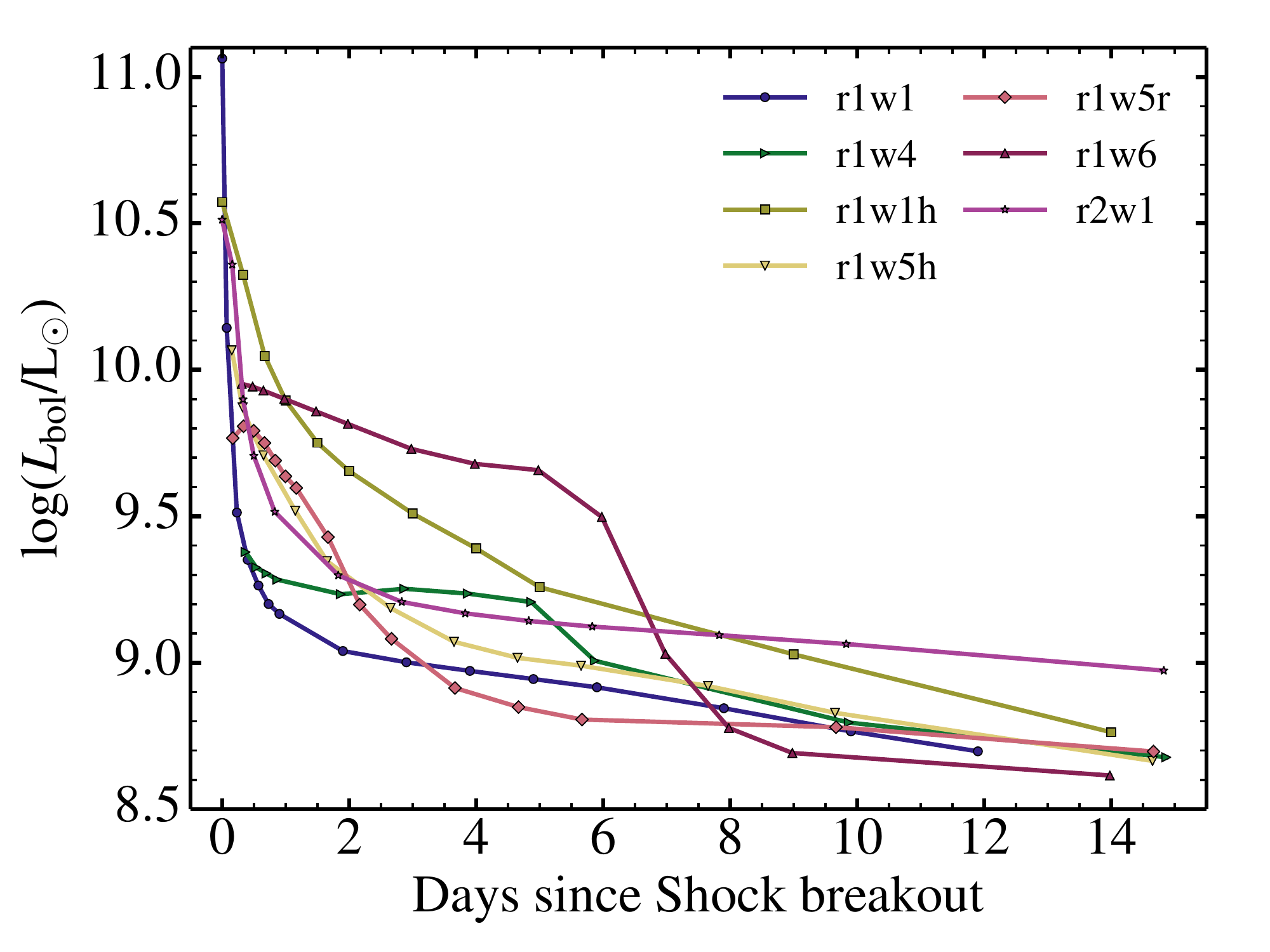,width=16cm}
\caption{
Bolometric light curves obtained with \cmfgen\ for models
r1w1, r1w4, r1w1h, r1w5h, r1w5r, r1w6, and r2w1.
The time origin corresponds to shock breakout.
The \cmfgen\ models are computed adopting the radius, the velocity,
the density, and the gas temperature from the \heracles\ simulations at each
epoch. The luminosity offset between the \heracles\ and the \cmfgen\ simulations
is about 10--20\%, and the resulting light curves are very similar (compare with
Fig.~\ref{fig_lbol_heracles}). There is a conceptual difference between the two codes
since \heracles\ takes explicit account of time dependence while \cmfgen\ computes
the spectrum in steady state but adopting the time-dependent structure from \heracles).
\heracles\  treats the gas in LTE while \cmfgen\ uses a non-LTE approach.
\label{fig_lbol_cmfgen}
}
\end{figure*}

\section{Montage of \cmfgen\ spectra for all epochs and for all models}

\begin{figure*}
\epsfig{file=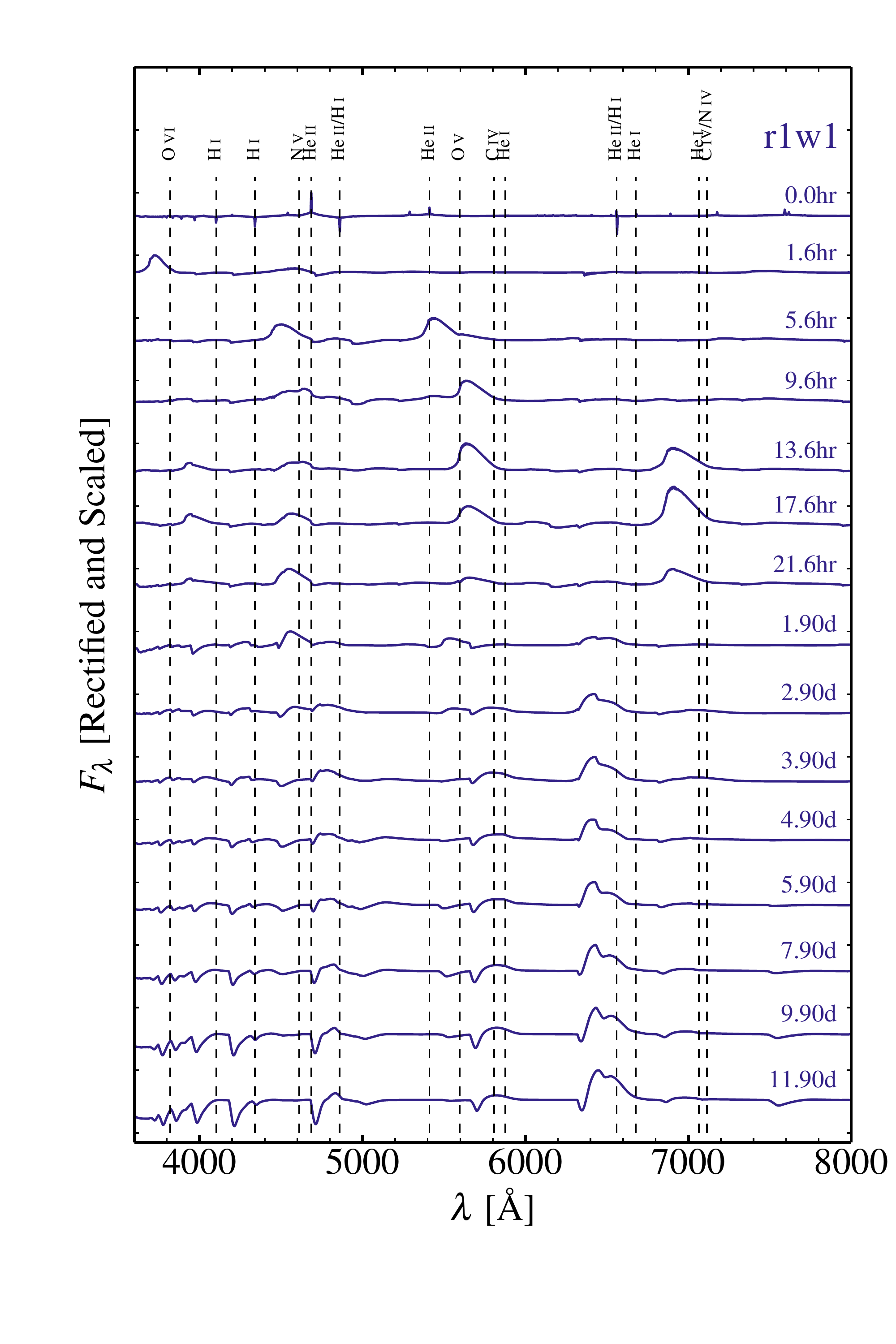,width=17cm}
\vspace{-2cm}
\caption{Montage of spectral for model r1w1 shown at all epochs computed.
The time origin corresponds to shock breakout. The spectra are rectified and
scaled (and finally stacked vertically).
For the spectra, we actually show the quantity
[$(F_\lambda/F_{\rm c} -1) \times \alpha + 1]$, where $F_{\rm c}$ is the continuum flux.
To better show the weak lines, we take $\alpha=$\,3.
\label{fig_spec_r1w1_all}
}
\end{figure*}

\begin{figure*}
\epsfig{file=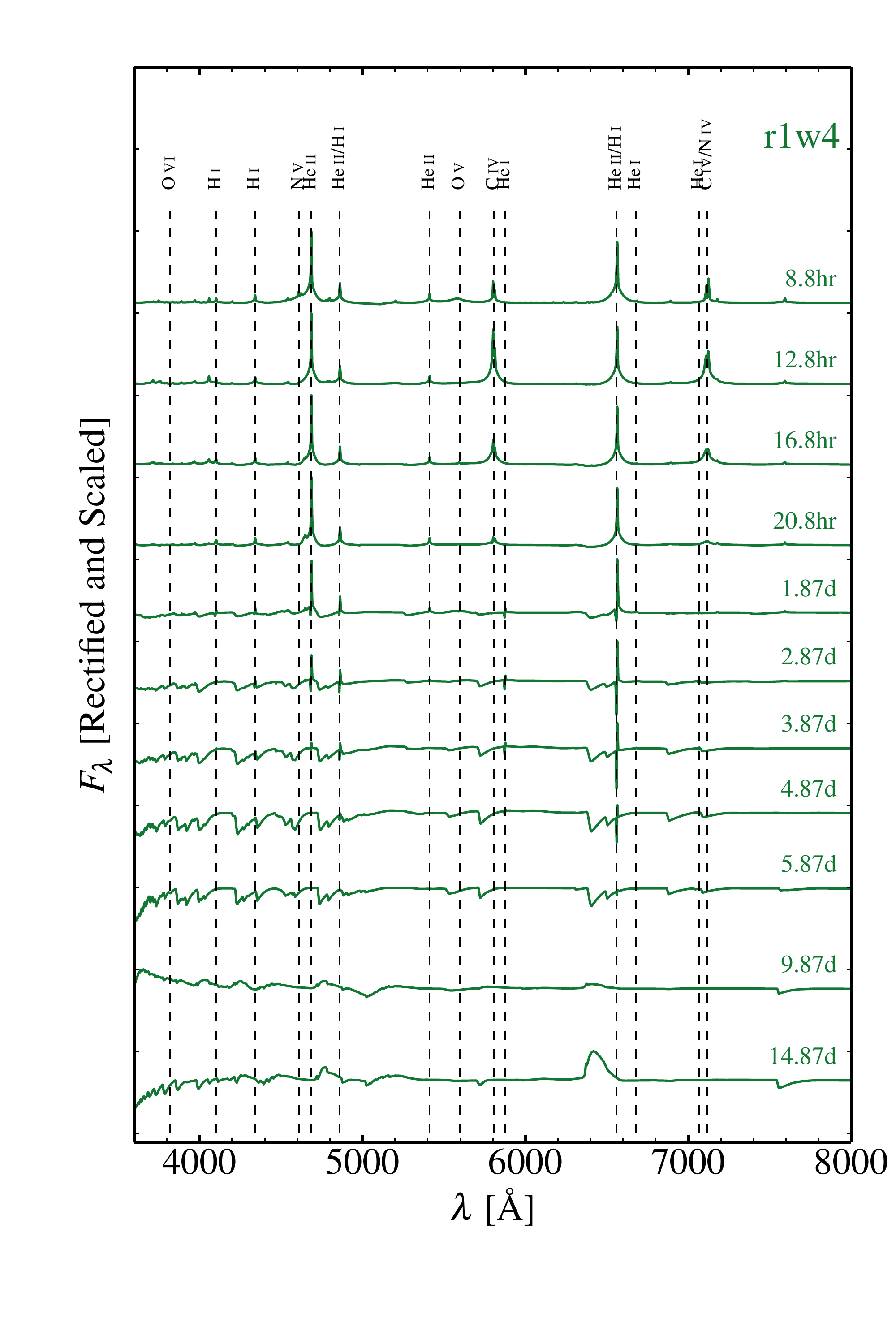,width=17cm}
\vspace{-2cm}
\caption{Same as Fig.~\ref{fig_spec_r1w1_all} but now for model r1w4
\label{fig_spec_r1w4_all}
}
\end{figure*}

\begin{figure*}
\epsfig{file=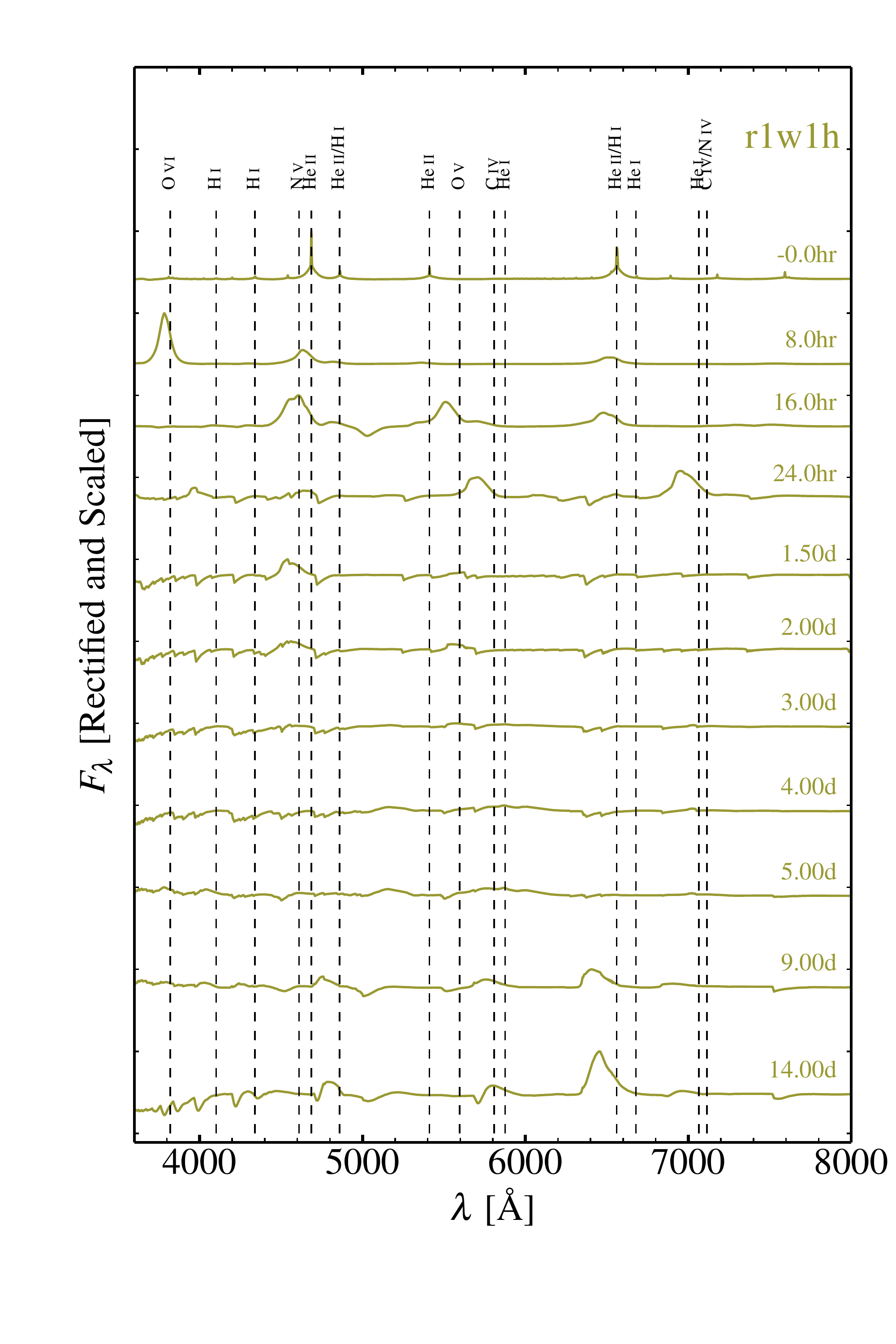,width=17cm}
\vspace{-2cm}
\caption{Same as Fig.~\ref{fig_spec_r1w1_all} but now for model r1w1h
\label{fig_spec_r1w1h_all}
}
\end{figure*}

\begin{figure*}
\epsfig{file=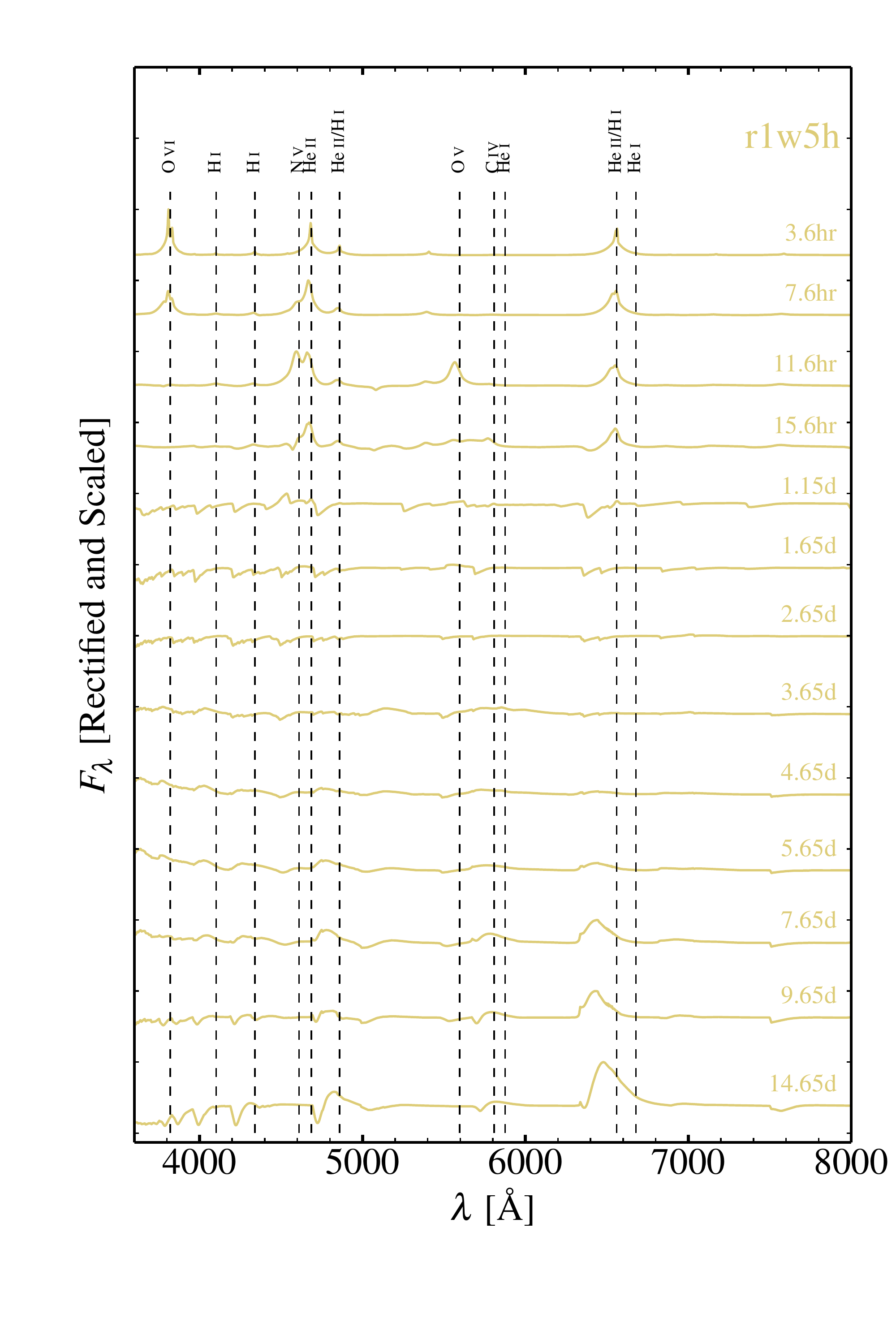,width=17cm}
\vspace{-2cm}
\caption{Same as Fig.~\ref{fig_spec_r1w1_all} but now for model r1w5h
\label{fig_spec_r1w5h_all}
}
\end{figure*}

\begin{figure*}
\epsfig{file=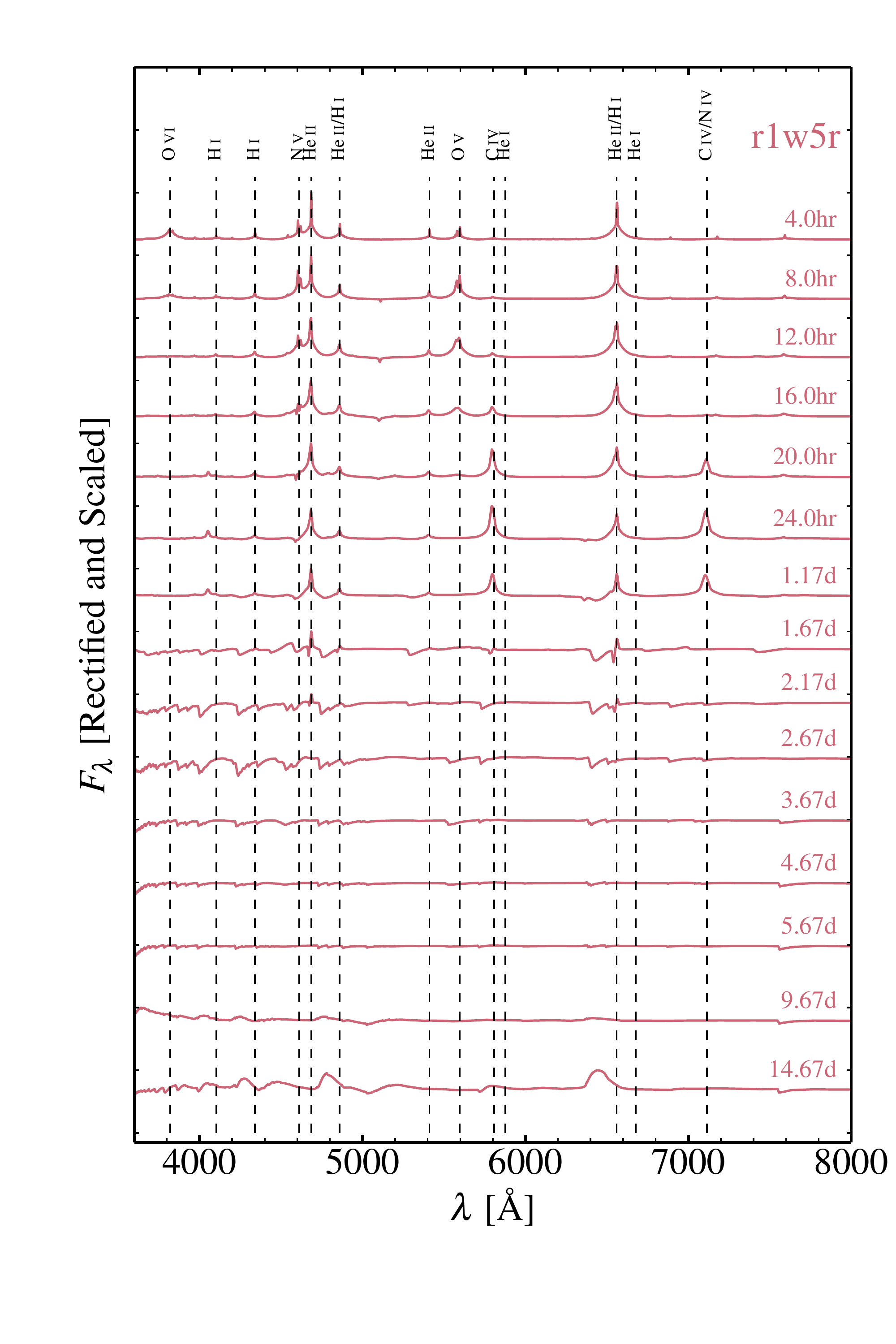,width=17cm}
\vspace{-2cm}
\caption{Same as Fig.~\ref{fig_spec_r1w1_all} but now for model r1w5r
\label{fig_spec_r1w5r_all}
}
\end{figure*}

\begin{figure*}
\epsfig{file=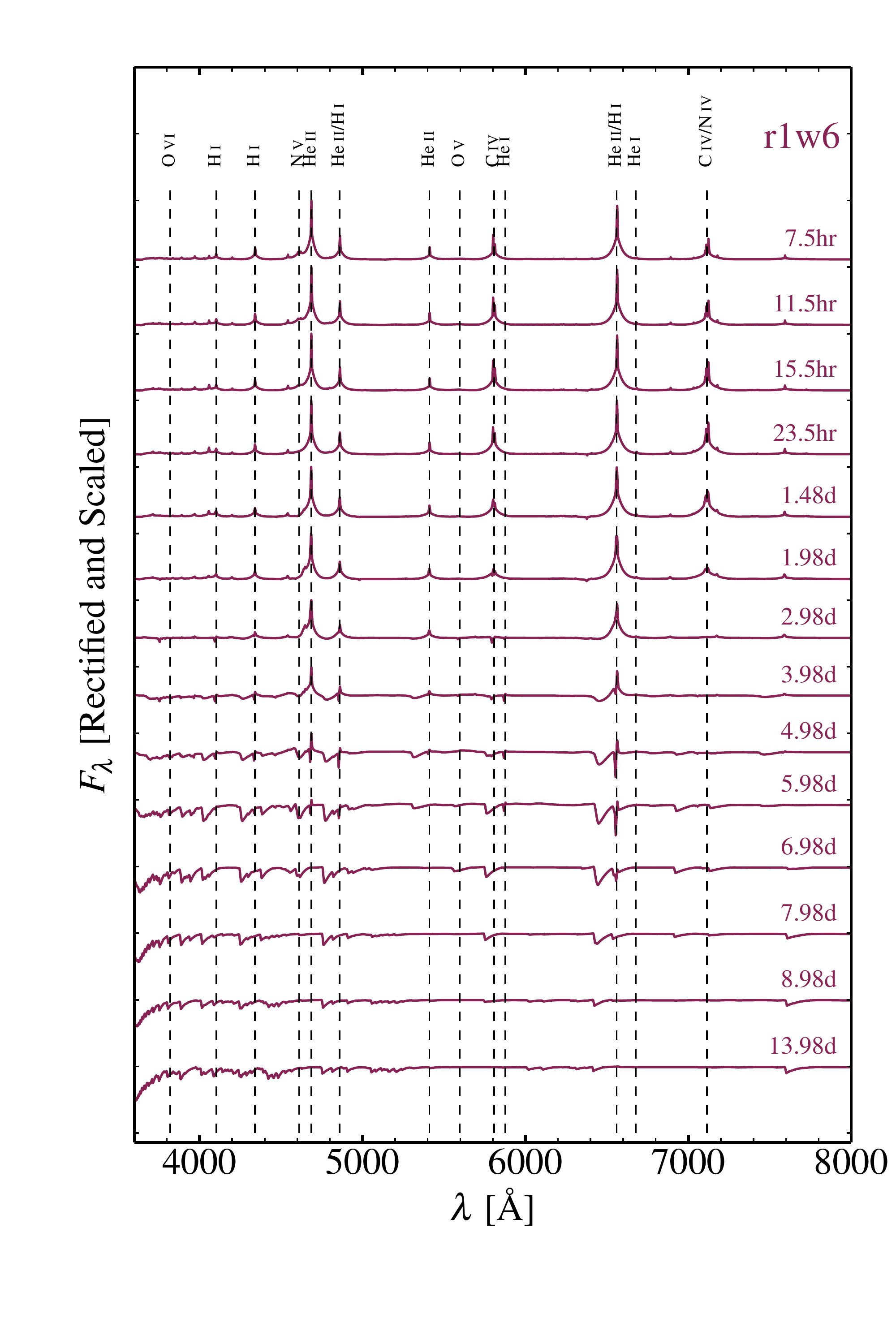,width=17cm}
\vspace{-2cm}
\caption{Same as Fig.~\ref{fig_spec_r1w1_all} but now for model r1w6
\label{fig_spec_r1w6_all}
}
\end{figure*}

\begin{figure*}
\epsfig{file=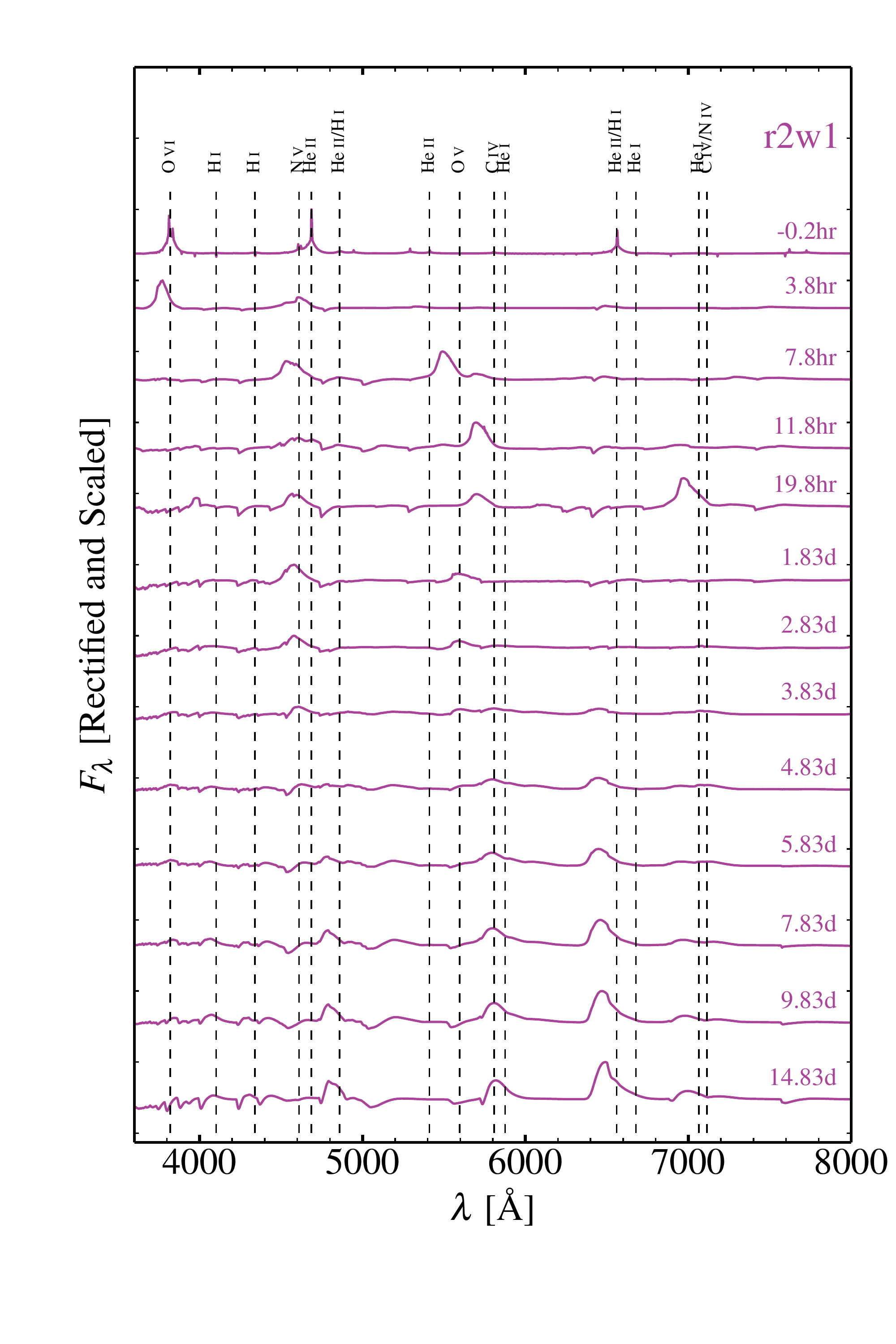,width=17cm}
\vspace{-2cm}
\caption{Same as Fig.~\ref{fig_spec_r1w1_all} but now for model r2w1
\label{fig_spec_r2w1_all}
}
\end{figure*}

\end{document}